\documentclass[
]{ceurart}

\usepackage[margin=25mm]{geometry}

\usepackage{graphicx}  %%% for including graphics
\usepackage{url}       %%% for including URLs
\usepackage{natbib}
\usepackage{xspace}
\usepackage{amsmath,amssymb}
\usepackage{enumerate}
\usepackage[defblank]{paralist}
\usepackage{booktabs}
\usepackage{array}
\newcolumntype{P}[1]{>{\centering\arraybackslash}p{#1}}
\usepackage{marginnote}
\usepackage{color}
\usepackage{stmaryrd}
\usepackage{soul}
\usepackage[framemethod=tikz]{mdframed}

\usepackage{listings}
\usepackage{tikz}
\usetikzlibrary{arrows, fit, calc}
\usetikzlibrary{arrows.meta}
\usetikzlibrary{shapes,shadows}

\newif\ifdraft
\drafttrue
\draftfalse

\newcommand{\perms}[2][]{\mathcal{S}#1(#2#1)}
\newcommand{\asbag}[1]{\mathsf{bag}(#1)}

\newcommand{\idle}[1]{\mathsf{idle}(#1)}

\newcommand{\rightsplit}[1]{#1^{\rightarrow}}
\newcommand{\mversion}{\ensuremath{8.0.12}\xspace}

\ifdraft
  \newcommand{\nb}[1]{\textcolor{red}{\bf!}%
  \marginpar[\parbox{\marginparwidth}{\raggedleft\scriptsize\textcolor{red}{#1}}]%
   {\parbox{\marginparwidth}{\raggedright\scriptsize\textcolor{red}{#1}}}}
\else
  \newcommand{\nb}[1]{}
\fi

\newtheorem{definition}{Definition}

\newtheorem{example}{Example}

\makeatletter
\newcommand{\nobrackettag}[0]{\def\tagform@##1{\maketag@@@{##1}}}
\makeatother

\newcommand{\ignore}[1]{}

\definecolor{mintgreen}{HTML}{ADEBB3}

\mdfdefinestyle{deltamongo}{%
  backgroundcolor=mintgreen!40, %
  linecolor=gray!50,%
  linewidth=1pt, %
  shadow=true, %
  shadowsize=5pt, %
  leftmargin=1cm, %
  innertopmargin=10pt, %
  innerbottommargin=10pt, %
  innerleftmargin=10pt, %
  innerrightmargin=10pt, %
  roundcorner=8pt, %
}
\newcommand{\deltabox}[1]{%
  \begin{mdframed}[style=deltamongo]
    \textbf{Delta with MongoDB's implementation (version \mversion).}
    \small #1
  \end{mdframed}
}

\mdfdefinestyle{core}{%
  backgroundcolor=cyan!10, %
  linecolor=gray!50,%
  linewidth=1pt, %
  shadow=true, %
  shadowsize=5pt, %
  leftmargin=1cm, %
  innertopmargin=10pt, %
  innerbottommargin=10pt, %
  innerleftmargin=10pt, %
  innerrightmargin=10pt, %
  roundcorner=8pt, %
}

\mdfdefinestyle{abstractmquery}{%
  backgroundcolor=SkyBlue!20, %
  linecolor=SkyBlue!20,%
  linewidth=1pt, %
  innerleftmargin=5pt, %
  innerrightmargin=5pt, %
}
\newcommand{\mquerybox}[1]{%
  \begin{mdframed}[style=abstractmquery]
    \small #1
  \end{mdframed}
}

\mdfdefinestyle{raquery}{%
  backgroundcolor=RedOrange!20, %
  linecolor=RedOrange!20,%
  linewidth=1pt, %
  innerleftmargin=5pt, %
  innerrightmargin=5pt, %
}
\newcommand{\rabox}[1]{%
  \begin{mdframed}[style=raquery]
    \small #1
  \end{mdframed}
}

\newcommand{\A}{\mathcal{A}}

\newcommand{\C}{\mathcal{C}}

\newcommand{\K}{\mathcal{K}}
\renewcommand{\L}{\mathcal{L}}

\renewcommand{\O}{\mathcal{O}}
\renewcommand{\P}{\mathcal{P}}

\renewcommand{\S}{\mathcal{S}}

\newcommand{\V}{\mathcal{V}}

\newcommand{\X}{\mathcal{X}}

\newcommand{\dvalue}[1][]{d-value{#1}\xspace}
\newcommand{\dvalues}{\dvalue{s}\xspace}

\newcommand{\bag}[2][]{#1\{\!\!#1\{#2#1\}\!\!#1\}}

\newcommand{\lobject}{\langle}%{\{\!\!\{}
\newcommand{\robject}{\rangle}%{\}\!\!\}}
\newcommand{\objectbr}[2][]{\ensuremath{#1\lobject #2#1\robject}}
\newcommand{\larray}{\lfloor}
\newcommand{\rarray}{\rfloor}
\newcommand{\arraybr}[2][]{#1\larray\!\,#2#1\rarray}

\newcommand\cond[3]{{#1}?\,{#2}{:}\,{#3}}
\newcommand\mapz{\mathsf{map}}
\newcommand\map[3]{\mapz({#1},\: \lambda {#2}\colon{#3})}
\newcommand\filterz{\mathsf{filter}}
\newcommand\filter[3]{\filterz({#1},\: \lambda {#2}\colon{#3})}

\newcommand{\nullvalue}{\ensuremath{\mathbf{null}}\xspace}
\newcommand{\truevalue}{\ensuremath{\mathbf{true}}\xspace}
\newcommand{\falsevalue}{\ensuremath{\mathbf{false}}\xspace}

\newcommand{\eval}[2][o]{\llbracket#2\rrbracket^{#1}}

\newcommand{\evald}[2][o]{\llbracket#2\rrbracket^{#1}}

\newcommand{\evalq}[2][I]{{#2}^{#1}}

\newcommand{\evalc}[2][I]{{#2}^{#1}}
\newcommand{\op}{\mathop{\mathsf{op}}}

\newcommand{\pipeline}{\triangleright}
\newcommand{\q}{\boldsymbol{q}}

\newcommand{\valuefont}[1]{{\small\textup{\texttt{#1}}}}

\newcommand{\id}{\valuefont{\_\!\!\;id}\xspace}

\newcommand{\DEF}{\textcolor{red!30!black}{::=}}

\newcommand{\match}[1]{\mathsf{ma}_{#1}}
\newcommand{\unwind}[1]{\mathsf{uw}_{#1}}
\newcommand{\project}[1]{\mathsf{pr}_{#1}}
\newcommand{\group}[2]{\mathsf{gr}_{#1\text{\bf\,:\,}#2}}
\newcommand{\lookup}[2]{\mathsf{lo}^{#1}_{#2}}
\newcommand{\glookup}[2]{\mathsf{gl}^{#1}_{#2}}
\newcommand{\unionwith}[1]{\mathsf{un}_{#1}}

\newcommand{\sort}[1]{\mathsf{so}_{#1}}
\newcommand{\limit}[1]{\mathsf{li}_{#1}}
\newcommand{\skipstage}[1][m]{\mathsf{sk}_{#1}}
\newcommand{\countstage}[1][k]{\mathsf{co}_{#1}}

\newcommand{\mongodb}{MongoDB\xspace}

\newcommand{\mquery}{\text{MQuery}\xspace}

\newcommand{\mqueries}{\text{MQueries}\xspace}
\newcommand{\te}[1]{\text{ #1 }}

\newcommand{\nn}{\mathbb{N}}

\newcommand{\ei}{\emph{(i)~}}
\newcommand{\eii}{\emph{(ii)~}}
\newcommand{\eiii}{\emph{(iii)~}}
\newcommand{\eiv}{\emph{(iv)~}}

\newcommand{\aggz}{\mathsf{agg}}
\newcommand{\agg}[3][]{\aggz_{#2}#1(#3#1)}
\newcommand{\groups}[2]{\mathsf{group}_{#1}(#2)}

\newcommand{\ground}[2]{\mathsf{ground}_{#1}(#2)}

\newcommand{\sorted}[1]{\mathsf{sorted}(#1)}
\newcommand{\evalcdb}[2][O,\,I]{ ({#1}) \blacktriangleright {#2}}
\newcommand{\evalcdbl}[2][a,\,I]{ ({#1}) \mathop{\arraybr{\blacktriangleright}} {#2}}

\newcommand{\paths}[1]{\mathsf{paths}(#1)}
\newcommand{\prefixes}[1]{\mathsf{prefixes}(#1)}
\newcommand{\extensions}[1]{\mathsf{extensions}(#1)}

\begin{document}
%%
%% Rights management information.
%% CC-BY is default license.
\copyrightyear{2025}
\copyrightclause{Copyright for this paper by its authors.
  Use permitted under Creative Commons License Attribution 4.0
  International (CC BY 4.0).}

%%
%% This command is for the conference information
\conference{Technical Report}

%%
%% The "title" command
\title{Towards a Standard for JSON Document Databases}

%%
%% The "author" command and its associated commands are used to define
%% the authors and their affiliations.
\author[1]{Elena Botoeva}[%
orcid=0000-0001-5881-0258,
email=e.botoeva@kent.ac.uk,
]
\address[1]{University of Kent, Canterbury, UK}
\author[2]{Julien Corman}[%
%orcid=0000-0003-1097-2965,
email=corman@inf.unibz.it,
]
\address[2]{Free University of Bozen-Bolzano, Bolzano, Italy}
\author[3]{Norman Townsend}[%
email=ntownsen@amazon.com,
]
\address[3]{Amazon Web Services, Toronto, Canada}

\maketitle

\begin{abstract}
  In this technical report, we present a formalisation of the MongoDB aggregation framework. Our aim is to identify a fragment that could serve as the starting point for an industry-wide standard for querying JSON document databases. We provide a syntax and formal semantics for a set of selected operators, We show how this fragment relates to known relational query languages. We explain how our semantics differs from the current implementation of MongoDB, and justify our choices. We provide a set of algebraic transformations that can be used for query optimisation.
\end{abstract}


\section{Introduction}
\label{sec:introduction}

JavaScript Object Notation (JSON) is the leading format for data exchange
and representation on the Web (notably for Web APIs), 
but is also 
% and not only: 
% it is the standard format for Web
% APIs, 
widely used for storing application data, and, importantly for our concern,
adopted as a data model for % popular
document databases such as MongoDB~\cite{BradshawBC19},
Couchbase~\cite{Brown12}, ArangoDB, Amazon DocumentDB (with MongoDB Compatibility),
 Microsoft's Azure Cosmos DB For MongoDB, and The Linux Foundation's DocumentDB.
JSON~\cite{bourhis17json} organizes data into human- and machine-readable textual documents composed of key-value pairs and arrays, and
is commonly considered a lightweight alternative to XML for data exchange and storage.

An example of a JSON document
is shown in
Figure~\ref{fig:json-document}.
It represents information about the 
%rock 
band
Queen, using key-value pairs (such as name and year of formation) and arrays
(such as their albums and band members).
\begin{figure}[h]
  \centering
\begin{lstlisting}
{   "_id": 2,
    "name": "Queen",
    "formation": 1970,
    "albums": [
        { "title": "Queen", "release": 1973 },
        { "title": "A Night at the Opera", "release": 1975, "length": "43:08" },
        { "title": "News of the World", "release": 1977, "labels": ["EMI", "Elektra"] }
    ],
    "members": [
        { "name": "Freddie Mercury", "role": ["lead vocals", "piano"] },
        { "name": "Brian May", "role": ["guitar", "vocals"] },
        { "name": "Roger Taylor", "role": ["drums", "vocals"] },
        { "name": "John Deacon", "role": "bass" }
    ]
}
\end{lstlisting}
% \begin{lstlisting}
% { "_id": 4,
%   "awards": [
%     { "award": "Rosing Prize", "year": 1999, "by": "Norwegian Data Association" },
%     { "award": "Turing Award", "year": 2001, "by": "ACM" },
%     { "award": "IEEE John von Neumann Medal", "year": 2001, "by": "IEEE" } ],
%   "birth": "1926-08-27",
%   "contribs": [ "OOP", "Simula" ],
%   "death": "2002-08-10",
%   "name": { "first": "Kristen", "last": "Nygaard" } }
% \end{lstlisting}
  \caption{A JSON document about the band Queen.}
  \label{fig:json-document}
\end{figure}

JSON document databases belong to the \emph{NoSQL} family of database
management systems~\cite{DavoudianCL18,MKMK21}, which relax the strict requirement of
the data complying with a predefined schema.
In contrast to XML, which is supported by the standardised XQuery
language~\cite{Koch06}, the JSON ecosystem lacks a universally adopted query
standard.
Several proposals have emerged--such as JMESPath\footnote{\url{https://jmespath.org/}}, or the JSON-supporting extensions of XPath\footnote{\url{https://www.w3.org/TR/xpath-31/}} and XQuery\footnote{\url{https://www.w3.org/TR/xquery-31/}}--but these have seen little traction within the JSON database community. Vendor-specific solutions have also appeared: N1QL by Couchbase and AQL by ArangoDB, both SQL-inspired languages adapted to handle the nested structure of JSON documents (e.g., nested arrays). However, these languages remain confined to their respective platforms.
The notable exception is the \emph{\mongodb aggregation
  framework}\footnote{\url{https://www.mongodb.com/docs/manual/core/aggregation-pipeline/}},
a widely adopted query language for collections of
JSON documents. Owing to its expressive pipeline model,
% \textcolor{red}{broad ecosystem support},
large developer base, and popularity among users, it has effectively become a
de-facto standard for querying JSON. The fact that major vendors--including
Amazon, Microsoft, Oracle (Autonomous JSON), and Google (Firestore)--as well
as smaller players such as FerretDB, seek to provide compatibility with MongoDB
API further underscores the recognition of its query model as a common lingua
franca. %, even if not officially standardised.
%
% \nb{NT: This and the subsequent statements are perhaps too strong (at least without a reference to back it up). 
% There are a few different languages out there. I cannot speak to how popular they are. But later versions of XPath and XQuery support JSON.
% Here are some examples: https://jmespath.org/specification.html, https://www.rfc-editor.org/rfc/rfc6901, 
% https://www.w3.org/TR/xquery-31/, https://www.w3.org/TR/xpath-31/.
% Maybe say something like: 
% "Unlike XML which has XQuery as its standard query language, the JSON ecosystem lacks a widely adopted, vendor-neutral standard query language, despite several proposals existing."
% and maybe cite some of the proposals.
% }
%
% In terms of adoption, the \emph{\mongodb aggregation
%   framework}\footnote{\url{https://www.mongodb.com/docs/manual/core/aggregation-pipeline/}}
% is currently the most prominent language providing rich querying capabilities
% over collections of JSON documents.

% As such, the language is very expressive and rich in features, but it has been
% developed without theoretical foundations. This resulted in some
% counter-intuitive behaviours, even in simple cases.
% %
% The example below illustrates one such case.
% \nb{NT: This example feels a bit too complex for the introduction and could confuse readers before they get the basic concepts. Consider moving it to Section 4 
% and replace it with a simpler motivating example that builds on the rock band example before.}
%
% \nb{NT: Maybe this as a motivating example instead?}

Despite the popularity of the MongoDB aggregation framework, currently there is
no agreed vendor-neutral standard that would ensure interoperability and
portability of document data stores, meaning that every vendor has its own
ad-hoc implementation. This situation mirrors the early years of relational
databases, before the emergence of SQL as a unifying standard, when systems
were fragmented and tightly coupled to individual vendors. As a definite sign
that the industry is ready to define such a common standard for JSON document
databases, Microsoft has recently announced an open-source release of their
DocumentDB~\cite{Rameesh25}, explicitly highlighting compatibility with MongoDB
APIs, followed by an announcement that DocumentDB joined the Linux Foundation
as an open source project that would provide ``an open standard for document
based applications''~\cite{Stokes25}.

The creation of a standard, however, is impeded by several factors.
First, the authors have not found evidence of a published theoretical
foundation for the MongoDB aggregation framework.
Its query language is modeled on the flexible notion of a data
processing pipeline, where a query consists of multiple stages, each defining a
specific transformation applied to the set of documents produced by the
previous stage.
While this design is pragmatic and powerful,\nb{E: developer friendly?}
it has not been systematically
formalized in the way relational algebra underpins SQL.
Second, the semantics of MongoDB pipelines is procedural rather than purely
declarative, in contrast to SQL. The result is a rich language that can model
complex data manipulations, including a number of map-reduce-style
transformations and user-defined functions, whose exact semantics is
implementation specific.
%
% For such a language, different vendors may introduce subtle variations in
% operator semantics.
%
% This makes achieving consistent behaviour across implementations more challenging.
%
Third, MongoDB's
documentation\footnote{\url{https://www.mongodb.com/docs/manual/}}, while
providing a comprehensive description of syntax, omits crucial semantic
details, which may lead to ambiguity when interpreting various constructs.
We illustrate this on an example.
\begin{example}\em
  \label{ex:semantic-complexity}
  Assume that we have a collection of documents about music bands, similar to
  the one in Figure~\ref{fig:json-document}, and we want to find all bands
  formed before 1975. In MongoDB, this can be achieved with a single match
  stage that conditions the values of the \valuefont{formation} field:
\begin{lstlisting}
db.bands.aggregate([
    { $match: { "formation": { $lt: 1975 } } }
])
\end{lstlisting}

Syntactically, this query is very simple. Semantically, it may also seem
straightforward to compute its output, especially for collections of documents
where the value of \valuefont{formation} is a number. However, considering that
JSON databases are schemaless, what if this value is a string or, more
generally, a complex value? What is the expected semantics?

We can take this further.  What if want to find all bands that released an
album before 1975? By analogy with the query above, MongoDB allows to write the
following query:
\begin{lstlisting}
db.bands.aggregate([
    { $match: { "albums.release": { $lt: 1975 } } }
])
\end{lstlisting}
Now, what does the path \valuefont{albums.release} refer to in a document? A
possible value would be an array. In that case, what is the result of comparing
an array to a number? Should all the values in the array satisfy the comparison
or at least one of them?

% These are some of the possible questions that may arise already when dealing
% with such simple queries.
%
%While the syntax of this query appears straightforward,
This example illustrates that the underlying semantics must necessarily take
into account complex interactions between document structure, path evaluation,
and type comparisons.
Unfortunately, these are not formally specified in MongoDB's documentation.
%
% This highlights the need for a
% proper formalisation of the semantics of the aggregation framework.
This lack of formal foundations creates several problems: unpredictable query
behavior in edge cases, difficulty in query optimization, limited
interoperability with other systems, and barriers to creating industry
standards.
\qed
\end{example}

% One can reverse-engineer the semantics through comprehensive testing, or rely
% on forums, such as Stack Overflow or Reddit\footnote{\url{Example of
%     questions?}}.

The challenges mentioned above motivate the need for a rigorous mathematical
framework underpinning MongoDB aggregation queries.
%
% \nb{NT: This transition feels abrupt. Consider some bridging text. Maybe something like:
% "The lack of formal foundations creates several problems: unpredictable query behavior in edge cases, 
% difficulty in query optimization, and barriers to creating industry standards.
% These challenges motivate the need for a rigorous mathematical framework."}
%
Without a canonical underlying formalism, reasoning about query equivalence,
development of query optimisation techniques, and achieving interoperability
becomes difficult, limiting the ability of standardisation bodies to propose a
vendor-independent specification.

The purpose of this report is twofold. First, provide a formalisation of the
\mongodb data model and the aggregation framework. % in a way that is accessible to
% non theoreticians.
Second, identify a fragment of the aggregation framework that could serve as the
starting point for an industry-wide standard for JSON document databases.
Our approach is inspired by the foundational work of Edgar Codd, whose
relational algebra provided the theoretical underpinning for SQL. Just as
relational algebra enabled rigorous reasoning about SQL queries, optimization,
and correctness, we aim to provide similar foundations for JSON document
databases.

We present in this report an abstract syntax for the MongoDB aggregation framework,
define a formal semantics for it,
and discuss how its expressive power relates to more traditional query languages, notably the ones underpinning SQL.
In some places, our semantics slightly differs from the one implemented in \mongodb.
To explain and motivate these differences,
we provide examples that justify it.
%highlight why the behaviour of \mongodb may be counter-intuitive.
%
Our formalisation is largely based on an article published at ICDT \cite{BCCX18}.
We extend it in several directions, 
notably to
\ei include more pipeline operators,
\eii relax the assumption that the JSON documents stored in the database comply to a predefined schema,
\eiii allow  objects that are either ordered or unordered sets of key-value pairs.
%well-typedness, 
%in line with the schemaless nature of JSON document DBs\nb{E: finish here}

The rest of this document is structured as follows.
In Section~\ref{sec:data-model} we present a formalisation of the JSON data model. 
In Section~\ref{sec:mquery} we present an abstraction of the MongoDB aggregation framework query language,
which we call \emph{MQuery}.
It comes equipped with an algebra-like syntax and a well-defined semantics over collections of JSON objects.
Then in Section~\ref{sec:differences}, 
we illustrate the few cases where this semantics differs from the one implemented 
within MongoDB.
Section~\ref{sec:expressivity} focuses on the expressivity of this language:
we explain how (fragments of) MQuery correspond to fragments or extensions of 
%in an accessible way 
% the correspondence between 
relational algebra,
which is the formalism underpinning SQL-like query languages.
% fragments and
% supersets.
In Section~\ref{sec:optimisation} we present a number of algebraic
equivalences between MQueries, which could be used for query optimisation.

%%% Local Variables:
%%% mode: latex
%%% TeX-master: "main-tr"
%%% fill-column: 79
%%% End:

\section{Data Model}
\label{sec:data-model}
In this section, we present a data model for database management systems that
store collections of JSON documents.
We start by introducing in Section~\ref{sec:dvalues} a more abstract syntax for
JSON values that is better-suited for a theoretical analysis.
Then, in Section~\ref{sec:path-evaluation} we define how a \emph{path}--a
sequence of \emph{keys}--is associated with a value inside a JSON object, and
in Section~\ref{sec:unordered-interpretation} we briefly mention an
alternative--unordered--semantics for JSON objects.

Before defining our model,
we briefly summarize how it differs from the 
% highlight the differences between our data model and the
MongoDB implementation.
First, our formalisation abstracts some low-level typing details of JSON
values, essential in practice, but less relevant for the more foundational
perspective of this technical report.
More precisely,
MongoDB serialises JSON documents in the so-called BSON (for binary JSON) format for
efficient storage and retrieval.
%
%In BSON, each scalar value is annotated with its datatype\nb{Why scalar? Every value?}.
%
%The primitive datatypes that are supported in BSON include standard ones such as
BSON supports various datatypes, where the primitive ones include
double, string, boolean, %but also additional ones such as
date, and timestamp, and the complex ones include object, array, JavaScript code and binary data.\nb{E: binary data was a primitive datatype}
Instead, in our simplified data model, we assume that all primitive values have
the same type, regardless of whether they are doubles, strings, timestamps,
etc, and simply refer to them as \emph{literals}, and consider only objects and arrays as complex values.
%
% we only distinguish three types of JSON values:
% literals (a.k.a. values of a primitive datatype), arrays and objects.
% 
% This model, in particular, endows  with a formal semantics the collections of BSON (for binary JSON) documents used by \mongodb.
%
% We also abstract away from the primitive datatypes (such as \texttt{decimal} or \texttt{date})
% used in JSON/BSON documents,
% as well the natural orders used by MongoDB to compare such values within and across datatypes.\footnote{However, 
% we specify how they can be combined to order objects or arrays.}
% %
% This is once again for the sake of flexibility.
% Besides, the main focus of this document is the interpretation of query operators,
% which is mostly independent of primitive datatypes and orderings.

Second, we offer two alternative treatments of objects.
MongoDB interprets an object as a \emph{sequence} of key-value pairs (with distinct keys).
This implies for instance that the JSON objects
\valuefont{\{"title": "News of the World", "release": 1977\}}
and
\valuefont{\{"release": 1977, "title": "News of the World"\}}
represent two different entities.
We call this first treatment the \emph{ordered} semantics for objects.
%
%This is indeed a practical requirement in some scenarios
However, in many applications, the data represented by a JSON object is instead meant to be interpreted as a \emph{set} of key-value pairs
(similarly to the way objects and associative arrays are treated in most programming languages).
Under this view, these two objects represent the same entity.
We also consider this treatment, which we call the \emph{unordered} semantics,
as an alternative to the ordered one.

\subsection{Document Databases}
\label{sec:dvalues}

In this section, we present the main building blocks of a document database. We
start by introducing an abstraction of JSON values in the form of \dvalues. Then,
we define collections and  database instances.

\paragraph{D-value} We use the generic term \emph{document value}, or
\emph{\dvalue} for short, to refer to a unit of data, which can be a
\emph{literal} (a primitive value), an \emph{array} or an \emph{object}.
We distinguish between the \emph{syntactic} representation of a d-value (how it is written using symbols) 
and its \emph{semantic interpretation} (what computational value it denotes). %it means computationally).
This distinction will allow us later to define the unordered semantics.
% For instance, 
% syntactically different objects like \valuefont{\{a:1, b:2\}} 
% and \valuefont{\{b:2, a:1\}} become semantically equivalent under the unordered semantics.

Syntactically, \dvalues are built inductively out of two (infinite)
% and disjoint\nb{JC: consider relaxing this disjointness condition, so that readers do not take it as a syntactic constraint.})
sets $\K$ of \emph{keys} and $\L$ of \emph{literals}, as per the definition below.

% We assume countably infinite disjoint sets
% $\K$ of \emph{keys},
% $\I$ of \emph{indexes} (non-negative integers) and 
% $\L$ of \emph{literals}.

\begin{definition}[\dvalue]
  \label{def:d-value}
  A \emph{\dvalue} is a finite\footnote{
The only purpose of this finiteness requirement is to rule out 
% and \ref{def:object} are
% mutually recursive. 
% We include the finiteness condition to avoid
\dvalue{s} with infinite nesting (since infinite width is already ruled out by the
definitions of arrays and objects).
}
    structure.
  It can be either:
  \begin{itemize}\itemsep 0cm
  \item a \emph{literal} (i.e., an element of $\L$), or
    
  \item an \emph{array} $\arraybr{v_1, \dots, v_n}$ of \dvalue{s}, or
    
  \item an \emph{object}--a finite sequence of \emph{key-value} pairs of the form $k \mapsto v$ where $k\in\K$,
    $v$ is a \dvalue,
    and keys are distinct.
  \end{itemize}
\end{definition}

We us $a[i]$ to refer to the $i^{\text{th}}$ elements of an array $a$, asssuming that indices start at 1
(i.e.~$a[1]$ is the first element of $a$ if nonempty).
When enumerating the key-value pairs of an object, we enclose them
in~$\objectbr{\cdot}$.
For instance, $\objectbr{\valuefont{title}\mapsto \textit{``News of the
  World''},\ \valuefont{release}\mapsto 1977}$ is
the \dvalue corresponding to the JSON object \valuefont{\{"title": "News of the
  World", "release": 1977\}}.
Within our set $\K$ of keys, we distinguish a special key $\id$,
which we use for document identifiers.
It corresponds to the field \valuefont{\_id} used by MongoDB (for instance in Figure~\ref{fig:json-document}).
We also distinguish a special value $\nullvalue$ within our set $\L$ of literals,
to represent missing fields or undefined values.
%
% \nb{NT: Say why. Maybe:
% "Within $\K$, we distinguish a special key $\id$ used for document identifiers 
% (corresponding to MongoDB's \valuefont{\_id} field). 
% Within $\L$, we distinguish 
% a special literal $\nullvalue$ representing explicit null values (as opposed to missing fields)."}
% %
We use $\V$ for the universe of all \dvalue{s},
i.e., all expressions that can be built (inductively) from $\K$ and $\L$,
according to Definition~\ref{def:d-value}.
Among these, we use $\A$ for the arrays and $\O$ for the objects.
% We observe that by construction, $\V$ and $\K$ are disjoint.\nb{JC: again, consider dropping disjointness (I am not sure it serves any purpose), to avoid readers taking it as a constraint.}

% \nb{NT: The distinction between syntax and interpretation of d-values may require more explanation 
% before we move on to more definitions.
% Consider adding something between here like this:
% "We distinguish between the \emph{syntactic} representation of a d-value (how it appears textually) 
% and its \emph{semantic interpretation} (what it means computationally). This distinction becomes 
% crucial when considering object ordering: syntactically different objects like \valuefont{\{a:1, b:2\}} 
% and \valuefont{\{b:2, a:1\}} may be semantically equivalent under certain interpretations."}
%
% Note also that these definitions define \dvalue{s} and objects as
% \emph{syntactic} entities.

We illustrate the syntax of \dvalues and our terminology in the example below.
\begin{example}\ \\
  \label{ex:dvalues}\em
 \quad The array $\arraybr{1, 2, 1}$ consists of the literals $1, 2$ and $1$ (in this order).

  The object $o_1 = \objectbr{\valuefont{age} \mapsto 27, \ \valuefont{name} \mapsto \textit{``Alex Doe''}}$ consists of two key-value pairs.
  It maps the key $\valuefont{age}$ to the literal $27$ and the key $\valuefont{name}$ to the literal \textit{``Alex Doe''}.
  
  The object
  $o_2 = \objectbr{\valuefont{name} \mapsto \textit{``Alex Doe''}, \ \valuefont{age} \mapsto 27}$ is different from $o_1$.
  
  The object 
  $\objectbr{\valuefont{age} \mapsto 27, \ \valuefont{name} \mapsto
    \objectbr{\valuefont{first} \mapsto \textit{``Alex''}, \ \valuefont{last} \mapsto
      \textit{``Doe''}}}$,
      differently from $o_1$, maps the key $\valuefont{name}$
  to a complex \dvalue (an object in this case).
   
 The array
  $\arraybr[\big]{ 2,\ 
    \objectbr{\valuefont{age} \mapsto 27, \ \valuefont{name} \mapsto
      \textit{``Alex Doe''}}}$ consists of two \dvalues.
  The first one is a literal, whereas the second one is an object.
  \qed
\end{example}
For a more complex example of a \dvalue, in Figure~\ref{fig:queen_dvalue} we provide the \dvalue corresponding to the JSON document in Figure~\ref{fig:json-document}.
\begin{figure}[t]
  \mquerybox{\centering
  \[\small
    \begin{array}{ll}
      \big\lobject
      &\valuefont{\_id} \mapsto  2,\\
      & \valuefont{name} \mapsto \textit{``Queen''},\\
      & \valuefont{formation} \mapsto 1970,\\
      & \valuefont{albums} \mapsto \larray\\
      & \qquad \objectbr[\big]{
        \valuefont{title} \mapsto \textit{``Queen''},\
        \valuefont{release} \mapsto 1973
        },\\
      & \qquad \objectbr[\big]{
        \valuefont{title} \mapsto \textit{``A Night at the Opera''},\
        \valuefont{release} \mapsto 1975,\
        \valuefont{length} \mapsto \textit{``43:08''}
        },\\
      & \qquad \objectbr[\big]{
        \valuefont{title} \mapsto \textit{``News of the World''},\
        \valuefont{release} \mapsto 1977,\ 
        \valuefont{labels} \mapsto \arraybr{\textit{``EMI''}, \textit{``Elektra''}}
        }\\
      &\rarray\\
      & \valuefont{members} \mapsto \larray\\
      & \qquad \objectbr[\big]{
        \valuefont{name} \mapsto \textit{``Freddie Mercury''},\
        \valuefont{role} \mapsto \arraybr{\textit{``lead vocals''}, \textit{``piano''}}
        },\\
      & \qquad \objectbr[\big]{
        \valuefont{name} \mapsto \textit{``Brian May''},\
        \valuefont{role} \mapsto \arraybr{\textit{``guitar''}, \textit{``vocals''}}
        },\\
      & \qquad \objectbr[\big]{
        \valuefont{name} \mapsto \textit{``Roger Taylor''},\
        \valuefont{role} \mapsto \arraybr{\textit{``drums''}, \textit{``vocals''}}
        },\\
      & \qquad \objectbr[\big]{
        \valuefont{name} \mapsto \textit{``John Deacon''},\
        \valuefont{role} \mapsto \textit{``bass''}
        },\\
      &\rarray\\
      \big\robject
      % &&\robject&&
    \end{array}
  \]}
  \caption{The \dvalue that corresponds to the JSON document of Figure~\ref{fig:json-document}.}
  \label{fig:queen_dvalue}
\end{figure}

The standard semantics we assume for \dvalues essentially interprets a \dvalue
as itself. This corresponds to the \emph{ordered} semantics since it interprets
objects as sequences of key-value pairs, as opposed to (unordered) sets of
key-value pairs.

% \deltabox{
%   Objects in MongoDB are \emph{ordered} sets of key-value pairs with no repeated key.
%   For instance, 
%   $\{k_1 \mapsto v_1, k_2 \mapsto v_2\}$ and
%   $\{k_2 \mapsto v_2, k_1 \mapsto v_1\}$ are different objects.
%
%   TODO: update the formalization accordingly (inclusing the formal semantics of the language).
%  Not taht we can still use the notation $o(k)$ (in some places) for the function (from keys to d-values) induced by $o$. 
% }

\paragraph{Database}
Document databases store collections of JSON documents. We now define all terminology required to talk about database instances.

Collections in document databases are analogous to relations in relational
databases, with objects serving as the counterparts of relational tuples.
% As a reminder, we assume that we deal with interpreted objects,
% as opposed to syntactic ones
% (even when we talk about objects stored in a database instance).

\begin{definition}[Collection]
  A \emph{collection} is a finite bag of objects.
\end{definition}
We emphasize the generic meaning of the term `collection' in this definition: a
collection may not only be stored in a database instance, but also be the
output of a (sub)query.

As in SQL, a \mongodb aggregate query refers to a stored collection via its
name.  Accordingly, we assume an infinite set $\C$ of collection names.
Finally, we can define a database instance as a way to assign collections to
such names:
\begin{definition}[Database instance]
  A \emph{database instance} $I$ maps each collection name $C \in \C$ to a collection $\evalc{C}$,
  and finitely many to a non-empty collection.
\end{definition}

We conclude with definitions that are specific to how MongoDB stores objects in
a database instance. Namely, every stored object necessarily has the \id field,
whose purpose is to uniquely identify this object within the stored collection.
We refer to such objects as documents: a \emph{document} is an object $d$ that
contains a key-value pair of the form $\id\mapsto v$.
We say that two documents $d_1$ and $d_2$ have \emph{distinct identifiers} if
$d_1$ and $d_2$ contain key-value pairs $\id \mapsto \mathit{id}_1$ and
$\id\mapsto\mathit{id}_2$, respectively, and $\mathit{id}_1\neq\mathit{id}_2$.
MongoDB always treats the field \id as the primary key of a stored collection. Therefore, for MongoDB a collection in a database instance--\emph{source collection}--is a finite set of documents with distinct identifiers.

Note that our definition of a database instance does not impose such a
requirement on stored collections, so they may in general contain multiple
occurrences of the same object.

% \nb{NT: Transitions between sections are abrupt without connecting material. Possible text for 
% bridgin section 2 to 3:
% "Having established our data model for JSON documents and collections, 
% we now turn to defining a query language that operates over these structures. 
% The next section introduces MQuery, our formalization of the MongoDB aggregation framework."}

\subsection{Objects as Interpretation Structures}
\label{sec:path-evaluation}

A \dvalue object is a complex value; in general, it has a hierarchical
structure with nested objects and arrays.
In this subsection we go into more technical depth and provide definitions that
allow us to ``look inside'' an object.
These will be useful when formalising the semantics of \mquery.

\smallskip
We start by viewing an object as a function. Recall that an object is a
sequence of key-value pairs where all keys are distinct. Therefore, each key in
an object provides access to the corresponding \dvalue (e.g., in the object of
Figure~\ref{fig:queen_dvalue}, the key \valuefont{formation} allows us to
access the value 1975).
More generally, an object \emph{induces} a function from keys to \dvalues.
\begin{definition}
  The \emph{function $f_o$ induced by} an object $o = \objectbr{k_1\mapsto v_1,\dots,k_n\mapsto v_n}$ is defined by $f_o(k_i) = v_i$,
  for each $i \in \{1,\dots, n\}$.
  % $o=\objectbr{k_1\mapsto v_1,\dots,k_n\mapsto v_n}$, the
  % \emph{function $f_o$ induced} by $o$ is defined as
  % % \julien{\st{the set of key-value pairs} simply} 
  % \[\{k_1\mapsto v_1,\dots,k_n\mapsto v_n\}.\]
\end{definition}
For convenience, we often write $o(k)$ to refer to $f_o(k)$,
i.e., the value associated with key $k$ in object $o$.
For instance, if $o$ is the object of Figure~\ref{fig:queen_dvalue},
then $o(\valuefont{formation}) = 1975$.

% By abuse of notation, we use often write $o(k)$ instead of $f_o(k)$.
% \nb{NT: Grammatically incorrect. Maybe: 
% "For convenience, we write $o(k)$ to mean $f_o(k)$ when accessing the value associated with key $k$ in object $o$."}
% Note that two objects that induce the same function may be interpreted
% differently under the ordered semantics (for instance $o_1$ and $o_2$ in
% Example~\ref{ex:dvalues}).

\medskip

The function induced by an object allows us to access the \dvalue associated to a key.
To access values that are nested ``deeper'' in an object, we consider
concatenations of keys and indices into what we call a \emph{path}
(\emph{field} in MongoDB terminology).
Paths allow us to navigate nested JSON structures, as is done in JSONPath. %similarly to file system paths. 
% \nb{NT: The path evaluation definition jumps immediately to formal notation
% without intuitive explanation.  Consider something like: "Paths allow us to
% navigate nested JSON structures, similar to file system paths.  For example,
% in a document representing a band, the path \valuefont{albums.0.title} would
% navigate to the \valuefont{albums} array, select the first element (index 0),
% and access its \valuefont{title} field.  We define path evaluation formally
% as follows:"}
For example, in the object of Figure~\ref{fig:queen_dvalue},
the path $\valuefont{albums.0.title}$ would navigate to the array \valuefont{albums}, select the first element (index \valuefont{0}), 
and access the value of its \valuefont{title} key (in this case \textit{"A Night at the Opera"}).
Formally, paths are defined as follows:
\begin{definition}[Path]
  A \emph{path} is a finite sequence of keys and indices in $\K \cup \nn$.
  
  The \emph{empty path} is denoted with $\varepsilon$.
\end{definition}
We represent a path as the concatenation of its elements, separated with a dot.
For instance, \valuefont{albums.0.title} represents the path $(\valuefont{albums}, \valuefont{0}, \valuefont{title})$.

% To refer to \dvalue{s} that appear ``deeper'' in the object, we introduce the notion of \emph{path} (a \emph{field} in MongoDB terminology).
% This notion allows us to navigate an object not only via keys, but also via array indices. Therefore, we consider a set $\I$ of \emph{indices} (non-negative integers), disjoint from $\K$ and $\L$.
% \nb{NT: I would then move this little paragraph directly after the definition but before "Now, we use the notion...".}
%
% \begin{definition}[Path]
%   A \emph{path} is a finite sequence of keys and indices in $\K \cup \I$.
%
%   The \emph{empty} path is denoted by $\varepsilon$.
%
%   Two (distinct) paths are \emph{compatible} if none of them is a
%   prefix of the other.
% \end{definition}

Now, to specify what a path refers to in an object, we define an
\emph{evaluation function~$\eval{\cdot}$} that extends the function~$f_o$ from
keys to paths and maps a path~$p$ to the corresponding \dvalue in the object
$o$ (intuitively, the \dvalue inside $o$ that $p$ points to).
Furthermore, we generalise this evaluation function from objects to \dvalues,
in particular, to establish what a path starting with an index refers to in an
array. This will be useful when we define the semantics of \mquery in
Section~\ref{sec:mquery}.
%
% \nb{JC: since this function extends the function $f_o$, shall we use the notation $o(p)$ instead of $\eval{p}$?}

\begin{definition}[Evaluation function]
  \label{def:eval-function}
  For a \dvalue $v$, the \emph{evaluation function~$\eval[v]{\cdot}$} from paths
  to \dvalues is defined inductively as:
\begin{itemize}\itemsep 0cm
\item $\eval[v]{\varepsilon} = v$
\item $\eval[v]{p.k} = o'(k)$ \quad if $\eval[v]{p}$ is the object $o'$ and $o'(k)$ is defined,
\item $\eval[v]{p.i} = a[i]$ \qquad if $\eval[v]{p}$ is the array $a$ and $a[i]$ is defined,
\item otherwise, $\eval[v]{p} = \nullvalue$.
\end{itemize}
\end{definition}

For example, let $o$ be the object in Figure \ref{fig:queen_dvalue}. Then
$\eval{\valuefont{albums.0}}$ is the object
\[o' = \objectbr[\big]{ \valuefont{title} \mapsto \textit{``A Night at the
    Opera"},\ \valuefont{release} \mapsto 1975,\ \valuefont{length} \mapsto
  \textit{``43:08''}}.\]
As for the fallback case $\eval[v]{p} = \nullvalue$, intuitively, the evaluation
function is not defined for paths $p$ that do not appear in $o$ (when starting
from the top level). To reflect this, we define $\eval[v]{p}$ as
\nullvalue. Note that an explicit \nullvalue may be already present in an
object, in which case there would be a path that is present and evaluates to
\nullvalue. This is in line with our treatment of \nullvalue as an undefined
value.
In particular, $\eval{\valuefont{albums.0.labels}}$ is undefined since
\valuefont{labels} is not a key in $o'$ and $\eval{\valuefont{title}}$ is
undefined since \valuefont{title} is only a key in $o'$, but not in $o$.

% \medskip

% Now, we use the notion of path to define an alternative syntactic description
% of an object, which maps paths to \dvalues: %
% a \emph{path/\dvalue pair} is a pair $p\mapsto v$, where $p$ is a path and $v$
% a \dvalue.
% %
% For sequences of path/\dvalue pairs we use the same enclosing delimiters
% $\objectbr{\cdot}$ as for objects.

% \begin{definition}[Compatible paths]
%   Two (distinct) paths are \emph{compatible} if none of them is a
%   prefix of the other.
% \end{definition}

% \nb{NT: This definition is fairly dense. Consider an intuitive explanation first to motivate. Something like:
% "Sometimes it's convenient to specify an object by listing the paths it contains and their values, 
% rather than describing its nested structure. For example, instead of writing the nested object structure explicitly, 
% we can specify an object as a flat list of path-value mappings."}
% \begin{definition}[path/\dvalue specification of an object]
%   \label{def:path_value_spec}
%   Let $r$ be a finite sequence of path/\dvalue pairs where paths are
%   compatible and non-empty. Then $r$ \emph{specifies} the
%   object~$o$ defined as the smallest object such that
%   $p \mapsto v \in r$ implies $\eval{p} = v$. 
%   % Additionally, under the
%   % ordered semantics, $o$ is required to comply with the order over $\paths{o}$
%   % induced by $r$ (defined in the appendix).
% \end{definition}

\subsection{The  Unordered Interpretation of \dvalues}
\label{sec:unordered-interpretation}
% \nb{NT: Consider moving the detailed discussion about ordered/unordered until after the
% definition of path.
% }
% \nb{NT:
% Maybe also replace this intro with something more motivating, like: 
% "Objects in JSON can be interpreted in two ways: as ordered sequences 
% of key-value pairs (preserving insertion order) or as unordered mappings 
% (like hash tables in programming languages). 
% We formalize both interpretations 
% to accommodate different application needs."

As we discussed already, a JSON object can be interpreted in two ways:
as a sequence of key-value pairs,
or as a map from keys to values (like a hash table in a programming language),
where the order of keys is irrelevant.
The former interpretation is the standard one, also referred to as
\emph{ordered} semantics.
In this subsection, we formalise the latter interpretation as the
\emph{unordered} semantics.
% to accommodate different application needs.
% We call the former interpretation the \emph{ordered} semantics,
% and the latter the \emph{unordered} semantics.

\begin{definition}[Unordered semantics]
  For a \dvalue $v$, denote by $v^*$ the interpretation of $v$.  The
  \emph{unordered semantics} interprets
 \begin{itemize}\itemsep 0cm
 \item a literal  as itself,
   
 \item an array $\arraybr{v_1, \dots, v_n}$ as $\arraybr{v_1^*, \dots, v_n^*}$,
   the interpretations of its elements (in the same order).
   
 \item an object $\objectbr{k_1\mapsto v_1,\dots,k_n\mapsto v_n}$ as the
   function $\{k_1\mapsto v_1^*,\dots,k_n\mapsto v_n^*\}$.
\end{itemize}

\end{definition}

\begin{example}
  \label{ex:dvalues-semantics}\em
  Under the unordered semantics, the two objects $o_1$ and $o_2$ from
  Example~\ref{ex:dvalues} are interpreted as the same map
  % set of
  % key-value pairs
  $\{\valuefont{name} \mapsto \textit{``Alex Doe''}, \ \valuefont{age} \mapsto 10\}$.
  Whereas under the ordered semantics, they are interpreted as
  different sequences of key-value pairs.
\end{example}

% Note that one programming analogy for the two semantics is two different
% implementations of a function testing deep equality between two class
% instances. In our model, distinct \dvalues correspond to distinct instances of
% a class. The ordered semantics corresponds to the implementation that takes
% into account the order of key-value pairs in the instances, while the unordered
% one does not. Hence, e.g., \valuefont{eq\_ordered(o1, o2)} returns false, while
% \valuefont{eq\_unordered(o1, o2)} returns true.  \nb{JC: I am not aware of a
%   programming languages where instances of the same class have ordered
%   attributes.  Where does this paragraph come from?  }

\medskip

Now that we have established our data model for JSON documents and collections, 
we turn to defining a query language that operates over these structures. 
The next section introduces MQuery, our formalization of the \mongodb aggregation framework.

%%% Local Variables:
%%% mode: latex
%%% TeX-master: "main-tr"
%%% fill-column: 79
%%% End:

\section{The Syntax and Semantics of MQuery}
\label{sec:mquery}
In this section we present a formalisation of the query language of the MongoDB
aggregation framework, which we call \mquery. We first define various building
blocks of MQuery, such as terms and Boolean expressions, in
Section~\ref{sec:mquery-building-blocks}.
Then we define the syntax and the semantics of \mquery stages (representing
top-level constructs), starting with core stages in Section~\ref{sec:stages}
and additional ones in Section~\ref{sec:additional-stages}.

An \mquery is a pipeline of stages evaluated one after the other.
\mquery allows for seven core types of \emph{stages}:
% \footnote{We suggest readers unfamiliar with \mongodb to read the following
% paragraphs in parallel to the respective subsections in
% App.~\ref{sec:mongodb-queries-examples}, which contain additional examples
% and the actual syntax of \mongodb.}:
\begin{enumerate}[\it (i)]
\item \emph{match}, which selects documents according to a filtering
  criterion. Such criterion is a Boolean combination of atomic conditions
  expressing, e.g., the existence of a path or its equality to a \dvalue.
\item \emph{unwind}, which flattens an array by creating a separate document
  for each element of the array.
\item \emph{project}, which can modify each document by removing, renaming or introducing new paths.
  \nb{Think if we want to have syntax for excluding paths (exclusion project is different from inclusion projection. When excluding paths other than \id, the project become exclusion project, so we cannot specified any paths that should be included. Currently we have inclusion project). TODO: check if exclusion project can have an impact of the complexity?}
\item \emph{group}, which can aggregate multiple documents into a
  single one.
\item \emph{lookup}, which performs a left join with an external
  collection.
\item \emph{graphLookup}, which allows to recursively traverse a collection and
  store all reachable documents in an array.
\item \emph{union}, which extends the current collection with the result of a
  query evaluated over an external collection.
\end{enumerate}
Other stages include those integral to any practical query language such as
\emph{limit}, \emph{skip}, \emph{sort} and \emph{count}. They are not part of
the \mquery core, we present their formalisation in
Section~\ref{sec:additional-stages}.

% To define the stages formally, we first define various expressions, building blocks of \mquery. Then we define the syntax and the semantics of the stages.

\subsection{\mquery Expressions}
\label{sec:mquery-building-blocks}

In this 
% subsection 
section,
we define the syntax and the semantics of expressions used
in \mquery.  Such an expression can be % a \emph{path iterator} $p$,
a \emph{term} $\tau$ (an \emph{expression} in MongoDB terminology) or a
\emph{Boolean expression} $\varphi$.

\medskip

\newpage
Terms are expressions that evaluate to \dvalues and can be built from simpler
components.
Boolean expressions are used in filtering conditions and can combine basic
comparisons using standard logical operators.
The syntax of a \emph{term} $\tau$ and a \emph{Boolean expression} $\varphi$ is
specified in the grammar below, where $\ell$ denotes a literal, $p$ a path, $k$
a key, %
$f$ the name of a computable function with fixed interpretation
$f^*\colon\V\to \L$ (such as \texttt{toUpperCase} for a string, or \texttt{max}
for an array), and `$\lambda x\colon {}$' starts the description of an
anonymous function.
% where the variable $x$ is bound to the context where it is used.
%
\begin{eqnarray}
  \tau &\DEF& \ell ~\mid~ p ~\mid~
              \arraybr{\tau,\dots,\tau} ~\mid~ 
              \objectbr{k\mapsto\tau,\dots,k\mapsto\tau} ~\mid~
              f(\tau) ~\mid~
              % \mathsf{map}(p, \lambda x.\tau) ~\mid~ 
              \cond{\varphi}{\tau}{\tau} ~\mid~ \nonumber\\
       &    & \map{p}{x}{\tau} ~\mid~ 
              \filter{p}{x}{\varphi} 
              \label{gramm:term}
  \\
   \varphi &\DEF&  
   \exists p
   ~\mid~ \tau = \tau
   ~\mid~ \tau \leq \tau
   ~\mid~ \tau\in \tau
   ~\mid~ \neg \varphi
   % ~\mid~ \varphi \lor\varphi
   ~\mid~ \varphi \land \varphi
                  \label{gramm:boolean}
\end{eqnarray}

Note that our terms and Boolean expressions do not cover all expressions
present in MongoDB, many of which are convenient in practice, but could
equivalently be expressed with other basic constructs--thus in essence being
syntactic sugar--e.g., \valuefont{anyElementTrue} can be expressed as a
combination of $\map{p}{x}{\tau}$ and $\tau\in\tau$.
We include only those expressions that can be considered to constitute a
reasonable core.
We now discuss terms and Boolean expressions, each in more detail.

\medskip
\noindent\textbf{Term}. %
A term $\tau$ is an expression that evaluates to a \dvalue, and may be used in
\mquery in a Boolean expression or to set the value of a new path.
Terms are considerably more powerful than paths in that they allow for
construction of new \dvalues, which may involve evaluation of conditional
statements and computable functions.
A term can be seen as a template for a \dvalue built 
inductively
% starting 
from constant
values~($\ell$) and paths
% iterators 
($p$),
and then applying to them an
\emph{array} or \emph{object constructor} (resp., $\arraybr{\tau,\dots,\tau}$
and $\objectbr{k\mapsto\tau,\dots,k\mapsto\tau}$), a named function
($f(\tau)$), a \emph{conditional term definition}
($\cond{\varphi}{\tau}{\tau}$), or a \emph{map} ($\map{p}{x}{\tau}$) or
\emph{filter} ($\filter{p}{x}{\varphi}$) function.

We extend the evaluation function 
(see Definition~\ref{def:eval-function})
from paths to terms:
\begin{definition}[Evaluation function for terms]
  Let $v$ be a \dvalue. The evaluation function interprets a literal $\ell$ as itself,
  i.e.~$\eval[v]{\ell}= \ell$, while the evaluation of complex terms is defined inductively as
  follows:
\begin{itemize}
\item
  $\eval[v]{\arraybr{\tau_1,\dots,\tau_m}}= \arraybr[\Big]{\eval[v]{\tau_1},
    \dots, \eval[v]{\tau_m}}$, i.e., the array constructor
  $\arraybr{\tau_1,\dots,\tau_m}$ evaluates to the array with $i$-th
  element $\eval[v]{\tau_i}$;

\item
  $\eval[v]{\objectbr{k_1\mapsto\tau_1,\dots,k_m\mapsto\tau_m}}=\objectbr[\Big]{k_i\mapsto\eval[v]{\tau_i}
    \Bigm\vert \eval[v]{\tau_i}\neq\nullvalue, \ i=1,\dots,m}$, i.e., the
  object constructor $\objectbr{k_1\mapsto\tau_1,\dots,k_m\mapsto\tau_m}$
  evaluates to the object where the value for $k_i$ is $\eval[v]{\tau_i}$;

\item $\eval[v]{f(\tau)}= f^*(\eval[v]{\tau})$, where $f^*$ is the function
  whose name is $f$;

\item $\eval[v]{\cond{\varphi}{\tau_1}{\tau_2}} =
  \begin{cases}
    \eval[v]{\tau_1} & \te{if} v \models \varphi\\
    \eval[v]{\tau_2} & \te{otherwise}
  \end{cases}$, i.e., a conditional term definition
  $\cond{\varphi}{\tau_1}{\tau_2}$ evaluates to $\eval[v]{\tau_1}$ if the
  Boolean condition $\varphi$ is satisfied in $v$ (defined below), and to
  $\eval[v]{\tau_2}$ otherwise;

\item
  $\eval[v]{\map{p}{x}{\tau}} =
  \begin{cases}
  \arraybr[\Big]{\evald[\mu_1]{\tau}, \dots, \evald[\mu_n]{\tau}} & \text{if } \eval[v]{p} = \arraybr{v_1,\dots,v_n},
  \te{with} \mu_i = \{x\mapsto v_i,\ \varepsilon\mapsto v\},\\
  \nullvalue \te{otherwise.}
  \end{cases}
  $
  % otherwise $\eval[v]{\map{p}{x}{\tau}} = \nullvalue$.
  In other words, $\map{p}{x}{\tau}$ returns an array of the same size as the array $\eval[v]{p}$,
  where each element is the evaluation of $\tau$ in the evaluation context $\mu_i = \{x\mapsto v_i,\ \varepsilon\mapsto v\}$.
  In this context, the variable $x$ is assigned the value $v_i$,
  while the empty path (referring to the whole value) is assigned the current \dvalue $v$.
  The evaluation function in a context $\mu$ is defined similarly 
  as in a \dvalue $v$, the only difference being the base case,
 with $\eval[\mu]{x.p} = \eval[\mu(x)]{p}$.
  This definition can be extended to nested $\mapz$ and/or $\filterz$ expressions,
  by defining an evaluation function for richer evaluation contexts.
  % \nb{E: fix the semantics
  %   considering that some paths in $\tau$ may not be bound to $\eval[v]{p}$}

\item
  $\eval[v]{\filter{p}{x}{\varphi}} =
  \begin{cases}
  \arraybr[\Big]{v_i \in \eval[v]{p} \Bigm\vert v_i\models_{\mu_i} \varphi} &\text{if } \eval[v]{p} \te{is an array},
  \te{with} \mu_i = \{x\mapsto v_i,\ \varepsilon\mapsto v\},\\
  \nullvalue \te{otherwise.}
  \end{cases}
  $

% \arraybr[\Big]{v_i \in \eval[v]{p} \Bigm\vert v_i\models_{\mu_i} \varphi}$ if $\eval[v]{p}$ is an array,
%   otherwise $\eval[v]{\filter{p}{x}{\varphi}} = \nullvalue$.
  In other word, $\filter{p}{x}{\varphi}$ returns a subsequence of the array $\eval[v]{p}$,
  which consists of the elements for which the Boolean condition $\varphi$ holds,
  in the same order as they appear in $\eval[v]{p}$.
  The role of the context $\mu_i$ is identical to the case of $\mapz$.
\end{itemize}

\end{definition}

Note that the term $f(\tau)$ abstracts various computable functions, including
arithmetic functions, functions on strings (such as \valuefont{concat},
\valuefont{substr}, \valuefont{toLowerCase}) and aggregate functions (such as
\valuefont{min}, \valuefont{max}, \valuefont{sum}, \valuefont{count}). Here we
only assume that we know how to compute $f^*$. In practice, in \mongodb, $f$
would need to be one of the supported function names.
For readability, 
% for arithmetic functions we may use an 
%arithmetic expression,
we may use an infix operator for an arithmetic function. 
For instance, we may write $p + 1$ instead of
%a function name 
$\mathit{incrementByOne}(p)$.

Note that the expression $\map{p}{x}{\tau}$ produces an array by
evaluating $\tau$ over each element of the array referred to by $p$, similarly
to the \emph{map} function in functional programming languages.
For example, $\map{p}{x}{x+1}$ increases by $1$ the value of each element in
the array referred to by $p$, while
$\map{p}{x}{\cond{x \geq 10}{\truevalue}{\falsevalue}}$ outputs a Boolean value
% \nb {JC: Can these Boolean value interact with the semantics of Boolean expressions?
% My understanding is that they cannot.
% If so, then maybe use 0 and 1 here instead to avoid confusions.
% E: I don't understand. I think there is no confusion.
% }
for each element of the referenced array, depending on whether it satisfies the
comparison or not.
In these examples $x$ refers directly to the elements of the referenced array.
The expression $\map{p}{x}{x.p'}$, on the other hand, returns the value of~$p'$
in each element of the referenced array, in other words, it allows to
\emph{access values nested inside an array}; while the expression
$\map{p}{x}{x.p' + 1}$ not only accesses nested values, but also transforms
them, increasing each of them by $1$.

Note that an array constructor $\arraybr{\tau_1,\dots,\tau_m}$ always produces
an array with $m$ elements, even if 
% one of 
some
$\tau_i$ evaluates to
\nullvalue. Similarly, a map function $\map{p}{x}{\tau}$ produces an array of
the same size as~$\eval[v]{p}$, possibly containing \nullvalue values.  Such
\nullvalue values, when not useful, can be removed with a $\filterz$ function such
as $\filter{p'}{x}{x\neq\nullvalue}$.
However, an object constructor
$\objectbr{k_1\mapsto\tau_1,\dots,k_m\mapsto\tau_m}$ may produce a sequence of
fewer than $m$ key-value pairs. In particular, there would be no pair
$k_i\mapsto\eval[v]{\tau_i}$ if $\eval[v]{\tau_i}$ evaluates to \nullvalue.

% In the case where the expression $\tau$ specifies an object
% (i.e. is of the form $\objectbr{k_1\mapsto\tau_1,\dots,k_m\mapsto\tau_m}$),
% we used $\S$ to emphasize that the evaluation $\evald[o]{\tau}$ may be either a sequence (under the ordered semantics)
% or a function (under the unordered semantics). 
% \nb{JC: this is technically not needed, because we said earlier that throughout the paper,
%   an objects stands for its interpretations.
% We only use it here to avoid misunderstandings.
% Shall we write this?
% Or nothing (and remove the $\S$ from the definition)?
% }
% \nb{Stopped here}

\medskip
\noindent\textbf{Boolean expression}. %
A Boolean expression may be used as a filtering condition (in a match stage or
a $\filterz$ function), or in the conditional definition of the \dvalue associated
to a new path (in a project stage).
As specified in Grammar~\eqref{gramm:boolean} above,
a Boolean expression $\varphi$ is built using the standard Boolean operators
$\neg$, $\land$ from atomic formulas of the form:
\begin{itemize}\itemsep 0cm
\item $\exists p$:\  path $p$ evaluates to some non-\nullvalue \dvalue.
  
\item $\tau_1 = \tau_2$ or $\tau_1 \leq \tau_2$:\ a comparison of (the
  evaluations of) two terms, e.g., $p = 5$, $p_1 \leq p_2$,
  $\min(p_1) = \max(p_2)$.

\item $\tau_1 \in \tau_2$:\ the evaluation of $\tau_2$ as an array contains the
  evaluation of $\tau_1$.
\end{itemize}
In what follows, we may use $\neq$, $<$, $\ge$ and $>$ as syntactic sugar for
comparison operators, and $\varphi_1\lor\varphi_2$ as syntactic sugar for
$\neg(\neg\varphi_1\land\neg\varphi_2)$.

\deltabox{\mongodb allows using in a Boolean context expressions that strictly
  speaking do not evaluate to Boolean values. Specifically, 
  % \dvalue{s} 
  0 and
  \nullvalue are interpreted as false, whereas all other values
  % \dvalue{s}
  (including
  the empty string ``'', empty array $[]$ and empty object $\{\}$) are interpreted
  as true.
  However,
  the evaluation of an arbitrary term $\tau$ in a Boolean context can be
  simulated with our syntax,
  using the Boolean expressions
  $\neg(\tau\in[0,\falsevalue,\nullvalue])$.
}% \nb{JC: The content of this box seems ok (I slightly updated it), but the meaning of "Expressivity/core language" is unclear.}

We define the semantics of Boolean expressions in terms of what it means for a
\dvalue to satisfy a Boolean expression.

\begin{definition}
  Let $v$ be a \dvalue and $\varphi$ a Boolean expression. The
  \emph{satisfaction} of $\varphi$ by $v$, denoted $v \models \varphi$, is
  defined inductively as follows:
  \[
    \begin{array}{lll}
      v \models \exists p
      & \te{if} & \eval[v]{p} \neq \nullvalue
      \\
      v \models \tau_1 \op \tau_2
      & \te{if} & \evald[v]{\tau_1} \op{} \evald[v]{\tau_2} \text{\quad for }\op\in\{=,\leq\}
      \\
      v \models \tau \in \tau'
      & \te{if} & \evald[v]{\tau'}\text{ is an array \quad and \quad $\evald[v]{\tau}$ appears in $\evald[v]{\tau'}$}
      \\
      v \models \neg \varphi 
      % & \te{iff it is not the case that } v \models \varphi
      & \te{if} & v \not\models \varphi
      \\
      v \models \varphi_1 \land \varphi_2
      & \te{if} & v \models \varphi_1 \te{and} v \models \varphi_2
      \\
      % v \models \varphi_1 \lor \varphi_2
      % & \te{if} & v \models \varphi_1 \te{or} v \models \varphi_2
      % \\
    \end{array}
  \]
\end{definition}
For the semantics of $\le$, we assume a total order over the set $\V$ of
\dvalues and refer to it as the \emph{natural order} over $\V$.  In
Appendix~\ref{sec:appendix_order}, we show how such an order can be defined
inductively out of an (underspecified) order over the set $\L$ of literals.

% \deltabox{Our semantics of Boolean expressions is always defined. In particular,
%   $v\models \tau\in\tau'$ does not hold when $\tau'$ does not evaluate to an
%   array. MongoDB, instead, throws an exception if $\tau'$ does not evaluate to
% an array (see Section~\ref{sec:data-independent} for a discussion on this topic).}
%

\subsection{Core \mquery Stages}
\label{sec:stages}
% We now define the syntax of \mquery, then the semantics of each stage, and
% finally the evaluation of a complete \mquery. 
We define the syntax and semantics of the core \mquery stages.

We start with the syntax of \mquery.
Let $\X$ be a countably infinite set of variable names, disjoint from $\K$ and $\L$. We first define some building blocks of \mquery stages:
 \begin{itemize}
\item a \emph{path definition} has the form $p/\tau$ where $p$ is a nonempty index-free path and $\tau$ is a term,
% \item a \emph{path renaming} has the form $p/e$, where $p$ is a nonempty path and $e$ a nonempty path iterator,
\item a \emph{variable definition} has the form $x/p$, where $x\in\X$ is a variable name and $p$ a nonempty path.
 \end{itemize}
 We say that two paths are \emph{compatible} if none is a prefix of the other.

The \mquery language is defined according to the following grammar:
\begin{equation}
  \begin{array}{r@{~}r@{~}l}
    \toprule
    \mquery &\DEF& C \pipeline s \pipeline \cdots \pipeline s
    \\
    s &\DEF& \match{\varphi}
             \mid \unwind{p}
             % \mid \unwind[+]{p}
             \mid \project{D}
             % \mid \project[\noid]{P}
             \mid \group{g}{A}
             \mid \lookup{X,t}{p}
             \mid \glookup{p,C,p,p}{p}
             \mid \unionwith{q}
    \\
    \bottomrule
  \end{array}
  \label{eq:mquery-grammar}
\end{equation}
%          s
where $C$ is a collection name,
$\varphi$ a Boolean expression,
$p$ an index-free path,
$g$ an optional key,
$D$ a nonempty sequence of path definitions,
$A$ a sequence of keys, 
$X$ a set of variable definitions,
$q$ a query, 
and $t$ a query template (defined below).
In addition we require the variables defined in $X$ to be distinct,
and all paths defined in $D$ to be pairwise compatible.
Each expression for $s$ in this grammar is called a \emph{stage}.

\medskip

We now define the semantics of \mquery.
We use $\evalcdb{s}$ to denote the evaluation of a stage $s$ \emph{over} a collection $O$ and \emph{with respect to} a database instance $I$.
% \nb{JC. Maybe we can keep $I$ implicit, and introduce it as the "current database instance" when defining the semantics of lookup and graphLookup.}
This evaluation is itself a collection.
Then we use this notion to define inductively the evaluation $\evalq{q}$ of a query $q$ over $I$,
as follows:
\begin{equation*}
  \evalq{q} = \begin{cases} 
    \evalc{C} & \te{if} q = C\\
        \big(\evalq{q'},\ I\big) \blacktriangleright  s  & \te{if} q = q' \pipeline s 
      \end{cases}\\
% \label{eq:sem-query}
\end{equation*}
The evaluation of $s$ over $O$ with respect to $I$, in symbols $\evalcdb{s}$, is defined below, for each type of stage.
We observe that for match, unwind, project and group stages, the collection $\evalcdb{s}$ is determined by $O$ and $s$ only (in other words, the database instance $I$ in these cases is irrelevant).
Intuitively, these stages can be viewed as unary operators (taking as operand a collection $O$, and returning another).
In contrast, lookup, graphLookup and union stages can be understood as binary operators (taking $O$ as first operand, and the evaluation in $I$ of another query as second operand).

% We use $\eval[O]{s}$ to denote the evaluation of a stage $s$ over a collection $O$.
% This evaluation is itself a collection.

% A stage $s$ takes as
% input a collection (i.e. a bag of objects) $O$ and returns another collection
% $O\pipeline s$.

\begin{figure}
  \centering
\begin{lstlisting}
{   "_id": 3,
    "name": "ABBA",
    "formation": 1972,
    "albums": [
        { "title": "Waterloo", "release": 1974, "length": "38:09" },
        { "title": "ABBA", "release": 1975, "labels": ["Polar", "Epic", "Atlantic"] }
    ],
    "members": [
        { "name": "Agnetta Faltskog", "role": "lead vocals" },
        { "name": "Björn Ulvaeus", "role": ["guitar", "vocals"] },
        { "name": "Benny Andersson", "role": ["keyboard", "vocals"] },
        { "name": "Anni-Frid Lyngstad", "role": "vocals" }
    ]
}
\end{lstlisting}
  \caption{A JSON document about the band ABBA.}
  \label{fig:json-document-abba}
\end{figure}

\subsubsection{Project}
A project stage $\project{D}$ is parameterized with a sequence $D$ of 
path definitions where all defined paths are pairwise compatible.

% If $D$ consists of a single path definition $p/\tau$ (where $p$ is a path and $\tau$ a term),
% then $\project{D}$ intuitively
Intuitively, one path definition $p/\tau \in D$, for a path $p$ and a term
$\tau$, transforms an object $o$ into the object $o'$ associated with the
\emph{path-value} pair $p\mapsto \eval{\tau}$.
%\nb{JC: I rephrased this sentence, because it was talking about \emph{any} $p/\tau \in D$, which was incorrect (I think). E: no, I only wanted to talk about one path definition, not about a project with one path definition.}
%
% \nb{JC: This paragraph is only meant to describe the base case of the definition of $\uplus$.
% But this is not explained.
% Instead, "Intuitively" suggests that we are giving here an informal overview here of the whole definition, before going into details.
% E: intuition is for one path definition.
% }
For instance, if $p=k_1.k_2.k_3$, then $o'$ is the object
\[\objectbr{k_1\mapsto\objectbr{k_2\mapsto\objectbr{k_3\mapsto\eval{\tau}}}}.\]
In other words, the term $\tau$ \emph{defines} the value of the path $p$ in the
transformed object $o'$.

Before we move to multiple path definitions, we fix some useful notation.
\begin{definition}
  \label{def:path-overriding}
Let $o$ be an object, $p$ an index-free path and $v$ a \dvalue.

We denote by $\mathsf{obj}_{p\mapsto v}$ the \emph{object associated with the
path-value pair~$p\mapsto v$}.

Assume that $o\models\exists p$ and $p = k_1\dots k_m$.  The result of
\emph{overriding the value of $p$ by $v$ in $o$}, denoted $o[p/v]$, is the
object defined as a copy of $o$ where the key-value pair $k_m\mapsto \eval{p}$
inside the object $o(k_1)\cdots(k_{m-1})$ is replaced by the key-value pair
$k_m\mapsto v$.
\end{definition}
For instance, if
\[o=\objectbr{\valuefont{album} \mapsto \objectbr{\valuefont{title} \mapsto
    \textit{``Queen''},\ \valuefont{release} \mapsto 1973}},\] then
$o[\valuefont{album.title}/\objectbr{\valuefont{abbr} \mapsto
  \textit{``SHA''}}]$ is the object
\[o_{\text{sha}}=\objectbr{\valuefont{album} \mapsto \objectbr{\valuefont{title} \mapsto
    \objectbr{\valuefont{abbr} \mapsto \textit{``SHA''}},\
    \valuefont{release}\mapsto 1973}}.\]

To understand how a sequence of path definitions transforms an object, we
define what it means to merge a path-value pair into an existing object.
Let $o$ be an object and $p\mapsto v$ a path-value pair such that $p$ does not
exist in $o$, i.e., $o\not\models\exists p$. Then the object $o'$ resulting
from \emph{merging} $p\mapsto v$ into $o$, denoted $o \oplus (p\mapsto v)$, is
defined as follows:
\begin{itemize}\itemsep 0cm
\item if $v=\nullvalue$, then $o'=o$;
  
\item if there is no non-empty prefix $p_1$ of $p$ such that
  $o\models\exists p_1$, then $o'$ is the sequence
  $o\cdot\mathsf{obj}_{p\mapsto v}$, where `$\cdot$' stands for the
  concatenation of two sequences;
  
\item otherwise, assume that $p=p_1.p_2$, where $p_1$ is the longest prefix of
  $p$ such that $o\models\exists p_1$, and $\eval[o]{p_1} = u$ is an object value.
%   \nb{JC: Not sure I am parsing this correctly.
% My reading is
% "such that ($o\models\exists p_1$, and $\eval[o]{p_1} = u$ is an object value)".
% If so, then the second condition seems redundant (this should automatically follow from the fact that the paths defined in $D$ are compatible).
% E:the definition is for merging a path into an object, so in general the second condition is not redundant.
%   }
  %
  Define $u'$ as the extension of $u$ with $p_2 \mapsto v$, i.e.,
  $u' = u \cdot \mathsf{obj}_{p_2\mapsto v}$.  Then $o'$ is defined as
  $o[p_1/u']$.
\end{itemize}
Intuitively, if path $p$ is not present in $o$, then merging $p\mapsto v$ into
$o$ simply appends the new information at the end of $o$. But if a prefix of
$p$ is present in $o$, then we insert the new information in the correct place.
For instance,
$o_{\text{sha}}\oplus(\valuefont{album.title.full}\mapsto\textit{``Sheer Heart
  Attack''})$ is the object
\[\objectbr{\valuefont{album} \mapsto \objectbr{%
      \valuefont{title} \mapsto \objectbr{%
        \valuefont{abbr} \mapsto \textit{``SHA''},\
        \valuefont{full} \mapsto \textit{``Sheer Heart Attack''}},\
      \valuefont{release}\mapsto 1973}}.\]

Note that the operation $\oplus$ is left-associative. This means that we can omit
parentheses, so we write $o \oplus (p_1\mapsto v_1) \oplus (p_2\mapsto v_2)$
instead of $\Big(o \oplus (p_1\mapsto v_1)\Big) \oplus (p_2\mapsto v_2)$.
Also note that this operation is not symmetric as the left argument is an
object and the right argument is a path-value pair.

% Now, two path definitions $p_1/\tau_1$ and $p_2/\tau_2$ transform an
% object $o$ into an object $o'$ associated with the path-value pairs
% $p_1\mapsto \eval{\tau_1}$ and $p_2\mapsto \eval{\tau_2}$. For instance, if
% $p_1=k_1.k_2.k_3$ and $p_2=k_4.k_5$, where each $k_i$ is distinct, then $o'$ is
% the object
% %
% \[\objectbr[\Big]{k_1\mapsto\objectbr{k_2\mapsto\objectbr{k_3\mapsto\eval{\tau_1}}},\
%     k_4\mapsto\objectbr{k_5\mapsto\eval{\tau_2}}}\]
% %
% while if
% $p_1=k_1.k_2.k_3$ and $p_2=k_1.k_4$, where each $k_i$ is distinct, then $o'$ is
% the object
% %
% \[\objectbr{k_1\mapsto\objectbr[\Big]{k_2\mapsto\objectbr{k_3\mapsto\eval{\tau_1}},\ k_4\mapsto\eval{\tau_2}}}.\]

\medskip

With a slight abuse of notation, for an object $o$, we use $\project{D}(o)$ to
denote the object specified as:
\begin{equation*}
%\project{D}(o) = \objectbr[\Big]{p\mapsto \evald[o]{\tau} \Bigm\vert p/\tau \in D},
\project{p_1/\tau_1,\dots,p_n/\tau_n}(o) =  \objectbr{} \oplus (p_1\mapsto \evald[o]{\tau_1}) \oplus \cdots \oplus (p_n\mapsto \evald[o]{\tau_n}),
% \label{eq:sem-project-object}
\end{equation*}
where each path-value pair $p_i\mapsto \evald[o]{\tau_i}$ is merged into the
result of merging the previous $i-1$ path-value pairs, and $\objectbr{}$
denotes the empty object.
%
% For instance, the sequence
% $\objectbr{ \valuefont{name}\mapsto \textit{``Queen''},\
%   \valuefont{formation.year} \mapsto 1975 }$ specifies the object
% $\objectbr{ \valuefont{name}\mapsto \textit{``Queen''},\
%   \valuefont{formation}\mapsto \objectbr{ \valuefont{year} \mapsto 1975} }$.
\nb{
  JC: we need examples (this is really hard to parse).
  Maybe reuse or adapt the one we had already (it was illustrating different possible cases).
}

Now, for an input collection $O$ and database instance $I$,
the evaluation of a project stage is defined as:
\begin{equation*}
  \evalcdb{\project{D}} = \bag{\ \project{D}(o) \mid o\in O\ }\\
% O \pipeline \project{P} = \bag{\ \project{P}(o) \mid o\in O\ }\\
  % \label{eq:sem-project}
\end{equation*}
In the examples below, we abbreviate a path definition of the form $p/p$ simply as $p$. 

% Note that due to the semantics of $\eval[o]{\cdot}$ and $\evald[o]{\cdot}$, if
% $d$ is a path that is not present in $o$, then $\evald[o]{d}$ is equal to
% $\nullvalue$. Specifically, if $p/q \in P$, and the path $q$ is not present in
% $o$, then the resulting document will have a path-value pair
% $p\mapsto \nullvalue$.

\begin{example}\label{ex:project-1}
  \em
  In the subsequent examples of this section, we assume a database instance $I$
  that assigns to the collection name \valuefont{bands} the collection
  consisting of the two JSON documents in Figures~\ref{fig:json-document} and
  \ref{fig:json-document-abba}.

Assume that for each band, we only want to output its name and its year of formation.
We can express this as the following project-only query, presented below in the MongoDB and our abstract syntaxes.

\begin{minipage}[t]{0.49\linewidth}
  \begin{lstlisting}
db.bands.aggregate([
    { $project: {
        "name": 1, "formation": 1
    } },
])
  \end{lstlisting}%$
\end{minipage}
\hfill%
\begin{minipage}[t]{0.49\linewidth}
  \mquerybox{$
    \begin{array}{l}
      \texttt{bands}\\[-0.5mm]
      \hspace{0.7cm}{}\pipeline \project{\id,\ \texttt{name},\ \texttt{formation}}\\
    \end{array}\\[5.8mm]
$}
\end{minipage}
Note how MongoDB by default keeps the \id field, but all other fields that are
not mentioned in $D$ are projected away. 
In \mquery, to keep the \id, it must be included in $D$.

If we want to have the year of formation under the key
\valuefont{year\_formed}, and to omit the \id, we may achieve it with the
following queries:

\begin{minipage}[t]{0.49\linewidth}
  \begin{lstlisting}
db.bands.aggregate([
    { $project: {
        "_id": 0, "name": 1,
        "year_formed": "$formation"
    } },
])
  \end{lstlisting}%$
\end{minipage}
\hfill%
\begin{minipage}[t]{0.49\linewidth}
  \mquerybox{$
    \begin{array}{l}
      \texttt{bands}\\[-0.5mm]
      \hspace{0.7cm}{}\pipeline \project{\scriptsize
      \begin{array}[t]{@{}l}
        \texttt{name},\\
        \texttt{year\_formed}/\texttt{formation}
      \end{array}
      }\\
    \end{array}\\[6mm]
$}
\end{minipage}

Evaluating this query over the \valuefont{bands} collection, we obtain:

\begin{lstlisting}
{   "name": "Queen", "year_formed": 1970   }
{   "name": "ABBA",  "year_formed": 1972   }
\end{lstlisting}
\end{example}

In the next example, we illustrate complex term definitions. 
\begin{example}\label{ex:project-2}
  \em %
  Assume that for each band, we want to retrieve its name and the years when an
  album was released. This can be done as follows:

  \noindent
\begin{minipage}[t]{0.5\linewidth}
  \begin{lstlisting}
db.bands.aggregate([
    { $project: {
        "_id": 0, "name": 1,
        "albums_released": { $map: {
            input: "$albums",
            as: "x",
            in: { $trunc: "$$x.release" }
        } }
    } },
])
  \end{lstlisting}%$$
\end{minipage}
\hfill%
\begin{minipage}[t]{0.5\linewidth}
  \mquerybox{$
    \begin{array}{l}
      \texttt{bands}\\[-0.5mm]
      \hspace{0.6cm}{}\pipeline \project{\scriptsize
      \begin{array}[t]{@{}l}
        \texttt{name},\\
        \texttt{albums\_released}/\map{\texttt{albums}}{x}{x.\texttt{release}}
      \end{array}
      }\\
    \end{array}\\[2cm]
$}
\end{minipage}
Note how for the MongoDB query we had to use the function \valuefont{\$trunc}
inside the \valuefont{in} argument, as there is no available identity
function. We also note that there is an alternative method to compute the same
result in MongoDB: 

\begin{minipage}[t]{0.49\linewidth}
  \begin{lstlisting}
db.bands.aggregate([
    { $project: {
        "_id": 0, "name": 1,
        "albums_released": "$albums.release"
    } },
])
  \end{lstlisting}%$
\end{minipage}

\noindent This is not allowed in \mquery, though. We discuss this 
%difference 
choice in
detail in Section~\ref{sec:diff-interp-paths}.

The above queries evaluate to:

\begin{lstlisting}
{   "name": "Queen", "albums_released": [ 1973, 1975, 1977 ]   },
{   "name": "ABBA",  "albums_released": [ 1974, 1975 ]   }
\end{lstlisting}

More generally, the $\mapz$ construct allows 
% for a function application 
applying a function
to each
element of an array. We may, for instance, compute how many years 
passed between each album release and the formation of the band.

\noindent
\begin{minipage}[t]{0.5\linewidth}
  \begin{lstlisting}
db.bands.aggregate([
   { $project: {
      "_id": 0, "name": 1,
      "diff": { $map: {
         input: "$albums",
         as: "x",
         in: { $subtract:
            ["$$x.release", "$formation"]
         }
      } }
   } }
])
  \end{lstlisting}%$
\end{minipage}
\hfill%
\begin{minipage}[t]{0.5\linewidth}
  \mquerybox{$
    \begin{array}{l}
      \texttt{bands}\\[-0.5mm]
      \hspace{0.6cm}{}\pipeline \project{\scriptsize
      \begin{array}[t]{@{}l}
        \texttt{name},\\
        \texttt{diff}/\map{\texttt{albums}}{x}{x.\texttt{release} - \texttt{formation}}
      \end{array}
      }\\
    \end{array}\\[2.67cm]
$}
\end{minipage}

These queries evaluate to:

\begin{lstlisting}
{   "name": "Queen", "diff": [ 3, 5, 7 ]   },
{   "name": "ABBA",  "diff": [ 2, 3 ]   }
\end{lstlisting}
\end{example}

\subsubsection{Unwind}
An unwind stage $\unwind{p}$ is parameterized with an index-free path $p$.  For
each object $o$ in the input collection~$O$, if $p$ evaluates in $o$ to an
array $\arraybr{v_1,\dots, v_n}$, then $\unwind{p}$ outputs $n$ copies of $o$,
but with values $v_1$, \dots, $v_n$, respectively, for $p$.

Formally, for an input collection $O$, the evaluation of an unwind stage is defined as:

\begin{equation*}
\evalcdb{\unwind{p}} = \bag[\Big]{o[p/v_i] \Bigm\vert o\in O,\ \eval[o]{p}=\arraybr{v_1,\dots,v_n} \text{ and }i \in \{1,\dots,n\}},
  % \label{eq:sem-unwind}
\end{equation*}
where $o[p/v]$ is the result of overriding the value of $p$ by $v$ (see
Definition~\ref{def:path-overriding}).

\begin{example}\label{ex:unwind-syn-sem}
\em%
Assume that we want to have the information about each album separately. This
can be achieved by unwinding the path \valuefont{albums}:

\begin{minipage}[t]{0.49\linewidth}
  \begin{lstlisting}
db.bands.aggregate([
    { $unwind: "$albums" }
])
  \end{lstlisting}%$
\end{minipage}
\hfill%
\begin{minipage}[t]{0.49\linewidth}
  \mquerybox{$
    \begin{array}{l}
      \texttt{bands}\\[-0.5mm]
      \hspace{0.7cm}{}\pipeline \unwind{\texttt{albums}}
    \end{array}
$ \hfill}
\end{minipage}

Evaluating this query over
\hyperref[fig:bands-collection]{$\valuefont{bands}^I$}, one of the output documents is as follows:

\begin{lstlisting}
{   "_id": 2,
    "name": "Queen",
    "formation": 1970,
    "albums": { "title": "A Night at the Opera", "release": 1975, "length": "43:08" },
    "members": [
        { "name": "Freddie Mercury", "role": ["lead vocals", "piano"] },
        { "name": "Brian May", "role": ["guitar", "vocals"] },
        { "name": "Roger Taylor", "role": ["drums", "vocals"] },
        { "name": "John Deacon", "role": "bass" }
    ]
}
\end{lstlisting}

We can add a project stage to this query to return only the name of the band
and the album information, and to project away \id, since it is not an
identifier anymore.

\begin{minipage}[t]{0.54\linewidth}
  \begin{lstlisting}
db.bands.aggregate([
    { $unwind: "$albums" },
    { $project: {
        "_id": 0, "name": 1, "album": "$albums"
    } },
])
  \end{lstlisting}%$
\end{minipage}
\hfill%
\begin{minipage}[t]{0.44\linewidth}
  \mquerybox{$
    \begin{array}{l}
      \texttt{bands}\\[-0.5mm]
      \hspace{0.7cm}{}\pipeline \unwind{\texttt{albums}}\\
      \hspace{0.7cm}{}\pipeline \project{\texttt{name},\ \texttt{album}/\texttt{albums}}\\
    \end{array}\\[0.58cm]
$}
\end{minipage}

Evaluating this query over the \valuefont{bands} collection, we get the following collection $O_{\text{band-album}}$:

\begin{lstlisting}
{   "name": "Queen",
    "album": { "title": "Queen", "release": 1973 } },
{   "name": "Queen",
    "album": { "title": "A Night at the Opera", "release": 1975, "length": "43:08" } },
{   "name": "Queen",
    "album": { "title": "News of the World", "release": 1977, "labels": ["EMI", "Elektra"] } },
{   "name": "ABBA",
    "album": { "title": "Waterloo", "release": 1974, "length": "38:09" } },
{   "name": "ABBA",
    "album": { "title": "ABBA", "release": 1975, "labels": ["Polar", "Epic", "Atlantic"] } }
\end{lstlisting}
\end{example}

\subsubsection{Match}
\label{sec:match-syn-sem}
A match stage $\match{\varphi}$ is parameterized with a Boolean expression $\varphi$.
For an input collection $O$, it outputs the objects in $O$ that satisfy~$\varphi$ (while preserving multiplicity of the objects):

\begin{equation*} \evalcdb{\match{\varphi}} = \bag{o \mid o\in O \text{ and }o\models\varphi}
  % O \pipeline \match{\varphi} = \bag{o \mid o\in O \text{ and }o\models\varphi}
  % \label{eq:sem-match}
\end{equation*}

\medskip

We illustrate the match stage with some examples.

\begin{example}\em
Assume that we want to find all bands that were formed before or in 1971.
We can express this as the following match-only query, presented below in the MongoDB and our abstract syntaxes.

\begin{minipage}[t]{0.49\linewidth}
  \begin{lstlisting}
db.bands.aggregate([
    { $match: {
        "formation": { $lte: 1971 }
    } }
])
  \end{lstlisting}%$
\end{minipage}
\hfill%
\begin{minipage}[t]{0.49\linewidth}
  \mquerybox{$
    \begin{array}{l}
      \texttt{bands}\\[-0.5mm]
      \hspace{0.7cm}{}\pipeline \match{\texttt{formation} \leq 1971}
    \end{array}\\[6mm]
$}
\end{minipage}

Evaluated over
\hyperref[fig:bands-collection]{$\valuefont{bands}^I$}, this query returns only the document about Queen, since
ABBA was formed later than 1971.
Now assume that we want to find bands that were formed no later than 1971,
or whose name is \textit{ABBA} or \textit{Beach Boys}.
This would correspond to the following representations:

\begin{minipage}[t]{0.49\linewidth}
  \begin{lstlisting}
db.bands.aggregate([
    { $match: {
        $or: [
            { "formation": { $lte: 1971 } },
            { "name": { $in:
                ["ABBA", "Beach Boys"]
            } }
        ]
    } }
])
  \end{lstlisting}%$
\end{minipage}
\hfill%
\begin{minipage}[t]{0.49\linewidth}
  \mquerybox{$
    \begin{array}{l}
      \texttt{bands}\\[-0.5mm]
      \hspace{0.7cm}{}\pipeline \match{\scriptsize
      \begin{array}[t]{@{}l}
        \texttt{formation} \leq 1971\\
        \qquad\lor\\
        \texttt{name} \in \arraybr{\textit{"ABBA"},\ \textit{"Beach Boys"}}
      \end{array}
      }
    \end{array}\\[1.66cm]
$}
\end{minipage}
This query returns both documents, since the document about ABBA satisfies the
second condition.
\end{example}

\begin{example}\em
Assume that we want to find all eponymous albums (i.e.~titled with their band's name).
This can be achieved by extending the queries from Example~\ref{ex:unwind-syn-sem}:

\begin{minipage}[t]{0.54\linewidth}
  \begin{lstlisting}
db.bands.aggregate([
    { $unwind: "$albums" },
    { $project: {
        "_id": 0, "name": 1, "album": "$albums"
    } },
    { $match: {
        $expr: {
            $eq: [ "name", "album.title" ]
        }
    } }
])
  \end{lstlisting}%$
\end{minipage}
\hfill%
\begin{minipage}[t]{0.44\linewidth}
  \mquerybox{$
    \begin{array}{l}
      \texttt{bands}\\[-0.5mm]
      \hspace{0.7cm}{}\pipeline \unwind{\texttt{albums}}\\
      \hspace{0.7cm}{}\pipeline \project{\texttt{name},\ \texttt{album}/\texttt{albums}}\\
      \hspace{0.7cm}{}\pipeline \match{\texttt{name}=\texttt{album.title}}
    \end{array}\\[1.9cm]
$}
\end{minipage}
Recall that the first two stages, over \hyperref[fig:bands-collection]{$\valuefont{bands}^I$}, evaluate to the
collection $O_{\text{band-album}}$. The final match stage selects two objects
from this collection:

\begin{lstlisting}
{   "name": "Queen",
    "album": { "title": "Queen", "release": 1973 } },
{   "name": "ABBA",
    "album": { "title": "ABBA", "release": 1975, "labels": [ "Polar", "Epic", "Atlantic" ] } }
\end{lstlisting}
\end{example}

\subsubsection{Group}

% \nb{J: Do we want to generalize the syntax and semantics of group to arbitrary aggregation functions?
% More precisely, use "path aggregation definitions", of the form $p/(e,f)$,
% where $p$ is an (output) path, 
% $e$ a path expression and $f:\A \mapsto \V$ a (possibly partial) function that "aggregates" an array of \dvalues.
% Alternatively, we can say that this is expressible with a "vanilla" group followed by a project stage.
% }

A group stage $\group{g}{A}$ is parameterised with an optional \emph{grouping
  key}~$g$ and a sequence $A$ of \emph{aggregation keys} such that $\id\notin A$.
The grouping key~$g$
specifies how the input collection should be partitioned into equivalence
classes. For each equivalence class, the stage outputs one document that
contains an array of aggregated values for each key in $A$.
Given a collection $O$ of objects, the semantics of a group stage is defined as:
\begin{equation*}
\evalcdb{\group{g}{A}} = \bag[\big]{\ \agg{A}{\gamma} \mid \gamma \in \mathsf{groups}_g(O)\ }
% O \pipeline \group{G}{A} = \bag[\big]{\ \agg{A}{\gamma} \mid \gamma \in \mathsf{groups}_G(O)\ }
  % \label{eq:sem-group}
\end{equation*}
where $\groups{g}{O}$ is the set of equivalence classes induced by $g$ over
$O$, and $\agg{A}{\gamma}$ aggregates according to $A$ the objects within the
equivalence class $\gamma$.
\nb{JC: the term "aggregates" is possibly misleading here, since we are not performing any aggregation.
  Maybe write something like "collects part of the objects within the equivalence class $\gamma$,
  as specified by $A$."
  I would also change the name of the function "agg".
  Maybe use "collate" or "collect" instead.
}

If $g$ is a key, then an \emph{equivalence class} $\gamma\in\groups{g}{O}$ is a
non-empty bag composed of all objects $o\in O$ whose value of the grouping
key~$g$ agrees with the class identifier $\gamma_{\text{id}}$, i.e.,
$o(g) = \gamma_{\text{id}}$.
If $g$ is nothing (\nullvalue in MongoDB), then all objects in $O$ are grouped
into one equivalence class $\gamma$ with $\gamma_{\text{id}} = \nullvalue$, and
$\groups{g}{O} = \{O\}$.
The object $\agg{A}{\gamma}$ is  specified as:
\[\agg{A}{\gamma} = \objectbr{\id \mapsto \gamma_{\text{id}}} \ \cdot \ \objectbr[\big]{a \mapsto \agg{a}{\gamma} \mid a \in A} \]
where `$\cdot$' stands for the concatenation of two sequences, and
$\agg{a}{\gamma} = \arraybr{o(a) \mid o\in\gamma}$ collects the values of $a$
in the objects of $\gamma$ (in any order).

\deltabox{
Our formalisation of the group stage departs from the one implemented in
MongoDB, in that we crystallise in the group stage the functionality pertaining to grouping only, while excluding the functionality associated or
realisable with the project stage.

In particular, in MongoDB the group stage may rename paths, construct objects
and arrays, evaluate arbitrary expressions (terms in our terminology), as well
as compute values of aggregate functions.
Instead, the \mquery group stage can only group documents according to the
value of the grouping key and collate the values of aggregation keys into
arrays.

While this affects the expressivity of the group stage, it does not change the
expressivity of \mquery, since every group stage can be preceded by a project
stage to rename paths and to construct values as needed. Furthermore, aggregate
functions such as \valuefont{count}, \valuefont{min}, \valuefont{max},
\valuefont{average}, etc, can be computed in a subsequent project stage.
}

\begin{example}\em
  Assume that we want to know the titles and the band names of all albums that
  were released in a calendar year.
  We can first unwind \valuefont{albums}, then rename the paths so as to nest
  the album title and the name of the band together under the key
  \valuefont{albums}, while storing the album release year 
  % to be 
  under the key
  \valuefont{year}. Then we can group by \valuefont{year} while aggregating
  \valuefont{albums}.

\begin{minipage}[t]{0.49\linewidth}
  \begin{lstlisting}
db.bands.aggregate([
    { $unwind: "$albums" },
    { $project: {
        "year": "$albums.release",
        "albums.title": "$albums.title",
        "albums.band": "$name"
    } },
    { $group: {
        "_id": { "year": "$year" },
        "albums": { $push: "$albums" }
    } }
])
  \end{lstlisting}%$
\end{minipage}
\hfill%
\begin{minipage}[t]{0.49\linewidth}
  \mquerybox{$
    \\[4mm]
    \begin{array}{l}
      \texttt{bands}\\[-0.5mm]
      \hspace{0.7cm}{}\pipeline \unwind{\texttt{albums}}\\
      \hspace{0.7cm}{}\pipeline \project{\scriptsize
      \begin{array}[t]{@{}l}
        \texttt{year}/\texttt{albums.release},\\
        \texttt{albums.title}/\texttt{albums.title},\\
        \texttt{albums.band}/\texttt{name}
      \end{array}
      }\\
      \hspace{0.7cm}{}\pipeline \group{\texttt{year}}{\{\texttt{albums}\}}\\
    \end{array}\\[1.05cm]
$}
\end{minipage}

Over \hyperref[fig:bands-collection]{$\valuefont{bands}^I$} this query returns:
  \begin{lstlisting}
{   "_id": { "year": 1973 },
    "albums": [ { "title": "Queen", "band": "Queen" } ]
},
{   "_id": { "year": 1975 },
    "albums": [ { "title": "A Night at the Opera", "band": "Queen" },
                { "title": "ABBA", "band": "ABBA" } ]
},
{   "_id": { "year": 1977 },
    "albums": [ { "title": "News of the World", "band": "Queen" } ]
},
{   "_id": { "year": 1974 },
    "albums": [ { "title": "Waterloo", "band": "ABBA" } ]
}
\end{lstlisting}

We now unpack the computations involved in this result.

First, 
$\evalq{(\valuefont{bands}\pipeline\unwind{\valuefont{albums}}\pipeline
  \project{\texttt{year}/\texttt{albums.release},\
        \texttt{albums.title}/\texttt{albums.title},\
        \texttt{albums.band}/\texttt{name}})}$ evaluates
to the following collection $O$:

\begin{lstlisting}
{   "_id": 2, "year": 1973, "albums": { "title": "Queen", "band": "Queen" } },
{   "_id": 2, "year": 1975, "albums": { "title": "A Night at the Opera", "band": "Queen" } },
{   "_id": 2, "year": 1977, "albums": { "title": "News of the World", "band": "Queen" } },
{   "_id": 3, "year": 1974, "albums": { "title": "Waterloo", "band": "ABBA" } },
{   "_id": 3, "year": 1975, "albums": { "title": "ABBA", "band": "ABBA" } }
\end{lstlisting}
Assume that the objects in $O$ are named $o_1,\dots,o_5$ in the order they are
listed, starting from the top.
Then $\groups{\texttt{year}}{O}$ produces 4 equivalence classes:
\begin{itemize}\itemsep 0cm
\item $\gamma_1 = \bag{o_1}$ with ${\gamma_1}_{\text{id}} = 1973$,
\item $\gamma_2 = \bag{o_2,o_5}$ with ${\gamma_2}_{\text{id}} = 1975$,
\item $\gamma_3 = \bag{o_3}$ with ${\gamma_3}_{\text{id}} = 1977$, and
\item $\gamma_4 = \bag{o_4}$ with ${\gamma_4}_{\text{id}} = 1974$.
\end{itemize}
Therefore, for instance,  the array
$\agg{\valuefont{albums}}{\gamma_1}$ consists of the single value
$o_1(\valuefont{albums})$%
% = \objectbr{\valuefont{title}\mapsto\textit{"Queen"},\
% \valuefont{band}\mapsto \textit{"Queen"}}$
, while the array
$\agg{\valuefont{albums}}{\gamma_2}$ contains two values,
$o_2(\valuefont{albums})$ and $o_5(\valuefont{albums})$.
\end{example}

\subsubsection{Lookup} A lookup stage $\lookup{X,t}{p}$ is parameterized with 
a set $X$ of variable definitions,
a so-called query template $t$ over $X$, and
an output path $p$.
%
% \begin{equation}
% O \pipeline \lookup{p=C.p_C}{y} = \bag[\big]{\ o \ \cup \ \left\{y\mapsto\sel{p=C.p_C}{o} \right\}  \mid o \in O\ }
%   \label{eq:sem-lookup}
% \end{equation}
% where $\sel{p=C.p_C}{o}$ selects all objects $o'$ from the collection $C$ whose value of path $p_C$ coincides with the value of $p$ in $o$: $\left(o'\in C\mid \eval[o]{p}=\eval[o']{p_C}\right)$.

A \emph{query template} $t$ over $X$ is a query that may contain variable names
from $X$ in place of terms (in path definitions and Boolean expressions); it
can be turned (`grounded') into an MQuery by substituting variables with
d-values.
For an object~$o$, the \emph{grounding} $\ground{X}{t, o}$ of a query template
$t$ \emph{according to} $X$ and $o$ is the query obtained by substituting in
$t$ the variable~$x$ with the d-value $\evald[o]{p}$, for each $x/p\in X$.

Given an input collection $O$, the lookup stage is evaluated as
\begin{equation*}
\evalcdb{\lookup{X,t}{p}} = \bag[\Big]{\ o\oplus^* \big(p\mapsto \agg{\varepsilon}{O_t}\big) \Bigm\vert o \in O\ }.
  % \label{eq:sem-lookup}
\end{equation*}
In other words, the lookup stage $\lookup{X,t}{p}$ extends each object~$o$ in
$O$ with the path-value pair $p\mapsto \agg{\varepsilon}{O_t}$, where
$O_t = \evalq{\big(\ground{X}{t, o}\big)}$ is the evaluation of $t$ grounded
according to $X$ and $o$, and
$o\oplus^* (p\mapsto v)$ stands for $o\oplus (p\mapsto v)$ when $p$ is not
present in $o$ and for $o[p/v]$ when $p$ is present in $o$.
Recall that $\varepsilon$ is the empty path, so $\agg{\varepsilon}{O_t}$
transforms the collection $O_t$ into an array.

The lookup stage allows us to query multiple collections, in a way that is
analogous to the so-called \emph{lateral join} of relational databases. We
illustrate it on an example.
\begin{example}\em
  Let us assume that in addition to the collection \hyperref[fig:bands-collection]{$\valuefont{bands}^I$},
  our database instance $I$ contains the collection $\valuefont{songs}^I$ presented in Figure~\ref{fig:coll-songs},
\begin{figure}
\centering
\begin{minipage}{0.58\linewidth}
  \begin{lstlisting}
{   "_id": 1,
    "title": "One night in Bangkok",
    "composers": ["Björn Ulvaeus"],
    "interprets": ["Murray Head"] 
},
{   "_id": 2,
    "title": "SOS",
    "composers": ["Björn Ulvaeus", "Benny Andersson"],
    "interprets": ["ABBA", "Portishead", "Cher"] 
},
{   "_id": 3,
    "title": "Gloria",
    "composers": ["Van Morisson"],
    "interprets": ["Them", "Patti Smith Group"] 
},
{   "_id": 4,
    "title": "Under pressure",
    "composers": ["David Bowie", "Freddy Mercury"],
    "interprets": ["Queen", "David Bowie"]
},
  \end{lstlisting}
\end{minipage}
\begin{minipage}{0.4\linewidth}
  \begin{lstlisting}[framexbottommargin=7mm]

{   "_id": 5,
    "title": "Ice Ice Baby",
    "interprets": ["Vanilla Ice"],
    "samples": 4
},

{   "_id": 6,
    "title": "Bambi",
    "interprets": ["BB TRickz"],
    "samples": 5
},

{   "_id": 7,
    "title": "7th Street",
    "interprets": ["Chinese Man"],
    "samples": 4
}
  \end{lstlisting}
\end{minipage}
\label{fig:coll-songs}
\caption{Collection $\valuefont{songs}^I$.}
\end{figure}
and that we want to retrieve all songs composed by a member of ABBA.

To do so, we may first construct a query $q_s$ 
over the source collection \valuefont{bands} that retrieves the members of ABBA,
using stages that we have already seen.

\begin{minipage}[t]{0.54\linewidth}
  \begin{lstlisting}
db.bands.aggregate([
    { $match: { "name": "ABBA" } },
    { $unwind: "$members" },
    { $project: {
        "_id": 0, "name": "$members.name"
    } }
])
\end{lstlisting}%$
\end{minipage}
\hfill%
\begin{minipage}[t]{0.44\linewidth}
  \mquerybox{$
    \begin{array}{l}
      \texttt{bands}\\[-0.5mm]
      \hspace{0.7cm}{} \pipeline \match{\texttt{name} = \textit{"ABBA"}}\\
      \hspace{0.7cm}{} \pipeline \unwind{\texttt{members}}\\
      \hspace{0.7cm}{} \pipeline \project{\texttt{name}\, /\, \texttt{members.name}}
    \end{array}\\[0.5cm]
$}
\end{minipage}
Evaluated over $I$,
this query $q_s$ outputs the collection
\begin{lstlisting}
{   "name": "Agnetta Faltskog"   },
{   "name": "Björn Ulvaeus"   },
{   "name": "Benny Andersson"   },
{   "name": "Anni-Frid Lyngstad"   }
\end{lstlisting}
Then we may construct a query template $t$ over the external collection \valuefont{songs}
that retrieves the titles of the songs composed by $x$, where $x$ is a placeholder variable:

\begin{minipage}[t]{0.49\linewidth}
  \begin{lstlisting}
  from : "songs",
  pipeline: [    
    { $match: {
        $expr: { $in: ["$$x", "$composers"] }
    } },
    { $project : {
        "_id": 0, "title": 1
    } }
  ]
\end{lstlisting}%$
\end{minipage}
\hfill
\begin{minipage}[t]{0.49\linewidth}
\mquerybox{$
  \begin{array}{l}
    \texttt{songs}\\[-0.5mm]
    \hspace{0.7cm}{} \pipeline \match{x\, \in\, \texttt{composers}}\\
    \hspace{0.7cm}{} \pipeline \project{\texttt{title}}
  \end{array}
  $\\[1.65cm]
}
\end{minipage}

Because
$t$ is a template, it cannot be evaluated by itself, however, a grounding ot
$t$ can.
For instance, if
$o$ is the object for Björn Ulvaeus, then the instantiated template
$\ground{x/\texttt{name}}{t, o}$ is the query

\begin{minipage}[t]{0.49\linewidth}
  \mquerybox{
    $\texttt{songs} \ \pipeline \ \match{\textit{"Björn Ulvaeus"}\, \in\,
      \texttt{composers}} \ \pipeline \ \project{\texttt{title}}$ }
\end{minipage}

\noindent which can be evaluated over an external collection
\hyperref[fig:coll-songs-appendix]{$\valuefont{songs}^I$}.

% Intuitively, for each object $o$ returned by $q_s$, the lookup stage injects
% into the template $t$ the value $\eval[o]{\texttt{name}}$ of the key
% $\text{\texttt{name}}$ in $o$, evaluates the resulting query
% $\ground{x/\texttt{name}}{t, o}$ over the external collection
% \texttt{songs}$^I$, and stores the result as an array, under the key
% \texttt{compositions}.

During the evaluation of the full lookup stage, the relevant value for $x$ is
injected into the template $t$. In our example, we want to inject the name of
an ABBA member and store the result of the grounded query as an array, under
the key \texttt{compositions}.  The complete lookup stage is as follows:

\begin{minipage}[t]{0.58\linewidth}
  \begin{lstlisting}
    { $lookup: {
        from : "songs",
        let : { x: "$name" },
        pipeline: [    
            { $match: {
                $expr: { $in: ["$$x", "$composers"] }
            } },
            { $project : {
                "_id": 0, "title": 1
            } }
        ],
        as : "compositions"
    } }
  \end{lstlisting}%$
\end{minipage}
\hfill
\begin{minipage}[t]{0.40\linewidth}
  \mquerybox{$ \lookup{x / \texttt{name},\ \texttt{songs} \pipeline \match{x\,
        \in\, \texttt{composers}}\pipeline \project{\texttt{title}}
    }{\texttt{compositions}}
    $\\[3.5cm]
  }
\end{minipage}

The final query $q$ is defined as $q = q_s \pipeline \lookup{x /
  \texttt{name},\ \texttt{songs} \pipeline \match{x\,
        \in\, \texttt{composers}}\pipeline \project{\texttt{title}}}{\texttt{compositions}}$, and evaluates in $I$ to:
\begin{lstlisting}
{   "name": "Agnetta Faltskog",
    "compositions": [ ]
},
{   "name": "Björn Ulvaeus",
    "compositions": [ { "title": "One night in Bangkok" }, { "title": "SOS" } ]
},
{   "name": "Benny Andersson",
    "compositions": [ { "title": "SOS" } ] 
}
{   "name": "Anni-Frid Lyngstad",
    "compositions": [ ]
}
\end{lstlisting}

\end{example}

\begin{example}\em
As a second illustration, over the same database instance,
the following query retrieves all songs composed by a member of some band but not interpreted by that band:
%   \mquerybox{$
%     \begin{array}{l}
%      q_s = \texttt{bands}\\[-0.5mm]
%       \hspace{1.1cm}{} \pipeline \match{\texttt{name} = \textit{"ABBA"}}\\
%       \hspace{1.1cm}{} \pipeline \unwind{\texttt{members}}\\
%       \hspace{1.1cm}{} \pipeline \project{\texttt{name}\, /\, \texttt{members.name}}
%     \end{array}\\[0.1cm]
%     $ \hfill}

\begin{minipage}[t]{0.54\linewidth}
  \begin{lstlisting}
db.bands.aggregate([
    { $unwind: "$members" },
    { $lookup: {
        from : "songs",
        let : { x: "$members.name" },
        pipeline: [    
            { $match: {
                $expr: { $in:
                    ["$$x", "$composers"]
                }
            } },
            { $project : {
                "_id": 0,
                "title": 1, "interprets": 1
            } }
        ],
        as : "compositions"
    } }
    { $unwind: "$compositions" },
    { $match: { $expr:
        { $not: { $in: 
            ["$name", "$compositions.interprets"]
        } }
    } },
    { $project: {
        "_id": 0,
        "composer": "$members.name",
        "title": "$compositions.title",
        "interprets": "$compositions.interprets"
    } }
])
\end{lstlisting}%$
\end{minipage}
\hfill%
\begin{minipage}[t]{0.44\linewidth}
  \mquerybox{$
    \begin{array}{l}
      \texttt{bands}\\
      \hspace{0.6cm}{}\pipeline \unwind{\texttt{members}} \\
      \hspace{0.6cm}{}\pipeline \lookup{x / \texttt{members.name},\
      {\scriptsize
      \begin{array}[b]{@{}l}
        \texttt{songs}\\
        \hspace{0.2cm}{}\pipeline \match{x\, \in\, \texttt{composers}}\\
        \hspace{0.2cm}{}\pipeline \project{\texttt{title},\ \texttt{interprets}}
      \end{array}}
      }{\texttt{composed}} \\
      \hspace{0.6cm}{}\pipeline \unwind{\texttt{compositions}} \\
      \hspace{0.6cm}{}\pipeline \match{\neg(\texttt{name} \in \texttt{compositions.interprets})} \\
      \hspace{0.6cm}{}\pipeline \project{\scriptsize
      \begin{array}[t]{@{}l}
        \texttt{composer}/\texttt{name},\\ \texttt{title}/\texttt{compositions.title},\\ \texttt{interprets}/\texttt{compositions.interprets}
      \end{array}
      }
    \end{array}\\[6.4cm]
$}
\end{minipage}

Over the database instance $I$, this query outputs a single document:

\begin{lstlisting}
{   "composer": "Björn Ulvaeus",
    "title": "One night in Bangkok",
    "interprets": ["Murray Head"]   }
\end{lstlisting}

\end{example}

% For instance, let us assume that in addition to our collection  our collection 
% \begin{lstlisting}
% {
%   { "_id": 1, "title": "One night in Bangkok", "composers": [ "Björn Ulvaeus", "Benny Andersson"], interprets: [ "Murray Head"] },
%   { "_id": 2, "title": "SOS", "composers": [ "Björn Ulvaeus", "Benny Andersson", "Stig Anderson"], interprets: ["ABBA", "Portishead", "Cher"] },
%   { "_id": 3, "title": "Under pressure", "composers": [ "David Bowie", "Freddy Mercury"], interprets: ["Queen", "David Bowie"] },
%   { "_id": 4, "title": "Ice Ice Baby", "interprets": [ "Vanilla Ice"], "samples": "3" },
%   { "_id": 5, "title": "Bambi", "interprets": [ "BB TRickz"], "samples": "4" }
% }
% \end{lstlisting}
% collection tjha we want to find all bands that were formed before or in 1971.
% %
% We can express this as the following match-only query, presented below in the MongoDB and our abstract syntaxes.
%
% \begin{minipage}[t]{0.49\linewidth}
%   \begin{lstlisting}
% db.bands.aggregate([
%     { $match: {
%         "formation": { $lte: 1971 }
%     } }
% ])
%   \end{lstlisting}%$
% \end{minipage}
% \hfill%
% \begin{minipage}[t]{0.49\linewidth}
%   \mquerybox{$
%     \begin{array}{l}
%       \texttt{bands}\\[-0.5mm]
%       \hspace{0.7cm}{}\pipeline \match{\texttt{formation} \leq 1971}
%     \end{array}\\[2mm]
% $ \hfill}
% \end{minipage}

\subsubsection{GraphLookup} 
The graphLookup stage performs a recursive traversal through a collection, following references between documents.
This is useful for connected data, like friend of a friend, reports to, or child of relationships, where one needs to find all documents (transitively) reachable from a given one.
The traversal is performed over objects of an external collection.

The graphLookup operator defines two binary relations: 
$S$ connects objects 
from the input collection to 
objects of the external collection that will act as “starting points” of the traversal,
whereas $R$ defines the traversal relationship itself, within the external collection.

Syntactically, a graphLookup stage $\glookup{p_s,C,p_f,p_t}{p_o}$ is parameterized with
a seed path $p_s$,
a \emph{from} path~$p_f$,
a \emph{to} path~$p_t$,
an output path $p_o$ and
a collection name $C$.
The “seed” and “to” paths $p_s$ and $p_t$ specify the binary relation
$S\subseteq O \times C^I$, where $O$ is the collection input to the graphLookup
stage and $C^I$ the “external” collection.
Similarly, the “from” and “to” paths $p_f$ and $p_t$ specify the binary
relation $R\subseteq C^I \times C^I$.
Let $T$ denote the transitive closure of $S\cup R$.  Then the graphLookup stage
extends each object~$o$ of $O$ with the path-value pair $p\mapsto a_{o}$,
where the array $a_{o}$ consists of the target documents associated to $o$
in $T$.

Formally,
the relations $S$ and $R$ are defined as
\begin{align*}
  S &= \left\{(o_s, o_t) \in O\,\,\times C^I \bigm\vert \eval[o_s]{p_s} = \eval[o_t]{p_t}\right\}\\
  R &= \left\{(o_f, o_t) \in C^I  \times C^I \bigm\vert \eval[o_f]{p_f} = \eval[o_t]{p_t}\right\}
\end{align*}

The transitive closure $T$ of $S\cup R$ is the least superset of $S\cup R$ that satisfies
\[\{(o_1, o_2), (o_2, o_3)\} \subseteq T \te{implies} (o_1, o_3) \in T. \]

% The transitive closure $R^*$ of $R$ is the least superset of $R$ that satisfies
% \[\{(o_1, o_2), (o_2, o_3)\} \subseteq R^* \te{implies} (o_1, o_3) \in R^* \]
% And the composition $S \mathbin; R^*$ of $S$ and $R^*$ is the relation
% \[S \mathbin; R^* = \{(o_i, o_t) \mid 
%     (o_i, o_s) \in S \te{and}
%   (o_s, o_t) \in R^*
% \te{for some} o_s\}
% \]

Then the output of the graphLookup stage is
\begin{equation*}
\evalcdb{\glookup{p_s,C,p_f,p_t}{p_o}}  = \bag[\Big]{\ o \oplus^* (p_o \mapsto a_o) \Bigm\vert o \in O\ },
% \label{eq:sem-glookup}
\end{equation*}
where $a_o$ denotes an array consisting of the target documents associated
to $o$ in the relation $T$ (in no specific order), i.e.,
$a_o = \arraybr{o_t \mid (o, o_t) \in T}$.

\begin{example}\em
As an illustration, we will reuse the database instance $I$ of the previous section,
and focus on the collection \hyperref[fig:coll-songs-appendix]{$\valuefont{songs}^I$}.
% But we will add four objects to the latter, namely:
%
Observe that each of the three last objects in \hyperref[fig:coll-songs-appendix]{$\valuefont{songs}^I$} contains a
new key \texttt{samples}, which indicates (the identifier of) another song that
this song sampled.

We construct a query that retrieves for each song the songs that sampled it,
as well as the songs that sampled them, in a recursive fashion.
% In other word, our relation $T$ will contain two songs iff one of them sampled the other, directly or transitively.
%
We use \hyperref[fig:coll-songs-appendix]{$\valuefont{songs}^I$} both as the current and external collection.
For each object $o$ in this collection,
the query retrieves the objects whose value for the key $\valuefont{sample}$ is the identifier of $o$.
So the ``seed'' and ``from'' paths in this example are both $\id$,
whereas the ``to'' path is \texttt{samples}.
Therefore this search for ``transitive samplers'' can be performed by evaluating over \hyperref[fig:coll-songs-appendix]{$\valuefont{songs}^I$} the
graphLookup stage 
\[\glookup{\ \id,\,\texttt{songs},\,\,\id,\,\texttt{samples}}{\texttt{sampledBy}}\]

In order to keep the output concise, we project away irrelevant information before evaluating this stage.
So our query becomes:

\begin{minipage}[t]{0.54\linewidth}
  \begin{lstlisting}
db.songs.aggregate([
    { $project: {
        "title": 1, "samples": 1
    } },
    { $graphLookup: {
         from: "songs",
         startWith: "$_id",
         connectFromField: "_id",
         connectToField: "samples",
         as: "sampledBy"
    } }
])
\end{lstlisting}%$
\end{minipage}
\hfill%
\begin{minipage}[t]{0.44\linewidth}
  \mquerybox{$
 {\renewcommand{\arraystretch}{1.5}
    \begin{array}{l}
     \texttt{songs}\\[-0.5mm]
      \hspace{0.7cm}{} \pipeline \project{\id,\, \texttt{title}, \texttt{samples}}\\
      \hspace{0.7cm}{} \pipeline \glookup{\ \id,\,\texttt{songs},\,\id,\,\texttt{samples}}{\texttt{sampledBy}}\\
    \end{array}
  }\\[2.05cm]
$}
\end{minipage}

% \[\texttt{songs} 
% \pipeline \project{\id,\, \texttt{title}, \texttt{samples}} \pipeline \glookup{\ \id,\,\texttt{songs},\,\id,\,\texttt{samples}}{\texttt{samplers}}\]

The output of this query extends each document of the input collection with a array for the key \texttt{sampledBy}.
We reproduce the only two objects of the output collection for which this array is nonempty:
\begin{lstlisting}
{   "_id": 4,
    "title": "Under pressure",
    "sampledBy": [
        { "_id": 5, "title": "Ice Ice Baby", "interprets": ["Vanilla Ice"], "samples": 4 },
        { "_id": 6, "title": "Bambi", "interprets": ["BB TRickz"], "samples": 5 },
        { "_id": 7, "title": "7th Street", "interprets": ["Chinese Man"], "samples": 4 }
    ]
},
{   "_id": 5, 
    "title": "Ice Ice Baby",
    "samples": 4, 
    "sampledBy": [
        { "_id": 6, "title": "Bambi", "interprets": ["BB TRickz"], "samples": 5 }
    ]
}
\end{lstlisting}

\end{example}

\subsubsection{Union}
A union stage (\valuefont{\$unionWith} in MongoDB) allows to extend the current
pipeline with the result of an arbitrary query, which in general can be over an external collection.

A union stage $\unionwith{q}$ is parameterized with a query $q$.
It outputs the (bag) union of the input collection~$O$ and the evaluation of $q$ over the current database instance $I$.
\begin{equation*}
\evalcdb{\unionwith{q}} = O \uplus q^I
\end{equation*}

\begin{example}\em
  Consider the collections \hyperref[fig:bands-collection]{$\valuefont{bands}^I$} and \hyperref[fig:coll-songs-appendix]{$\valuefont{songs}^I$}, and
  assume that we want to obtain the names of all people who are either band
  members or song composers.

  The query over \valuefont{bands} is as follows:
  
\begin{minipage}[t]{0.54\linewidth}
  \begin{lstlisting}
db.bands.aggregate([
    { $unwind: "$members" },
    { $project: {
        "_id": 0, "name": "$members.name"
    } },
])
  \end{lstlisting}%$
\end{minipage}
\hfill%
\begin{minipage}[t]{0.44\linewidth}
  \mquerybox{$
    \begin{array}{l}
      \texttt{bands}\\[-0.5mm]
      \hspace{0.7cm}{}\pipeline \unwind{\texttt{members}}\\
      \hspace{0.7cm}{}\pipeline \project{\texttt{name}/\texttt{members.name}}\\
    \end{array}\\[0.58cm]
$}
\end{minipage}

The query over \valuefont{songs} is similar:
  
\begin{minipage}[t]{0.54\linewidth}
  \begin{lstlisting}
db.songs.aggregate([
    { $unwind: "$composers" },
    { $project: {
        "_id": 0, "name": "$composers"
    } },
])
  \end{lstlisting}%$
\end{minipage}
\hfill%
\begin{minipage}[t]{0.44\linewidth}
  \mquerybox{$
    \begin{array}{l}
      \texttt{songs}\\[-0.5mm]
      \hspace{0.7cm}{}\pipeline \unwind{\texttt{composers}}\\
      \hspace{0.7cm}{}\pipeline \project{\texttt{name}/\texttt{composers}}\\
    \end{array}\\[0.58cm]
$}
\end{minipage}

We can take their (bag) union as follows:

\begin{minipage}[t]{0.54\linewidth}
  \begin{lstlisting}
db.bands.aggregate([
    { $unwind: "$members" },
    { $project: {
        "_id": 0, "name": "$members.name"
    } },
    { $unionWith: {
        coll: "songs",
        pipeline: [
            { $unwind: "$composers" },
            { $project: {
                "_id": 0, "name": "$composers"
            } },
        ]
    } }
])
  \end{lstlisting}%$
\end{minipage}
\hfill%
\begin{minipage}[t]{0.44\linewidth}
  \mquerybox{$
    \begin{array}{l}
      \texttt{bands}\\[-0.5mm]
      \hspace{0.7cm}{}\pipeline \unwind{\texttt{members}}\\
      \hspace{0.7cm}{}\pipeline \project{\texttt{name}/\texttt{members.name}}\\
      \hspace{0.7cm}{}\pipeline \unionwith{
        \texttt{songs}\pipeline \unwind{\texttt{composers}}\pipeline \project{\texttt{name}/\texttt{composers}}
      }\\
    \end{array}\\[3.2cm]
$}
\end{minipage}

Note that since the result is a bag, it contains multiple occurrences of the
same names.

\begin{minipage}[t]{0.49\linewidth}
\begin{lstlisting}
{   "name": "Freddie Mercury" },
{   "name": "Brian May" },
{   "name": "Roger Taylor" },
{   "name": "John Deacon" },
{   "name": "Agnetta Faltskog" },
{   "name": "Bjorn Ulvaeus" },
{   "name": "Benny Andersson" },
\end{lstlisting}
\end{minipage}
\begin{minipage}[t]{0.49\linewidth}
\begin{lstlisting}
{   "name": "Anni-Frid Lyngstad" },
{   "name": "Bjorn Ulvaeus" },
{   "name": "Bjorn Ulvaeus" },
{   "name": "Benny Andersson" },
{   "name": "Van Morisson" },
{   "name": "David Bowie" },
{   "name": "Freddy Mercury" }
\end{lstlisting}
\end{minipage}

To get rid of the duplicates, we can add the final stage that groups
the results by \valuefont{name}, $\group{\valuefont{name}}{}$.
\end{example}

\subsection{Additional Stages}
\label{sec:additional-stages}

In this section, we introduce additional stages of the \mquery language, which we excluded from the core fragment.
They are the counterparts of the \mongodb operations \texttt{\$count}, \texttt{\$sort}, \texttt{\$skip} and \texttt{\$limit}.

The stages of the core fragment, as well as the count stage,
operate on collections, which are bags (a.k.a. multisets) of objects.
In contrast,
the sort, skip and limit stages take as input or produce ordered collections.
This is why we first introduce the count stage (using bags, as we did for the core stages).
Then in Section~\ref{sec:ordered_collections}, we extend our semantics to accommodate for ordered collections,
and then only, in the sections that follow, we define the semantics of the three remaining stages.
\nb{JC: this structure feels a little bit odd.
Not sure what the best option is.
Maybe group the ordered collections and the three latter stages in a subsection.
}

\subsubsection{Count}
\nb{JC: Consider dropping this (dull) stage: we can simulate it with group and project if we have a function that returns the size of an array. }
A count stage $\countstage$ is parameterized with a key $k$.
It returns a single object with the size of the input collection as value for $k$, i.e. 
\[\evalcdb{\ \countstage} = \bag{ \objectbr{k \mapsto n}}\]
where $n$ is the cardinality of the collection $O$.

\subsubsection{Ordered Collections}
\label{sec:ordered_collections}

If $O$ is a collection (i.e., a bag of objects), 
we use $\perms{O}$ for the set of all permutations of $O$,
i.e., all possible ways to order the elements of $O$ while respecting their cardinalities.
We represent each of these permutations as an array of objects.
For instance, if $O = \bag{o_1, o_1, o_2}$, then 
$\perms{O} = \{
  \arraybr{o_1, o_1, o_2},
  \arraybr{o_1, o_2, o_1},
  \arraybr{o_2, o_1, o_1}
\}$.
We call an element of $\perms{O}$ an \emph{ordered collection}.

Some of the stages that we will introduce take as input or produce an ordered collection.
This contrasts with the stages that we have seen so far, which instead take as input and produce a ``regular'' collection--a bag of objects.
In order to provide a unified semantics for all (core and non-core) stages, 
we first reformulate the semantics of the stages seen so far, so that it can handle ordered collections.
The definition is equivalent, but non-deterministic (therefore a little bit harder to understand, which is why we have used bags so far).

First, if $a$ is an array of objects, we write $\asbag{a}$ for the bag that it induces.
For instance, if $a = \arraybr{o_1, o_1, o_2}$, then $\asbag{a} = \bag{o_1, o_1, o_2}$ (or equivalently $\bag{o_1, o_2, o_1}$ or $\bag{o_2, o_1, o_1}$).
Next, recall that the evaluation of a core stage $s$ over a collection $O$ and database instance $I$ is the bag denoted
$\evalcdb{s}$.
We adopt a nearly identical notation for ordered collections:
we write $\evalcdbl{s}$ for the \emph{evaluation of~$s$ over an ordered collection $a$} and database instance $I$.

We can now adapt the semantics of our stages accordingly.
If $s$ is a project, unwind, match, group, lookup, graphLookup, union or count stage,
then $\evalcdbl{s}$ is any permutation of the collection obtained by evaluating $s$ over (the bag induced by) $a$ and $I$.
Or in other words:
\[
\evalcdbl{s} \ \ \in \ \ \perms[\Big]{\ \evalcdb[\asbag{a},\ I]{s}\ }
\]
With a slight abuse of notation, in what follows, we may use $\perms{a}$ (instead of $\perms{\asbag{a}}$) to denote the set of permutations of an array $a$.

\subsubsection{Sort}

We call \emph{object comparator} an expression of the form $p^+$ or $p^-$, where $p$ is a path.
Intuitively, $p$ indicates the values of what path should be compared in two objects,
and the exponent ($+$ or $-$) determines whether the comparison should be performed in ascending or descending order.

A sort stage $\sort{S}$ is parameterized with a nonempty sequence $S$ of object comparators,
which are combined in a lexicographic fashion to sort the input collection.
For instance, the following \mquery (and its aggregate query counterpart on the left) sorts the bands of our collection with respect to their year of formation in descending order and, in case of a tie, with respect to their names in ascending order.

\begin{minipage}[t]{0.49\linewidth}
  \begin{lstlisting}
db.bands.aggregate([
  { $sort: { "formation": -1, "name": 1 } }
])
  \end{lstlisting}%$
\end{minipage}
\hfill%
\begin{minipage}[t]{0.49\linewidth}
 \mquerybox{$
    \begin{array}{l}
      \texttt{bands}\\[-0.5mm]
      \hspace{0.7cm}{}\pipeline \sort{\texttt{formation}^-,\ \texttt{name}^+}
    \end{array}
  $}
\end{minipage}

% \mquerybox{
% \[
%   \texttt{bands} \pipeline \sort{
%   \mathtt{formation}^-,\  \mathtt{name}^+
% }
% \]
% }

Formally, the sequence $S = p_1^{e_1}, \ldots,  p_n^{e_n}$ specifies a total preorder $\preceq_S$
over the universe $\O$ of all objects, defined inductively as 
\begin{align*}
o_1 \preceq_S o_2 \quad \te{iff} \quad & \eval[o_1]{p_1} < \eval[o_2]{p_1} \ \te{and} \ e_1 \te{is} +\\
  & \hspace{2cm}\te{or} \\
                          & \eval[o_1]{p_1} > \eval[o_2]{p_1} \ \te{and} \ e_1 \te{is} -\\
  & \hspace{2cm}\te{or}\\
                          & \eval[o_1]{p_1} = \eval[o_2]{p_1} \ \te{and} \ o_1 \preceq_{p_2^{e_2} \ldots,  p_n^{e_n}} o_2
\end{align*}

% \emph{object comparator} $\preceq$,
% Formally, a comparator is a so-called \emph{total preorder} over the universe $\O$ of all objects,
% i.e.~any binary relation over $\O$ such that for $ o, o_1, o_2 \in O $:
% \begin{itemize}
%   \item $o \preceq o$ (reflexivity)
%   \item $o_1 \preceq o_2 \te{and} o_2 \preceq o_3 \te{imply} o_1 \preceq o_3$ (transitivity)
%   \item $o_1 \preceq o_2$ or $o_2 \preceq o_1$ ("totality")
% \end{itemize}
% Note that $o_1 \preceq o_2$ and $o_2 \preceq o_1$ may both hold (for instance, if we sort a collection of people by age, then two of them may have the same age).
% If this cannot happen, then $\preceq$ is called a \emph{total order}.
% A specific case of total order is the natural order $\le$ over \dvalues (see Section XXX).

A sort stage $\sort{S}$ takes as input a collection, and outputs a permutation of this collection that complies with $\preceq_S$.
Or formally, for a database instance $I$, array $a$ of objects and object comparator $S$:
\[\evalcdbl{\ \sort{S}}\ \quad \te{is any array} a_s \in \perms{a} \te{that satisfies} i \le j \te{implies} a_s[i] \preceq_S a_s[j]\]
for $i,j \in \{1,\dots, n\}$, where $n$ is the length of $a_s$. 

Note that the semantics of the sort stage is non-deterministic when $\preceq_S$ is not an order, but only a preorder (for instance, if we sort our collection of bands by year of formation only, and two bands have been formed during the same year).
In other words, if two distinct objects in $a$ are equivalent w.r.t.~$\preceq_S$,
then they can appear in any order in the array $\evalcdbl{\ \sort{S}}$.
In particular, we do not require stable sorting (but this may be a feature of a specific implementation).

% \deltabox{
%   The \texttt{\$sort} operator of \mongodb is parameterized with a \emph{sequence} of object comparators (each specified as functions from $\O$ to $\{-1, 0, 1\}$, as is conventional in programming languages).
%   Then sorting is based on the lexicographic product of these comparators.
%   Our representation can be viewed as a slight abstraction: since the lexicographic product of two total preorders is itself a total preorder,
%   our object comparator $\preceq$ may stand for the product of several comparators.
% }

\begin{example}\em
  Assume that we want to list the year of release and title of albums, sorted
  by the year in ascending order and, in case of a tie, by the title in
  descending order. This can be achieved by the following query:
  
\begin{minipage}[t]{0.49\linewidth}
  \begin{lstlisting}
db.bands.aggregate([
  { $unwind: "$albums" },
  { $project: {
      "_id": 0,
      "year": "$albums.release",
      "title": "$albums.title"
  } },
  { $sort: {
      "year": 1, "title": -1
  } }
])
\end{lstlisting}
\end{minipage}
\hfill%
\begin{minipage}[t]{0.49\linewidth}
 \mquerybox{$
    \begin{array}{l}
      \texttt{bands}\\[-0.5mm]
      \hspace{0.7cm}{}\pipeline \unwind{\texttt{albums}}\\
      \hspace{0.7cm}{}\pipeline \project{\texttt{year}/\texttt{albums.release},\ \texttt{title}/\texttt{albums.title}}\\
      \hspace{0.7cm}{}\pipeline \sort{\texttt{year}^+,\ \texttt{title}^-}
    \end{array}\\[1.9cm]
  $}
\end{minipage}

Evaluated over \hyperref[fig:bands-collection]{$\valuefont{bands}^I$}, this query returns
\begin{lstlisting}
  { "year": 1973, "title": "Queen" },
  { "year": 1974, "title": "Waterloo" },
  { "year": 1975, "title": "ABBA" },
  { "year": 1975, "title": "A Night at the Opera" },
  { "year": 1977, "title": "News of the World" }
\end{lstlisting}

Instead, changing the last stage to $\sort{\texttt{year}^+,\ \texttt{title}^+}$ produces the following result:
\begin{lstlisting}
  { "year": 1973, "title": "Queen" },
  { "year": 1974, "title": "Waterloo" },
  { "year": 1975, "title": "A Night at the Opera" },
  { "year": 1975, "title": "ABBA" },
  { "year": 1977, "title": "News of the World" }
\end{lstlisting}
\end{example}

\subsubsection{Limit}

A limit stage $\limit{m}$ is parameterized with a positive integer $m$ that determines the maximal size of the output.
This stage takes as input an ordered collection $a$, and outputs the first $m$ elements of $a$,
or $a$ as a whole if the size of $a$ is less than or equal to $m$.
Or formally, for a database instance $I$, array $a$ of objects and positive integer $m$:
\[\evalcdbl{\ \limit{m}}\ = \arraybr[\Big]{\ a[i] \Bigm\vert i \in \{1,\dots, \min(m,n)\}\ }\]
where $n$ is the length of $a$.

% \nb{JC: we could relax the requirement on the order of the output,
%   i.e.~the limit stage could output any permutation of of the first $n$ elements of $a$.
% I have no preference.
% }

\begin{example}\em
  The following query outputs the band with the most number of albums:
  
\begin{minipage}[t]{0.49\linewidth}
  \begin{lstlisting}
db.bands.aggregate([
  { $project: {
      "_id": 0,
      "name": 1, "albums": 1,
      "albums_number": { $size: "$albums" }
  } },
  { $sort: {
      "albums_number": -1
  } },
  { $limit: 1 }
])
\end{lstlisting}
\end{minipage}
\hfill%
\begin{minipage}[t]{0.49\linewidth}
 \mquerybox{$
    \begin{array}{l}
      \texttt{bands}\\[-0.5mm]
      \hspace{0.7cm}{}\pipeline \project{\texttt{name},\ \texttt{albums},\ \texttt{albums\_number}/\texttt{size}(\texttt{albums})}\\
      \hspace{0.7cm}{}\pipeline \sort{\texttt{albums\_number}^-}\\
      \hspace{0.7cm}{}\pipeline \limit{1}
    \end{array}\\[1.9cm]
  $}
\end{minipage}

\noindent where the \valuefont{\$size} function returns the number of elements in the input array.

Evaluated over \hyperref[fig:bands-collection]{$\valuefont{bands}^I$}, this query returns
\begin{lstlisting}
{   "name": "Queen",
    "albums": [
        { "title": "Queen", "release": 1973 },
        { "title": "A Night at the Opera", "release": 1975, "length": "43:08" },
        { "title": "News of the World", "release": 1977, "labels": ["EMI", "Elektra"] }
    ],
    "albums_number": 3
} 
\end{lstlisting}
\end{example}

\subsubsection{Skip}

A skip stage $\skipstage$ is parameterized with a positive integer $m$ that determines how many elements of its input must be skipped.
This stage takes as input an ordered collection $a$, and outputs $a$ without its first $m$ elements.
Or formally, for a database instance $I$, array $a$ of objects and positive integer $m$:
\[\evalcdbl{\ \skipstage}\ = \arraybr[\Big]{\ a[i] \Bigm\vert i \in \{m+1,\dots, n \}\ }\]
where $n$ is the length of $a$.

% \nb{JC: Same observation as for limit: we could relax the requirement on the order of the output.
% Again, I have no preference.
% }

%%% Local Variables:
%%% mode: latex
%%% TeX-master: "main-tr"
%%% fill-column: 79
%%% End:

\section{Expressive Power of MQuery, Informally}
\label{sec:expressivity}
In this section we show an informal correspondence between MQuery and
well-known relational query languages~\cite{Codd90,ThomasFisher86,Vand01}. In
particular, we confirm the intuition that the MongoDB aggregation framework is
very expressive: at least as expressive as full \emph{relational algebra (RA)}.
To summarise:
\begin{itemize}\itemsep 0cm
\item the match, unwind, project and group fragment captures RA over a single
  relation.
\item the match, unwind, project, group, and lookup fragment captures arbitrary
  RA.
\end{itemize}
This means that joins are possible in MongoDB, even without lookup, the
explicit join operator.

Recall the RA operators:
\begin{itemize}\itemsep 0cm
\item \emph{Selection} ($\sigma$)  uses a Boolean condition to filter out tuples that do
  not satisfy the condition;
\item \emph{Projection} ($\pi$) can be used to project away attributes;
\item \emph{Join} ($\bowtie$) can be used to extend the tuples in one relation with tuples
  in another relation;
\item \emph{Union} ($\cup$) and \emph{difference} ($\setminus$) are standard set operators (under set semantics) or bag operators (under bag semantics).
\end{itemize}
In this report, we assume that the reader is familiar with the syntax and
semantics of relational algebra expressions.

We
illustrate
% establish 
the expressive power of MQuery by showing how each relational algebra operator can be simulated. Intuitively:

\begin{itemize}
\item \emph{Selection} ($\sigma$). The match stage directly
  corresponds to a selection, filtering documents based on a Boolean condition.

\item \emph{Projection} ($\pi$). The project stage provides projection capabilities. More precisely, it corresponds to an \emph{extended projection},
  as it allows to compute new values through expressions involving function calls, conditional expressions and complex value constructions. 
  % though it's more flexible than traditional relational projection as it can create new attributes through computed expressions.

\item \emph{Join} ($\bowtie$). Most interestingly, joins can be expressed in multiple ways:
  \begin{itemize}
  \item explicitly using lookup stages, or
  \item implicitly by combining the unwind and group operations.
  \end{itemize}

\item \emph{Union} (set union $\cup$ and bag union $\uplus$). Similarly to
  joins, unions can be expressed in multiple ways:
  \begin{itemize}
    \item with the union stage, which provides explicit union semantics, or
  \item through a combination of group and project.
  \end{itemize}

\item \emph{Difference} ($\setminus$). This operation can be expressed through a combination of group, project and match.

\item \textbf{Beyond Relational Algebra:} MQuery extends beyond traditional RA:
  \begin{itemize}
  \item \mquery subsumes \emph{nested relational algebra (NRA)} designed to
    operate over relations that are not in first normal form, i.e., where an attribute
    value can be a nested relation.
    \begin{itemize}
    \item Group corresponds to the \emph{nest ($\nu$)} operator, which allows
      creating nested relations.
    \item Unwind corresponds to the \emph{unnest ($\chi$)} operator, which allows
      flattening nested relations.
    \end{itemize}
    
  \item \mquery supports a form of linear recursion~\cite{Agrawal88} via graphLookup.
    % \item Complex value manipulation in projections
    % \item Array manipulation and aggregation
  \end{itemize}
\end{itemize}

We now illustrate the correspondence of \mquery with relational operators on
examples. Below we make use of the nested relation \textit{bands} representing
the same information as the JSON collection \hyperref[fig:bands-collection]{$\valuefont{bands}^I$} shown in
Figure~\ref{fig:relation-bands}.

\begin{figure}\small
  \rabox{\centering
  \begin{tabular}{@{}c@{~~}c@{~~}c@{~~}l@{~~}l@{}}
    \id
    & name
    & formation
    & \begin{tabular}{@{}l@{}}
        \\[-3.5mm]
        \begin{tabular}{@{}|@{~}P{3cm}@{~~}P{1cm}@{~~}P{1cm}@{~~}P{1.4cm}@{~}|@{}}
          \multicolumn{4}{l}{albums} \\
          \hline
          title & release & length &
                                     \begin{tabular}{@{}|@{~}P{1.2cm}@{~}|@{}}
                                       \multicolumn{1}{l}{labels} \\
                                       \hline
                                       name \\\hline
                                       \multicolumn{1}{l}{}\\[-3mm]
                                     \end{tabular}
          \\[2mm]\hline
        \end{tabular}
        \\[-3.5mm]~
      \end{tabular}

    & \begin{tabular}{@{}l@{}}
        \\[-3.5mm]
        \begin{tabular}{@{}|@{~}P{2.7cm}@{~~}P{1.8cm}@{~}|@{}}
          \multicolumn{2}{l}{members} \\
          \hline
          name & 
                 \begin{tabular}{@{}|@{~}P{1.6cm}@{~}|@{}}
                   \multicolumn{1}{l}{role} \\
                   \hline
                   title \\\hline
                   \multicolumn{1}{l}{}\\[-3mm]
                 \end{tabular}
          \\[2mm]\hline
        \end{tabular}
        \\[-3.5mm]~
      \end{tabular}
    
    \\[3mm]
    \toprule
    
    2
    & \textit{Queen}
    & 1970
    & \begin{tabular}{@{}l@{}}
        \\[-3.5mm]
        \begin{tabular}{@{}|@{~}P{3cm}@{~~}P{1cm}@{~~}P{1cm}@{~~}P{1.4cm}@{~}|@{}}
          \hline
          \textit{Queen} & 1973 &  & \\
          \textit{A Night at the Opera} & 1975 & \textit{43:08} & \\  
          \textit{News of the World} & 1977 &  & \begin{tabular}{@{}|@{~}P{1.2cm}@{~}|@{}}
                                                   \multicolumn{1}{l}{}\\[-2.8mm]
                                                   \hline
                                                   \textit{EMI} \\ 
                                                   \textit{Elektra} \\ \hline
                                                   \multicolumn{1}{l}{}\\[-2.8mm]
                                                 \end{tabular}
          \\[2mm] \hline 
        \end{tabular}
        \\[-3.5mm]~
      \end{tabular}
    
    & \begin{tabular}{@{}l@{}}
        \\[-3.5mm]
        \begin{tabular}{@{}|@{~}P{2.7cm}@{~~}P{1.8cm}@{~}|@{}}
          \hline
          \textit{Freddie Mercury} & \begin{tabular}{@{}|@{~}P{1.6cm}@{~}|@{}}
                                       \multicolumn{1}{l}{}\\[-2.8mm]
                                       \hline
                                       \textit{lead vocals} \\ 
                                       \textit{piano} \\ \hline
                                       \multicolumn{1}{l}{}\\[-2.8mm]
                                     \end{tabular}
          \\  
          \textit{Brian May} & \begin{tabular}{@{}|@{~}P{1.6cm}@{~}|@{}}
                                 \multicolumn{1}{l}{}\\[-2.8mm]
                                 \hline
                                 \textit{guitar} \\ 
                                 \textit{vocals} \\ \hline
                                 \multicolumn{1}{l}{}\\[-2.8mm]
                               \end{tabular}
          \\  
          \textit{Roger Taylor} & \begin{tabular}{@{}|@{~}P{1.6cm}@{~}|@{}}
                                    \multicolumn{1}{l}{}\\[-2.8mm]
                                    \hline
                                    \textit{drums} \\ 
                                    \textit{vocals} \\ \hline
                                    \multicolumn{1}{l}{}\\[-2.8mm]
                                  \end{tabular}
          \\  
          \textit{John Deacon} & \begin{tabular}{@{}|@{~}P{1.6cm}@{~}|@{}}
                                   \multicolumn{1}{l}{}\\[-2.8mm]
                                   \hline
                                   \textit{bass} \\ \hline
                                   \multicolumn{1}{l}{}\\[-2.8mm]
                                 \end{tabular}
          \\[2mm]\hline
        \end{tabular}
        \\[-3.5mm]~
      \end{tabular}
    \\[3mm]

    3
    & \textit{ABBA}
    & 1972
    & \begin{tabular}{@{}l@{}}
        \\[-3.5mm]
        \begin{tabular}{@{}|@{~}P{3cm}@{~~}P{1cm}@{~~}P{1cm}@{~~}P{1.4cm}@{~}|@{}}
          \hline
          \textit{Waterloo} & 1974 & \textit{38:09} & \\
          \textit{ABBA} & 1975 &  & \begin{tabular}{@{}|@{~}P{1.2cm}@{~}|@{}}
                                          \multicolumn{1}{l}{}\\[-2.8mm]
                                          \hline
                                          \textit{Polar} \\ 
                                          \textit{Epic} \\ 
                                          \textit{Atlantic} \\ \hline
                                          \multicolumn{1}{l}{}\\[-2.8mm]
                                        \end{tabular}
          \\[2mm] \hline 
        \end{tabular}
        \\[-3.5mm]~
      \end{tabular}
    
    & \begin{tabular}{@{}l@{}}
        \\[-3.5mm]
        \begin{tabular}{@{}|@{~}P{2.7cm}@{~~}P{1.8cm}@{~}|@{}}
          \hline
          \textit{Agnetta Faltskog} & \begin{tabular}{@{}|@{~}P{1.6cm}@{~}|@{}}
                                        \multicolumn{1}{l}{}\\[-2.8mm]
                                        \hline
                                        \textit{lead vocals} \\ \hline
                                        \multicolumn{1}{l}{}\\[-2.8mm]
                                      \end{tabular}
          \\  
          \textit{Björn Ulvaeus} & \begin{tabular}{@{}|@{~}P{1.6cm}@{~}|@{}}
                                     \multicolumn{1}{l}{}\\[-2.8mm]
                                     \hline
                                     \textit{guitar} \\ 
                                     \textit{vocals} \\ \hline
                                     \multicolumn{1}{l}{}\\[-2.8mm]
                                   \end{tabular}
          \\  
          \textit{Benny Andersson} & \begin{tabular}{@{}|@{~}P{1.6cm}@{~}|@{}}
                                       \multicolumn{1}{l}{}\\[-2.8mm]
                                       \hline
                                       \textit{keyboard} \\ 
                                       \textit{vocals} \\ \hline
                                       \multicolumn{1}{l}{}\\[-2.8mm]
                                     \end{tabular}
          \\  
          \textit{Anni-Frid Lyngstad} & \begin{tabular}{@{}|@{~}P{1.6cm}@{~}|@{}}
                                          \multicolumn{1}{l}{}\\[-2.8mm]
                                          \hline
                                          \textit{vocals} \\ \hline
                                          \multicolumn{1}{l}{}\\[-2.8mm]
                                        \end{tabular}
          \\[2mm] \hline 
        \end{tabular}
        \\[-3.5mm]~
      \end{tabular}
  \end{tabular}}

  \caption{Nested relation \textit{bands} about Queen and ABBA.}
  \label{fig:relation-bands}
\end{figure}

\subsection{Selection}
The correspondence between selection and match is quite straightforward. We saw
in Section~\ref{sec:match-syn-sem} that the match stage can select objects based
on atomic conditions of the form $p\op\ell$ or $p_1\op p_2$ (where $\op\in\{=,\leq\}$),
and Boolean combinations thereof.

For instance, we can select in both languages all bands that were formed in or before 1971:

\begin{minipage}[t]{0.49\linewidth}
  \rabox{
    $\sigma_{\text{formation}\leq 1971}(\textit{bands})$
  }
\end{minipage}
\hfill%
\begin{minipage}[t]{0.49\linewidth}
  \mquerybox{
    $\texttt{bands}\pipeline \match{\texttt{formation} \leq 1971}
    \phantom{\sigma_{\text{ft}\leq 19}(\textit{bd})}$
  }
\end{minipage}

Similarly, we can select all bands that were formed in or before 1971, or named
``ABBA''.

\begin{minipage}[t]{0.49\linewidth}
  \rabox{
    $\sigma_{(\text{formation}\leq 1971) \ \lor \ (\text{name}=\textit{``ABBA''})}(\textit{bands})$
  }
\end{minipage}
\hfill%
\begin{minipage}[t]{0.49\linewidth}
  \mquerybox{
    $\texttt{bands}\pipeline \match{(\texttt{formation} \leq 1971) \ \lor \ (\texttt{name} = \textit{``ABBA''})}
    \phantom{\sigma_{\text{f}\leq 1}(\textit{b}}$
    }
\end{minipage}

As expected, the first query returns only the tuple/document about Queen, while
the second query returns both bands.

\subsection{Projection}

There is a straightforward correspondence between projection and project as well.
For instance, for each band, we can output only its name and its year of formation with the following queries:

\begin{minipage}[t]{0.49\linewidth}
  \rabox{
    $\pi_{\text{name},\ \text{formation}}(\textit{bands})$
  }
\end{minipage}
\hfill%
\begin{minipage}[t]{0.49\linewidth}
  \mquerybox{
    $\texttt{bands}\pipeline \project{\texttt{name},\ \texttt{formation}}
    \phantom{\pi_{\text{ft}}(\textit{bd})}$
  }
\end{minipage}

Over the \textit{bands} relation this query evaluates to the relation:
\begin{center} 
  \begin{minipage}[t]{3cm}
    \rabox{
      \begin{tabular}{@{}c@{~~}c@{~~}c@{~~}l@{~~}l@{}}
        name           & formation \\\toprule
        \textit{Queen} & 1970      \\
        \textit{ABBA}  & 1972      \\
      \end{tabular}}
  \end{minipage}
\end{center}

\subsection{Join}

In traditional relational databases, join computation is recognized as one of
the primary reasons for slow query performance. To ensure efficient query
evaluation,
% at the beginning, 
many NoSQL databases (including \mongodb) were initially not offering an
explicit join operator. However, owing to the expressivity of the MongoDB
aggregation framework, it was already possible to compute joins over a same
collection even before the introduction the \texttt{\$lookup} stage (which is an
explicit join operator). In what follows, we show how to simulate joins in
\mquery without using lookup.

Assume that we want to find all pairs of bands that released an album in the
same year. A (nested) RA query would need to unnest the attribute “albums” and
join the resulting relation with itself. Before unnesting, to simplify
specification of the joining condition, we could rename the attributes “name” and
“albums”, and
% to reduce the computational overheads,
project away irrelevant information, such as the year of formation or the members of the
bands.\nb{JC: 
I commented out "to reduce the computational overheads" to be consistent with Section 5.
}

A possible RA query is shown below on the left.

\begin{minipage}[t]{0.44\linewidth}
  \rabox{
    $
    \begin{array}{l}
      \bowtie_{(\text{albums}_1.\text{release} = \text{albums}_2.\text{release}) \ \land \ (\text{name}_1 < \text{name}_2)}\\
      \hspace{0.7cm}\chi_{\text{albums}_1}\\
      \hspace{1.4cm}\pi_{\text{name}_1/\text{name},\ \text{albums}_1/\text{albums}}\\
      \hspace{2.1cm}\textit{bands}\\
      \hspace{0.7cm}\chi_{\text{albums}_2}\\
      \hspace{1.4cm}\pi_{\text{name}_2/\text{name},\ \text{albums}_2/\text{albums}}\\
      \hspace{2.1cm}\textit{bands}
    \end{array}
    $\\[1.3cm]
  }
\end{minipage}
\hfill
\begin{minipage}[t]{0.54\linewidth}
  \mquerybox{
    $
    \begin{array}{l}
      \texttt{bands}\\
      \hspace{0.7cm}{} \pipeline \project{\texttt{rel.name}/\texttt{name},\ \texttt{rel.albums}/\texttt{albums}}\\
      \hspace{0.7cm}{} \pipeline \unwind{\texttt{rel.albums}}\\[1mm]
      \hspace{0.7cm}{} \pipeline \group{}{\texttt{rel}}\\
      \hspace{0.7cm}{} \pipeline \project{\texttt{rel1}/\texttt{rel},\ \texttt{rel2}/\texttt{rel}}\\
      \hspace{0.7cm}{} \pipeline \unwind{\texttt{rel1}}\\
      \hspace{0.7cm}{} \pipeline \unwind{\texttt{rel2}}\\
      \hspace{0.7cm}{} \pipeline \match{\scriptsize
      \begin{array}[t]{@{}l}
        \texttt{rel1}.\texttt{albums.release} = \texttt{rel2}.\texttt{albums.release}\\
        \qquad\land \\
        \texttt{rel1}.\texttt{name} < \texttt{rel2}.\texttt{name}\\      
      \end{array}
      }\\
    \end{array}
    $
  }
\end{minipage}

An equivalent \mquery is shown to the right. We demonstrate now the
computations of the above \mquery step by step.
The first two stages are similar to a query in Example~\ref{ex:unwind-syn-sem}
and evaluate to:

\begin{lstlisting}
{ "rel": {
      "name": "Queen",
      "albums": { "title": "Queen", "release": 1973 }
} },
{ "rel": {
      "name": "Queen",
      "albums": { "title": "A Night at the Opera", "release": 1975, "length": "43:08" }
} },
{ "rel": {
      "name": "Queen",
      "albums": { "title": "News of the World", "release": 1977, "labels": ["EMI", "Elektra"] }
} },
{ "rel": {
      "name": "ABBA",
      "albums": { "title": "Waterloo", "release": 1974, "length": "38:09" }
} },
{ "rel": {
      "name": "ABBA",
      "albums": { "title": "ABBA", "release": 1975, "labels": ["Polar", "Epic", "Atlantic"] }
} }
\end{lstlisting}
Let us denote this collection by $O_{\text{rel}}$.

The next stage, $\group{}{\texttt{rel}}$, has no grouping key, so it groups all 5 objects in the same array:

\begin{lstlisting}
{ "_id": null,
  "rel": [
    { "name": "Queen",
      "albums": { "title": "Queen", "release": 1973 }
    },
    { "name": "Queen",
      "albums": { "title": "A Night at the Opera", "release": 1975, "length": "43:08" }
    },
    { "name": "Queen",
      "albums": { "title": "News of the World", "release": 1977, "labels": ["EMI", "Elektra"] }
    },
    { "name": "ABBA",
      "albums": { "title": "Waterloo", "release": 1974, "length": "38:09" }
    },
    { "name": "ABBA",
      "albums": { "title": "ABBA", "release": 1975, "labels": ["Polar", "Epic", "Atlantic"] }
    }
] }
\end{lstlisting}

The project stage
$\project{\texttt{rel}_1/\texttt{rel},\ \texttt{rel}_2/\texttt{rel}}$ creates
two copies of the array \valuefont{rel}, named \valuefont{rel1} and
\valuefont{rel2}.
Unwinding first \valuefont{rel1} and then \valuefont{rel2} creates a collection
of 25 objects, containing all possible pairs of objects from
$O_{\text{rel}}$. In other words, it computes the cross-product of
$O_{\text{rel}}$ with itself.

The final match stage, selects only those pairs that satisfy the joining
condition; its output is as follows:

\begin{lstlisting}
{ "_id": null,
  "rel1": {
      "name": "ABBA",
      "albums": { "title": "ABBA", "release": 1975, "labels": ["Polar", "Epic", "Atlantic"] } 
  },
  "rel2": {
      "name": "Queen",
      "albums": { "title": "A Night at the Opera", "release": 1975, "length": "43:08" }
  }
}
\end{lstlisting}

This trick of grouping all objects in the current collection in the same array
is essential to us being able to perform arbitrary joins over the same source
collection.  In this particular example, we could optimise the \mquery and
group by the album release year. By doing so we would avoid the quadratic
explosion of the size of an intermediate collection. However, this only works
for equality joining conditions. If the joining condition is an inequality,
then grouping the whole current collection in the same object is required.

The example considered above required computing a self-join, so we could simply
duplicate the collection \valuefont{rel}.
In general, in relational algebra, one can perform
%compute 
a join over the results of two
arbitrary queries, each of which may perform arbitrarily complex computations.
This is a
% represents one
mismatch with \mquery: relational algebra queries are tree
structured, while \mqueries are linear pipelines.
To overcome this limitation of \mquery, we demonstrate that it is possible to
\emph{linearise} relational algebra expressions in the next section on union.

\subsection{Union}
\label{sec:expressivity-union}
Consider a query $q_1$ that retrieves the titles of albums that were released
after 1975:

\begin{minipage}[t]{0.49\linewidth}
  \rabox{
    $
    \begin{array}{l}
      \hspace{0.0cm}\pi_{\text{ans}/\text{albums.title}}\\
      \hspace{0.7cm}\sigma_{\text{albums}.\text{release} > 1975}\\
      \hspace{1.4cm}\chi_{\text{albums}}\\
      \hspace{2.1cm}\textit{bands}\\
    \end{array}
    $
  }
\end{minipage}
\hfill
\begin{minipage}[t]{0.49\linewidth}
  \mquerybox{
    $
    \begin{array}{l}
      \texttt{bands}\\
      \hspace{0.7cm}{} \pipeline \unwind{\texttt{albums}}\\
      \hspace{0.7cm}{} \pipeline \match{\texttt{albums.release} > 1975}\\
      \hspace{0.7cm}{} \pipeline \project{\texttt{ans}/\texttt{albums.title}}\\
    \end{array}
    $
  }
\end{minipage}
and a query $q_2$ that retrieves the names of band members who play
guitar:

\begin{minipage}[t]{0.49\linewidth}
  \rabox{
    $
    \begin{array}{l}
      \hspace{0.0cm}\pi_{\text{ans}/\text{members.name}}\\
      \hspace{0.7cm}\sigma_{\text{members.role}=\textit{``guitar''}}\\
      \hspace{1.4cm}\chi_{\text{members.role}}\\
      \hspace{2.1cm}\chi_{\text{members}}\\
      \hspace{2.8cm}\textit{bands}\\
    \end{array}
    $
  }
\end{minipage}
\hfill
\begin{minipage}[t]{0.49\linewidth}
  \mquerybox{
    $
    \begin{array}{l}
      \texttt{bands}\\
      \hspace{0.7cm}{} \pipeline \unwind{\texttt{members}}\\
      \hspace{0.7cm}{} \pipeline \unwind{\texttt{members.role}}\\
      \hspace{0.7cm}{} \pipeline \match{\texttt{members.role} = \textit{``guitar''}}\\
      \hspace{0.7cm}{} \pipeline \project{\texttt{ans}/\texttt{members.name}}\\
    \end{array}
    $
  }
\end{minipage}

In relational algebra, we can compute their (bag) union simply as
$q_1 \uplus q_2$. In \mquery, it is possible to compute the union even without
the explicit \valuefont{\$unionWith} operator. Namely, we can compute the
results of $q_1$ and $q_2$ independently and then put them together in the same
collection.

We start by showing how to compute and store the results of $q_1$ and $q_2$
separately from each other, step by step. First, the following query creates
two copies of each source document with an auxiliary field \valuefont{actRel}.
Intuitively, every object with $\valuefont{actRel} = 1$ is used to compute the
result of $q_1$, while every object with $\valuefont{actRel} = 2$ to compute
$q_2$. If we denote by \valuefont{oqueen} and \valuefont{oabba} the documents
about Queen and ABBA, respectively, then over \hyperref[fig:bands-collection]{$\valuefont{bands}^I$}, this query
evaluates to the collection on the right:

\begin{minipage}[t]{0.49\linewidth}
\mquerybox{
  $
  \begin{array}{l}
    \texttt{bands}\\
    \hspace{0.7cm}{} \pipeline \project{\texttt{origDoc}/\varepsilon,\ \texttt{actRel}/[1,2]}\\
    \hspace{0.7cm}{} \pipeline \unwind{\texttt{actRel}}\\
    \hspace{0.7cm}{} \pipeline \project{\scriptsize
    \begin{array}[t]{@{}l}
      \texttt{actRel},\\
      \texttt{rel1}/\cond{(\texttt{actRel}=1)}{\texttt{origDoc}}{\nullvalue},\\
      \texttt{rel2}/\cond{(\texttt{actRel}=2)}{\texttt{origDoc}}{\nullvalue}
    \end{array}
}\\
  \end{array}
  $
}
\end{minipage}
\hfill
\begin{minipage}[t]{0.49\linewidth}
\begin{lstlisting}[xleftmargin=1mm, xrightmargin=1mm, framexbottommargin=2.5mm]
  
{ "actRel": 1, "rel1": oqueen },
{ "actRel": 1, "rel1": oabba },


{ "actRel": 2, "rel2": oqueen },
{ "actRel": 2, "rel2": oabba }
\end{lstlisting}
\end{minipage}

Second, we can compute the result of $q_1$ without affecting the objects with
$\valuefont{actRel}=2$. The stages below simulate the unwind stage in $q_1$. If
we denote by \valuefont{oqueen1}, \valuefont{oqueen2}, \valuefont{oqueen3} and
\valuefont{oabba1}, \valuefont{oabba2} the objects resulting from unwinding
\valuefont{albums} in \valuefont{oqueen} and \valuefont{oabba}, respectively,
then over the above collection, these stages evaluate to the collection on the
right:

\begin{minipage}[t]{0.49\linewidth}
\mquerybox{
  $
  \begin{array}{l}
    \hspace{0.7cm}{} \pipeline \project{\scriptsize
    \begin{array}[t]{@{}l}
      \texttt{actRel},\\
      \texttt{rel2},\\
      \texttt{rel1}/\cond{(\texttt{actRel}=1)}{\texttt{rel1}}{\objectbr{\texttt{albums}\mapsto \arraybr{0}}}
    \end{array}
    }\\
    \hspace{0.7cm}{} \pipeline \unwind{\texttt{rel1.albums}}\\
    \hspace{0.7cm}{} \pipeline \project{\scriptsize
    \begin{array}[t]{@{}l}
      \texttt{actRel},\\
      \texttt{rel1}/\cond{(\texttt{actRel}=1)}{\texttt{rel1}}{\nullvalue},\\
      \texttt{rel2}/\cond{(\texttt{actRel}=2)}{\texttt{rel2}}{\nullvalue}
    \end{array}
    }\\
  \end{array}
  $
}
\end{minipage}
\hfill
\begin{minipage}[t]{0.49\linewidth}
\begin{lstlisting}[xleftmargin=1mm, xrightmargin=1mm, framexbottommargin=2mm]
{ "actRel": 1, "rel1": oqueen1 },
{ "actRel": 1, "rel1": oqueen2 },
{ "actRel": 1, "rel1": oqueen3 },
{ "actRel": 1, "rel1": oabba1 },
{ "actRel": 1, "rel1": oabba2 },

{ "actRel": 2, "rel2": oqueen },
{ "actRel": 2, "rel2": oabba }
\end{lstlisting}
\end{minipage}
The trick for unwind here is to create a non-empty \valuefont{rel1.albums}
array for documents with $\valuefont{actRel}=2$. This ensures that those
documents would not get lost. The last project stage `normalises' the pipeline,
making sure that \valuefont{rel1} exists only for documents with
$\valuefont{actRel}=1$ and \valuefont{rel2} exists only for documents with
$\valuefont{actRel}=2$.

Next we can simulate the match and project stages of $q_1$. Their result over
the above collection is shown to the right:

\begin{minipage}[t]{0.54\linewidth}
\mquerybox{
  $
  \begin{array}{l}
    \hspace{0.7cm}{} \pipeline \match{\scriptsize
    \begin{array}[t]{@{}l}
      \texttt{rel1.albums.release} > 1975\\
      \quad\lor\\
      \texttt{actRel}=2
    \end{array}
    }\\
    \hspace{0.7cm}{} \pipeline \project{\scriptsize
    \begin{array}[t]{@{}l}
      \texttt{actRel},\\
      \texttt{rel2},\\
      \texttt{rel1.ans}/\cond{(\texttt{actRel}=1)}{\texttt{rel1.albums.title}}{\nullvalue}
    \end{array}
    }\\
  \end{array}
  $
}
\end{minipage}
\hfill
\begin{minipage}[t]{0.44\linewidth}
\begin{lstlisting}[xleftmargin=1mm, xrightmargin=1mm, framexbottommargin=1.6mm]
{ "actRel": 1,
  "rel1": {
      "ans": "News of the World"
  }
},
{ "actRel": 2, "rel2": oqueen },
{ "actRel": 2, "rel2": oabba }
\end{lstlisting}
\end{minipage}
The trick for match is to select not only the documents where
$\valuefont{rel1.albums.release} > 1975$, but also all documents with
$\valuefont{actRel} = 2$, so that the answers to $q_2$ are not affected.
While project, in addition to projecting the fields according to the project
stage from $q_1$, needs to preserve $\valuefont{actRel}$ and $\valuefont{rel2}$.

Similarly, we can compute the result of $q_2$ without affecting the objects with
$\valuefont{actRel}=1$. 

\begin{minipage}[t]{0.54\linewidth}
\mquerybox{
  $
  \begin{array}{l}
    \hspace{0.7cm}{} \pipeline \project{\scriptsize
    \begin{array}[t]{@{}l}
      \texttt{actRel},\\
      \texttt{rel1},\\
      \texttt{rel2}/\cond{(\texttt{actRel}=2)}{\texttt{rel2}}{\objectbr{\texttt{members}\mapsto \arraybr{0}}}
    \end{array}
    }\\
    \hspace{0.7cm}{} \pipeline \unwind{\texttt{rel2.members}}\\
    \hspace{0.7cm}{} \pipeline \project{\scriptsize
    \begin{array}[t]{@{}l}
      \texttt{actRel},\\
      \texttt{rel1},\\
      \texttt{rel2}/\cond{(\texttt{actRel}=2)}{\texttt{rel2}}{\objectbr{\texttt{members.role}\mapsto \arraybr{0}}}
    \end{array}
    }\\
    \hspace{0.7cm}{} \pipeline \unwind{\texttt{rel2.members.role}}\\
    \hspace{0.7cm}{} \pipeline \project{\scriptsize
    \begin{array}[t]{@{}l}
      \texttt{actRel},\\
      \texttt{rel1}/\cond{(\texttt{actRel}=1)}{\texttt{rel1}}{\nullvalue},\\
      \texttt{rel2}/\cond{(\texttt{actRel}=2)}{\texttt{rel2}}{\nullvalue}
    \end{array}
    }\\
    \hspace{0.7cm}{} \pipeline \match{\scriptsize
    \begin{array}[t]{@{}l}
      \texttt{rel2.members.role} = \textit{``guitar''}\\
      \quad\lor\\
      \texttt{actRel}=1
    \end{array}
    }\\
    \hspace{0.7cm}{} \pipeline \project{\scriptsize
    \begin{array}[t]{@{}l}
      \texttt{actRel},\\
      \texttt{rel1},\\
      \texttt{rel2.ans}/\cond{(\texttt{actRel}=2)}{\texttt{rel2.members.name}}{\nullvalue}
    \end{array}
    }\\

  \end{array}
  $
}
\end{minipage}
\hfill
\begin{minipage}[t]{0.44\linewidth}
\begin{lstlisting}[xleftmargin=1mm, xrightmargin=1mm, framexbottommargin=3.5mm]
{ "actRel": 1,
  "rel1": {
      "ans": "News of the World"
  }
},


{ "actRel": 2,
  "rel2": {
      "ans": "Brian May"
  }
},


{ "actRel": 2,
  "rel2": {
      "ans": "Bjorn Ulvaeus"
  }
}
\end{lstlisting}
\end{minipage}
At this point, the intermediate collection contains the results of $q_1$ in
documents with $\valuefont{actRel}=1$ and the results of $q_2$ in documents
with $\valuefont{actRel}=2$. That is, the query so far has \emph{linearised}
the computation of two queries into one pipeline.

Finally, we can take the (bag) union of the answers of $q_1$ and $q_2$ as
follows:

\begin{minipage}[t]{0.49\linewidth}
\mquerybox{
  $
  \begin{array}{l}
    \hspace{0.7cm}{} \pipeline \project{
      \texttt{ans}/\cond{(\texttt{actRel}=1)}{\texttt{rel1.ans}}{\texttt{rel2.ans}}
    }\\
    % \hspace{0.7cm}{} \pipeline \group{}{\texttt{ans}}\\
    % \hspace{0.7cm}{} \pipeline \unwind{\texttt{ans}}\\
    % \hspace{0.7cm}{} \pipeline \project{
    %   \texttt{ans}
    % }\\
  \end{array}
  $\\[0.1cm]
}
\end{minipage}
\hfill
\begin{minipage}[t]{0.49\linewidth}
\begin{lstlisting}[xleftmargin=1mm, xrightmargin=1mm, framexbottommargin=0mm]
{ "ans": "News of the World" },  
{ "ans": "Brian May" },  
{ "ans": "Bjorn Ulvaeus"  }
\end{lstlisting}
\end{minipage}

The complete MongoDB query is as follows.

\begin{lstlisting}[basicstyle=\linespread{0.9}\ttfamily\scriptsize]
db.bands.aggregate([
    { $project: {
        "origDoc": "$$ROOT",
        "actRel": [1, 2], "_id": 0
    } },
    { $unwind: "$actRel" },
    { $project: {
        "actRel": 1,
        "rel1": { $cond: { if: { $eq: ["$actRel", 1] }, then: "$origDoc", else: null } },
        "rel2": { $cond: { if: { $eq: ["$actRel", 2] }, then: "$origDoc", else: null } }
    } },
    
    { $project: {
        "actRel": 1,
        "rel1": { $cond: { if: { $eq: ["$actRel", 1] }, then: "$rel1", else: {"albums": [0]} } },
        "rel2": 1
    } },
    { $unwind: "$rel1.albums" },
    { $project: {
        "actRel": 1,
        "rel1": { $cond: { if: { $eq: ["$actRel", 1] }, then: "$rel1", else: null } },
        "rel2": { $cond: { if: { $eq: ["$actRel", 2] }, then: "$rel2", else: null } }
    } },
    { $match: { $or: [
        { "rel1.albums.release": { $gt: 1975 } }, 
        { "actRel": { $eq: 2 } }
    ] } },
    { $project: {
        "actRel": 1,
        "rel1": { $cond: { if: { $eq: ["$actRel", 1] }, then: { "ans": "$rel1.albums.title" }, else: null } },
        "rel2": 1
    } },

    { $project: {
        "actRel": 1,
        "rel1": 1,
        "rel2": { $cond: { if: { $eq: ["$actRel", 2] }, then: "$rel2", else: {"members": [0]} } }
    } },
    { $unwind: "$rel2.members" },
    { $project: {
        "actRel": 1,
        "rel1": 1,
        "rel2": { $cond: { if: { $eq: ["$actRel", 2] }, then: "$rel2", else: {"members": {"role": [0]}} } }
    } },
    { $unwind: "$rel2.members.role" },
    { $project: {
        "actRel": 1,
        "rel1": { $cond: { if: { $eq: ["$actRel", 1] }, then: "$rel1", else: null } },
        "rel2": { $cond: { if: { $eq: ["$actRel", 2] }, then: "$rel2", else: null } }
    } },
    { $match: { $or: [
        { "rel2.members.role": { $eq: "guitar" } }, 
        { "actRel": { $eq: 1 } }
    ] } },
    { $project: {
        "actRel": 1,
        "rel1": 1,
        "rel2": { $cond: { if: { $eq: ["$actRel", 2] }, then: { "ans": "$rel2.members.name" }, else: null } }
    } },

    { $project: {
        "ans": { $cond: { if: { $eq: ["$actRel", 1] }, then: "$rel1.ans", else: "$rel2.ans" } }
    } }
])
\end{lstlisting}%$

In this particular example, it does not make a difference if we take the set or
the bag union of $q_1$ and $q_2$.
But if we needed to make sure to remove duplicates, we could simply group all
documents in one array $\group{}{\valuefont{ans}}$, then in a project stage
apply the \valuefont{setUnion} function to remove duplicates from
\valuefont{ans}, and finally unwind \valuefont{ans}.

\subsection{Difference}

We conclude with showing how the results of one query can be removed from the
results of another query in MQuery, which does not have an explicit difference
operator.
For simplicity, we here consider two rather similar queries, so there may be an
easier way to compute the desired result. But for the sake of a more generic
exposition, we show the full construction that works for two arbitrary queries
that return results under the same key.

Consider a query $q_1$ that retrieves all band members that sing.

\begin{minipage}[t]{0.44\linewidth}
  \rabox{
    $
    \begin{array}{l}
      \hspace{0.0cm}\pi_{\text{ans}/\text{members.name}}\\
      \hspace{0.7cm}\sigma_{\text{members.role}=\textit{``vocals''}}\\
      \hspace{1.4cm}\chi_{\text{members.role}}\\
      \hspace{2.1cm}\chi_{\text{members}}\\
      \hspace{2.8cm}\textit{bands}\\
    \end{array}
    $
  }
\end{minipage}
\hfill
\begin{minipage}[t]{0.54\linewidth}
  \mquerybox{
    $
    \begin{array}{l}
      \texttt{bands}\\
      \hspace{0.7cm}{} \pipeline \unwind{\texttt{members}}\\
      \hspace{0.7cm}{} \pipeline \unwind{\texttt{members.role}}\\
      \hspace{0.7cm}{} \pipeline \match{\texttt{members.role} = \textit{"vocals"}\ \lor\ \texttt{members.role} = \textit{"lead vocals"}}\\
      \hspace{0.7cm}{} \pipeline \project{\texttt{ans}/\texttt{members.name}}\\
    \end{array}
    $
  }
\end{minipage}
and query $q_2$ from Section~\ref{sec:expressivity-union}, which retrieves all band members who play guitar.

Over \hyperref[fig:bands-collection]{$\valuefont{bands}^I$}, $q_1$ returns
\begin{lstlisting}
{ "ans": "Freddie Mercury" },
{ "ans": "Brian May" },
{ "ans": "Roger Taylor" },
{ "ans": "Agnetta Faltskog" },
{ "ans": "Bjorn Ulvaeus" },
{ "ans": "Benny Andersson" },
{ "ans": "Anni-Frid Lyngstad" }
\end{lstlisting}
while $q_2$ returns
\begin{lstlisting}
{ "ans": "Brian May" },
{ "ans": "Bjorn Ulvaeus" }
\end{lstlisting}
% \begin{minipage}[t]{0.49\linewidth}
%   \rabox{
%     $
%     \begin{array}{l}
%       \hspace{0.0cm}\pi_{\text{ans}/\text{members.name}}\\
%       \hspace{0.7cm}\sigma_{\text{members.role}=\textit{``guitar''}}\\
%       \hspace{1.4cm}\chi_{\text{members.role}}\\
%       \hspace{2.1cm}\chi_{\text{members}}\\
%       \hspace{2.8cm}\textit{bands}\\
%     \end{array}
%     $
%   }
% \end{minipage}
% \hfill
% \begin{minipage}[t]{0.49\linewidth}
%   \mquerybox{
%     $
%     \begin{array}{l}
%       \texttt{bands}\\
%       \hspace{0.7cm}{} \pipeline \unwind{\texttt{members}}\\
%       \hspace{0.7cm}{} \pipeline \unwind{\texttt{members.role}}\\
%       \hspace{0.7cm}{} \pipeline \match{\texttt{members.role} = \textit{"guitar"}}\\
%       \hspace{0.7cm}{} \pipeline \project{\texttt{name}/\texttt{members.name}}\\
%     \end{array}
%     $
%   }
% \end{minipage}

In relational algebra, we can get all vocalists who do not play guitar as
$q_1 \setminus q_2$.

In MQuery, we can follow the same strategy as when computing the union of two
queries to be able to compute and store the results of $q_1$ and $q_2$
independently of each other.
Assume that we already have the query that computes the collection such that
the result of $q_1$ is stored in the documents with $\valuefont{actRel}=1$
under the key $\valuefont{rel1}$ and the result of $q_2$ is stored in the
documents with $\valuefont{actRel}=2$ under the key $\valuefont{rel2}$.
Then we can compute all answers to $q_1$ that are not an answer to $q_2$ as follows:

\begin{minipage}[t]{0.49\linewidth}
\mquerybox{
  $
  \begin{array}{l}
    \hspace{0.7cm}{} \pipeline \project{
    \texttt{rel2},\
      \texttt{ans}/\cond{(\texttt{actRel}=1)}{\texttt{rel1.ans}}{\texttt{rel2.ans}}
    }\\
    \hspace{0.7cm}{} \pipeline \group{\texttt{ans}}{\texttt{rel2}}\\
    \hspace{0.7cm}{} \pipeline \match{\texttt{rel2}=\arraybr{}}\\
    \hspace{0.7cm}{} \pipeline \project{
      \texttt{ans}/\texttt{\_id}
    }\\
  \end{array}
  $\\
}
\end{minipage}
\hfill
\begin{minipage}[t]{0.49\linewidth}
\begin{lstlisting}[xleftmargin=1mm, xrightmargin=1mm, framexbottommargin=2.5mm]
{ "ans": "Freddie Mercury" },
{ "ans": "Roger Taylor" },
{ "ans": "Agnetta Faltskog" },
{ "ans": "Benny Andersson" },
{ "ans": "Anni-Frid Lyngstad" }
\end{lstlisting}
\end{minipage}

The complete MongoDB query is as follows. Note that this is a somewhat
simplified version that produces the desired result in MongoDB (some
normalisation stages have been omitted). Due to the differences with our
formalisation, this query makes use of a \valuefont{dummy} path in conditional
terms, that is a path that is known not to be present in the collection. The
corresponding MQuery instead uses the \nullvalue value.

\begin{lstlisting}[basicstyle=\linespread{0.9}\ttfamily\scriptsize]
db.bands.aggregate([
    { $project: {
        "origDoc": "$$ROOT",
        "actRel": [1, 2], "_id": 0
    } },
    { $unwind: "$actRel" },
    { $project: {
        "actRel": 1,
        "rel1": { $cond: { if: { $eq: ["$actRel", 1] }, then: "$origDoc", else: null } },
        "rel2": { $cond: { if: { $eq: ["$actRel", 2] }, then: "$origDoc", else: null } }
    } },
    { $project: {
        "actRel": 1,
        "rel1": { $cond: { if: { $eq: ["$actRel", 1] }, then: "$rel1", else: {"members": [0]} } },
        "rel2": 1
    } },
    { $unwind: "$rel1.members" },
    { $project: {
        "actRel": 1,
        "rel1": { $cond: { if: { $eq: ["$actRel", 1] }, then: "$rel1", else: {"members": {"role": [0]}} } },
        "rel2": 1
    } },
    { $unwind: "$rel1.members.role" },
    { $match: { $or: [
        { $or: [ { "rel1.members.role": "lead vocals" }, { "rel1.members.role": "vocals" } ] }, 
        { "actRel": { $eq: 2 } }
    ] } },
    { $project: {
        "actRel": 1,
        "rel1": { $cond: { if: { $eq: ["$actRel", 1] }, then: { "ans": "$rel1.members.name" }, else: "$dummy" } },
        "rel2": 1
    } },
    { $project: {
        "actRel": 1,
        "rel1": 1,
        "rel2": { $cond: { if: { $eq: ["$actRel", 2] }, then: "$rel2", else: {"members": [0]} } }
    } },
    { $unwind: "$rel2.members" },
    { $project: {
        "actRel": 1,
        "rel1": 1,
        "rel2": { $cond: { if: { $eq: ["$actRel", 2] }, then: "$rel2", else: {"members": {"role": [0]}} } }
    } },
    { $unwind: "$rel2.members.role" },
    { $match: { $or: [
        { "rel2.members.role": { $eq: "guitar" } }, 
        { "actRel": { $eq: 1 } }
    ] } },
    { $project: {
        "actRel": 1,
        "rel1": 1,
        "rel2": { $cond: { if: { $eq: ["$actRel", 2] }, then: { "ans": "$rel2.members.name" }, else: "$dummy" } }
    } },
    { $project: {
        "rel2": 1,
        "ans": { $cond: { if: { $eq: ["$actRel", 1] }, then: "$rel1.ans", else: "$rel2.ans" } }
    } },
    { $group: {
        "_id": "$ans", 
        "rel2": { $push: "$rel2" },
    } },
    { $match: {
        "rel2": [ ]
    } },
    { $project: {
        "_id": 0, "ans": "$_id"
    } }
])
\end{lstlisting}%$

%%% Local Variables:
%%% mode: latex
%%% TeX-master: "main-tr"
%%% fill-column: 79
%%% End:

\section{Differences with MongoDB's Implementation}
\label{sec:differences}

  We illustrate and motivate here the main differences between the semantics defined in the previous sections and the one applied by \mongodb version \mversion (for the aggregation framework).
  Our depiction of \mongodb's behavior's is only partially based on its documentation.
   In other cases, when the documentation was incomplete, we made hypotheses, tested them (via queries) and extrapolated the underlying rationale.
   But as with any (black-box) reverse-engineering process, some of these hypotheses may be incorrect.

  \nb{JC: TODO: Decide whether we want to talk once again about the distinction between ordered and unordered semantics (this has already been done in Section 2.
    This may be more an extension than a difference.
    For now the related content is commented out.
  }

  \subsection{A Unified Semantics for Boolean Expressions}
%   As many other query languages (e.g.~SQL based), 
%   the aggregation framework supports Boolean expression in a variety of contexts:
%   for instance, to specify a filter on a collection of documents (in a match stage),
% or a conditional statement in a complex projection (in a project stage).
Two concrete syntaxes for Boolean expressions coexist within the MongoDB aggregation framework,
called ``query predicates'' and \mongodb ``expression'' respectively,
with slightly different semantics.
This situation seems partly due to the historical development of the system.

Our proposal unifies these two syntaxes under a single grammar for Boolean expression.
This eliminates confusion about which semantics applies,
and provides a more principled foundation for query optimization.
We illustrate below these two syntaxes and their respective semantics, each in its own section.
In each case, we also explain how they compare to the Boolean expressions used in \mqueries,
and motivate our choices.

\subsubsection{Query Predicates}
\label{sec:diff_queryPredicates}
A Boolean expression written in the first syntax is called a \emph{query predicate}.
For instance, in the following query,
the expression 
 \valuefont{\{"origin": \{\$eq: "UK"\}\}} is a query predicate,
 used to filter documents where the key \texttt{origin} has value \textit{``UK''}
\begin{lstlisting}
db.bands.aggregate([
    { $match: { "origin": { $eq: "UK" } } }
])
\end{lstlisting}
Note that \mongodb's syntax also allows omitting the \texttt{\$eq} operator in this context (e.g.~writing \valuefont{\{"origin": "UK"\}}),
for a (seemingly) identical behavior.

Query predicates are borrowed from so-called \mongodb \emph{find} queries, which form a distinct,
less expressive query language that predates the introduction of the aggregation framework.
Within the aggregation framework, query predicates are used (by default) as filtering conditions in match stages,
as in this example, and possibly in lookup stages to express a joining condition.

As we can see in this query, a query predicate can make use of the operator \verb|$eq|,
which is \mongodb's syntax for equality.
In this case, the equality stands between the evaluation of the path \texttt{origin} and the value \textit{``UK''}.
So one may expect this query to correspond to the following \mquery:

\begin{minipage}[t]{0.5\textwidth}
\mquerybox{
$\texttt{bands} \pipeline \match{\texttt{origin} = \textit{"UK"}}$
}
\end{minipage}

This is not the case though, because query predicates rely on a looser notion of “equality” for paths that evaluate to an array.
For instance, let us assume that our collection \texttt{bands} provides values for \texttt{origin} that may be literals (e.g. \textit{``UK''} for the band Queen) or arrays (e.g. [\textit{``UK'', ``Japan'', ``US''}] for the band Gorillaz), as follows:
\begin{lstlisting}
{ "_id": 1, "name": "Queen", "origin": "UK" },
{ "_id": 2, "name": "Gorillaz", "origin": ["UK", "Japan", "US"] }
\end{lstlisting}

Over this collection,
the query above would retrieve the document for Queen, as expected.
But it would also retrieve the document for Gorillaz,
which may come as a surprise.
The reason is that the array [\textit{"UK", "Japan", "US"}] (which is the evaluation of \texttt{origin} in this document) contains \textit{"UK"}.

So a more appropriate translation in \mquery would be:

\begin{minipage}[t]{0.5\textwidth}
\mquerybox{
$ \texttt{bands} \pipeline \match{\ \texttt{origin} = \textit{"UK"}\ \lor\ \textit{"UK"}\, \in\, \texttt{origin}}$
}
\end{minipage}

In other words, \mqueries apply a stricter semantics than query predicates for the “equality” operator (denoted "=" in \mqueries),
when comparing the interpretation of a path to a value (as in this example) or comparing the interpretations of two paths:
the equality holds iff both values are identical.

The reason for this choice is algebraic: a relaxed notion of equality is not necessarily an issue,
but it should induces a so-called \emph{equivalence relation} over compared elements.
For instance, programming and query languages often provide additional comparison operators that relax strict equality,
such as case-insensitive equality for two strings,
or separator-insensitive equality for two paths.
Let us use $a \sim b$ to denote the fact that $a$ and $b$ are "equal" in this relaxed sense.
Three elementary algebraic properties are normally expected from such a relation,
which together form the definition of an equivalence relation:
reflexivity ($a \sim a$),
symmetry ($a \sim b$ implies $b \sim a$) and 
transitivity ($a \sim b$ and $b \sim c$ imply $a \sim c$).
These are elementary properties for algebraic optimization,
used (among others) to simplify an expression or determine whether two expressions are equivalent.
For instance, the expression
$a \sim b \land (c \sim a \lor c \sim b)$
can be simplified as 
$a \sim b \land c \sim a$ (due to transitivity),
and is equivalent to 
$a \sim b \land a \sim c$ (due to symmetry). 

Now observe that 
when used inside \mongodb query predicates,
the \verb|$eq| operator violates both symmetry and transitivity.
For instance, consider the following query, 
evaluated over the same collection as above:
\begin{lstlisting}
db.bands.aggregate([
    { $match: { "origin": { $eq: ["UK", "Japan", "US"] } } }
])
\end{lstlisting}
This query returns the document for Gorillaz (as expected), but not the document for Queen.
So according to this relaxed notion of equality, 
\textit{"UK"} (in the data) does not match [\textit{"UK", "Japan", "US"}] (in the query),
but,
as we saw with the previous query,
[\textit{"UK", "Japan", "US"}] (in the data) matches \textit{"UK"} (in the query),
which violates symmetry.

Transitivity is also violated.
For instance,
[[\textit{"UK"}]]
matches 
[\textit{"UK"}],
and the latter matches
\textit{"UK"}.
But 
\textit{[["UK"]]}
does not match 
\textit{"UK"}.
Note also that a symmetric version of \mongodb's relaxed equality would still violate transitivity.
Precisely, if we define it (in \mquery's syntax) as $a \sim b$ iff
$a = b \lor a \in b \lor b \in a$,
then 
\textit{"UK"} matches 
[\textit{"UK", "Japan"}],
and the latter matches
\textit{"Japan"}.
But 
\textit{"UK"}
does not match 
\textit{"Japan"}.
\nb{JC: this is the simpler example that Norman suggested last time.
Unfortunately, it cannot be used to illustrate non-transitivity for \mongodb's implementation,
due non-symmetry.}

This is why we opted for a stricter notion of equality,
which does induces an equivalence relation over compared elements.
But we emphasize that this is without loss of expressivity.
As we illustrated above, the behaviour of query predicates can still be simulated in \mquery,
albeit with a slightly more complex syntax.
So our proposal does not technically rule out query predicates, but treats them instead as syntactic sugar rather than primitives.
We believe that this approach is better-suited to a good understanding of the algebraic properties of the language.

Before describing the second syntax (and semantics) for Boolean expressions within \mongodb,
we briefly discuss the treatment of other comparison operators within \mongodb query predicates,
which is analogous to the treatment of equality.
Let us assume that our collection contains (starting) years of tours of each band, for instance:
\begin{lstlisting}
{ "_id": 1, "name": "Gorillaz", "tours": [2010, 2015] }
\end{lstlisting}
The following query retrieves the document above, because at least one value in the array [2010, 2015] is inferior to 2012.
\begin{lstlisting}
db.bands.aggregate([
    { $match: { "tours": { $lt: 2012 } } }
])
\end{lstlisting}

So the array [2010, 2015] (in the data) matches the condition < 2012 (in the query).
However, similarly to what we observed for equality,
2012 (in the data) does not match the condition > [2010, 2015] (in the query),
thus violating one of the elementary algebraic properties expected from a total order ($a < b$ implies $b > a$).

Besides, the following query also retrieves the document above, because 2015 is superior to 2012.
\begin{lstlisting}
db.bands.aggregate([
    { $match: { "tours": { $gt: 2012 } } }
])
\end{lstlisting}

So the array [2010, 2015] matches not only the condition < 2012, but also the condition > 2012.
At the algebraic level, this means that both $a < b$ and $a > b$ may hold. 

These two examples show that within query predicates,
similarly to the \verb|$eq| operator,
the behaviors of 
\verb|$lt| and
\verb|$gt| violate elementary properties of (total) orders (and similar observations can be made for 
\verb|$lte| and 
\verb|$gte|).
So once again,
in \mqueries,
we opted for a stricter interpretation of these operators.

\subsubsection{\mongodb Boolean “Expressions”}
\label{sec:diff_expr}
In addition to query predicates (discussed in the previous section), 
the aggregation framework supports a second syntax for Boolean expressions.
Precisely, a query can use so-called \mongodb “expressions”,
which are analogous to the \emph{terms} of Grammar~\eqref{gramm:term}
(in the sense that an expression evaluates to a literal, array or object).
Among these, expressions that evaluate to true or false are effectively Boolean expressions.
For instance, consider the following query:
\begin{lstlisting}
db.bands.aggregate([
    { $match: { $expr: { $eq: [ "$origin", "UK" ] } } }
])
\end{lstlisting}
In this query
\verb|{ $eq: [ "$origin", "UK" ] }| is a \mongodb expression,
and the keyword \verb|$expr| lets us use this expression in place of a query predicate.
\mongodb expressions are also used in other stages of the aggregation framework (e.g.~project, unwind or group),
without the need for the \verb|$expr| keyword.

At first sight, the query above seems very similar to the one that we used in the previous section (reproduced here for convenience):

\begin{lstlisting}
db.bands.aggregate([
    { $match: { "origin": { $eq: "UK" } } }
])
\end{lstlisting}

Intuitively, both queries retrieve bands with \textit{"UK"} as origin.
However, their syntaxes differ: the former uses a \mongodb “expression” in place of a query predicate,
whereas the latter uses a query predicate directly.
This difference is reflected in their respective semantics.
For instance, let us consider the same collection as above (reproduced here for convenience):
\begin{lstlisting}
{ "_id": 1, "name": "Queen", "origin": "UK" },
{ "_id": 2, "name": "Gorillaz", "origin": ["UK", "Japan", "US"] }
\end{lstlisting}
Over this collection, the first of the two queries (with a \mongodb “expression”) only retrieves the document for Queen,
whereas the second query (with a query predicate) retrieves both documents.
In other words, the first query corresponds to the \mquery

\begin{minipage}[t]{0.5\textwidth}
\mquerybox{
$\texttt{bands} \pipeline \match{\ \texttt{origin} = \textit{"UK"}}$
}
\end{minipage}

whereas the latter corresponds to

\begin{minipage}[t]{0.5\textwidth}
\mquerybox{
$\texttt{bands} \pipeline \match{\ \texttt{origin} = \textit{"UK"}\ \lor\ \textit{"UK"}\, \in\, \texttt{origin}}$
}
\end{minipage}

% Similarly, the following query only retrieves the document for Gorillaz:
% \begin{lstlisting}
% db.bands.aggregate([
%     { $match: { $expr: { $eq: [ "$origin", ["UK", "Japan", "US" ] } } }
% ])
% \end{lstlisting}
We already highlighted in the previous section how the semantics of \verb|$eq| in query predicates violates elementary algebraic properties expected from a (possibly relaxed) ``equality'' operator.
In contrast, in a \mongodb Boolean ``expression''
the operator \valuefont{\$eq} mostly behaves as one would expect.
This is why we used it as a blueprint for the semantics of “=” in \mqueries.
And an analogous observation can be made for the operators $<, >, \le$ and $\ge$ (written \valuefont{\$lt}, \valuefont{\$gt}, \valuefont{\$lte} and \valuefont{\$gte} in \mongodb's syntax).

\subsection{Interpretation of Paths}
\label{sec:diff-interp-paths}

\mongodb's interpretation of paths has subtle behaviors that can be hard to predict:
the intuitive meaning of a path may vary, depending on the document in which it is interpreted.
We already observed this behavior for atomic paths (i.e. paths that consist of a single key) in the previous section.
We saw with our running example that the path \texttt{origin} in a \mongodb “expression” (Section~\ref{sec:diff_expr}) was interpreted "literally", as the value that it points to in an object.
In contrast, in a \mongodb query predicate (Section~\ref{sec:diff_queryPredicates}),
we saw that this path could be interpreted in a more liberal way, either as the value that it points to, or, if it points to a array, as the set of values contained in this array.

We already explained above why this ambiguity may not be desirable in a Boolean condition.
This is why \mquery shifts away from the semantics of query predicates, 
and adopts instead a semantics for expressions that is closer to \mongodb “expressions”.
However, if atomic paths in \mongodb “expressions” are indeed non-ambiguous, this is not the case of composite ones.
To illustrate this, we will use a slight variation over the example above.
Let use assume that the collection \texttt{bands} contains the following document:
\begin{lstlisting}
 { "_id": 2, "name": "Gorillaz", "origin": { "country": ["UK", "Japan"] } }
\end{lstlisting}
The following query uses a \mongodb “expression” to select bands where the value of the path \texttt{origin.country} is [\textit{"UK", "Japan"}].
\begin{lstlisting}
db.bands.aggregate([
  { $match: { $expr: { $eq: [ "$origin.country", ["UK", "Japan"] ] } } }
])
\end{lstlisting}
As expected, this query retrieves the document above.
So a naive conversion of this query into \mquery would be

\begin{minipage}[t]{0.5\textwidth}
\mquerybox{
$\texttt{bands} \pipeline \match{\ \texttt{origin.country} = [\textit{"UK"}, \textit{"Japan"}]}$
}
\end{minipage}

However, this conversion is incorrect.
To see this, let us assume that our document has instead a slightly different shape:
\begin{lstlisting}
{ "_id": 2, 
  "name": "Gorillaz", 
  "origin": [
    { "country": "UK", "city": "London" },
    { "country": "Japan" } 
  ]
}
\end{lstlisting}
The value of \texttt{origin} is still an array, but this time it contains two objects, rather than two scalars.
Surprisingly, the query above also retrieves this document.
The reason is that \texttt{\$origin.country} evaluates to \texttt{["UK", "Japan"]} in this document,
even if the document does not contain such an array.

In other words, the meaning of \texttt{\$origin.country} in this query is overloaded.
It may be interpreted in two alternative ways, depending on the document $o$ under consideration:
\begin{enumerate}
  \item the first interpretation occurs when this path points to a \dvalue $d$ (i.e. an array, object or scalar) in $o$.
    In this case, \texttt{\$origin.country} evaluates to $d$.
  \item the second interpretation occurs when the prefix \texttt{\$origin} points to an array $a$ in $o$.
    In this case, \texttt{\$origin.country} evaluates to another array, 
    which consists of the evaluation of the suffix \texttt{\$country} in each element of $a$.
\end{enumerate}

In \mquery, we decided instead to shift away from this polysemy.
More precisely, we require using the $\mapz$ operator explicitly for the latter of these two meanings.
So a more appropriate conversion of the query above into \mquery is

\begin{minipage}[t]{0.7\textwidth}
\mquerybox{
$\texttt{bands} \pipeline \match{\ \texttt{origin.country} = [\textit{"UK"}, \textit{"Japan"}] \ 
   \lor\ \map{\texttt{origin}}{x}{x.\texttt{country}} = [\textit{"UK"}, \textit{"Japan"}] 
}$
}
\end{minipage}

There are three main reasons for this choice.
The first one is practical: from a software engineering perspective, the potential drawbacks of this polysemy outweighs its benefits.
To understand this, let us first focus on the benefits, namely the fact that \mongodb's overloaded interpretation offers a more concise syntax.
For instance, in this example, the \mongodb "expression"
\valuefont{\$eq: [ "\$origin.country", ["UK", "Japan"] ]}
is sufficient to encode (the disjunction of) the two \mquery Boolean expressions $\texttt{origin.country} = [\textit{"UK"}, \textit{"Japan"}]$ and
$\map{\texttt{origin}}{x}{x.\texttt{country}} = [\textit{"UK"}, \textit{"Japan"}]$.
However, it is hard to imagine a situation where on would want to leverage such a polysemy:
when an end-user writes \valuefont{\$origin.country} in a query, this is most likely with one of the two meanings in mind, not both.
And an analogous observation can be made if the query is automatically generated (either programmatically or by an agent).

The drawback of this overloaded semantics is that a query may produce results that correspond to the "other", unintended meaning, which, as any bug, can have a non-negligible cost.
Besides, when undesired, these results need to be filtered out explicitly, which eventually results in a syntactically more complex query.

Worse, this polysemy grows exponentially with the length of paths.
For instance, a seemingly innocuous expression of the form \texttt{\$a.b.c.d.e.f} has $2^{(6-1)} = 32$ alternative meanings, out of which very few (most likely one) will be the intended one(s).
This can make the evaluation of a query hard to explain or predict, especially when the queried data does not follows a regular structure.
As an (artificial) illustration, predicting the evaluation of \texttt{\$a.b.c} in the following object can be a puzzle (the answer is \texttt{[[1],[2,3]]}):
 \begin{lstlisting}
{ "a": [
      { "b": { "c" : [1] } },
      { "b": [
          { "c" : 2 },
          { "c" : 3 }
      ] }
] }
\end{lstlisting}

The second reason for our choice to deviate from \mongodb's interpretation of paths is computational.
As we just explained, the number of alternative meanings of a path grows exponentially with its length.
So a naive evaluation of a Boolean expression that contains such a path may perform exponentially many verifications within a single object.

The third reason has to do with coherence.
\nb{JC: This third argument is arguably weaker. TODO: decide whether we should drop it.}
Paths are used not only in expressions, but also as so-called \mongodb \emph{fields},
to specify where the output of an operation must be stored
(in stages like project, unwind, lookup, group, etc.).
For instance, consider the following query:

\begin{lstlisting}
db.bands.aggregate([
  { $project : { 
     "name" : "$name",
     "creation.year" : "$formation",
     "creation.location" : "$origin"
  } }
])
\end{lstlisting}
This query selects the name, year of formation and origin of each band,
grouping the latter two into a nested object with (up to) two attributes \texttt{year} and \texttt{location}.
It uses three \mongodb fields (\texttt{name}, \texttt{creation.year} and \texttt{creation.location}),
two of which are composite paths.

When a path is used as a \mongodb field, its meaning is non-ambiguous, which contrasts with what we observed above for paths used in expressions.
This discrepancy may easily results in queries with unexpected behaviors.
For instance, the following query projects the  name and country(ies) of each band in our collection.
Then it selects the bands with countries [ \textit{"UK", "Japan"} ].
\begin{lstlisting}
db.bands.aggregate([
    { $project : { 
        "name" : "$name",
        "origin.country" : "$origin.country"
    } },
    { $match : { 
         $expr: { $eq: [ "$origin.country", ["UK", "Japan" ] ] }
    } }
])
\end{lstlisting}
At first sight, the project stage in this query seems like a plain projection (similar to a projection in relational algebra,
whose only purpose is to select some attributes).
When optimizing a query, it is common practice to postpone such an operation.
For instance, in this example, let us assume that the collection \texttt{band} has an index over the field \texttt{origin.country}.
Then it would be tempting to swap the order of the two stages, so that the selection (a.k.a.~match stage) can take advantages of this index.
However, this would alter the semantics of the query.
Indeed, as we saw earlier, in a source document, \texttt{\$origin.country} may evaluate to an array that was not present.

We believe that a unified semantics for paths (regardless of whether they are used as expressions or fields) is less error-prone,
in particular when performing algebraic transformations.
% For instance, the \mquery above clearly indicates that the project stage cannot be naively postponed.

\subsection{Ordering}

\mongodb's documentation provides information about the way BSON atomic values, arrays and objects are naturally ordered.\footnote{
  \href{https://www.mongodb.com/docs/manual/reference/bson-type-comparison-order}
  {https://www.mongodb.com/docs/manual/reference/bson-type-comparison-order}
 }
This ordering is meant to determine the behavior of the sort stage,
as well as the comparison operators 
\texttt{\$lte},
\texttt{\$lt},
\texttt{\$gte} and 
\texttt{\$gt} (written $\le$, $<$, $\ge$ and $>$ in \mquery)
used in expressions.
More precisely, \mongodb's documentation specifies a so-called “comparison order” over atomic BSON values,
which is then lifted to arrays and objects.
However, this documentation is only partial, and (as we will see) sometimes contradicts \mongodb's behavior.

We discuss in what follows four aspects that distinguish \mongodb and \mqueries when it comes to ordering:
\ei how literals are ordered, \eii how this ordering is lifted to arrays and objects, 
and how \eiii comparison operators and \eiv sorting operations behave when evaluating a query.

\subsubsection{Literals}
As we already explained,
\mquery abstracts from concrete datatypes such as string, integer, float, etc.
When it comes to ordering, we only assume a total order $\le$ over literals, which should coincide (semantically) with the $\le$ operator used in \mquery expressions.

This requirement is already a slight deviation from (or more precisely, an abstraction of) what \mongodb's documentation describes.
To understand this, we need to discuss what "total order" precisely means.
First, a \emph{preorder} $\preceq$ is a transitive and reflexive binary relation over a set.
This relation is said to the \emph{total} if all elements of $S$ can be compared with each other (i.e.~if $e_1 \preceq e_2$ or $e_2 \preceq e_1$ holds for all $e_1, e_2 \in S$).
Intuitively, a total preorder is all we need to sort the elements of a set.

Note that a preorder does not forbid two distinct elements to be comparable (i.e. $e_1 \preceq e_2$ and $e_2 \preceq e_1$ may both hold for some $e_1 \neq e_2$).
For instance, 
one can sort a set of people by age, even if two of them have the same age.
An \emph{order} is a preorder that excludes this possibility, meaning that it also satisfies antisymmetry (in addition to reflexivity and transitivity),
i.e. $e_1 \preceq e_2$ and $e_2 \preceq e_1$ imply $e_1 = e_2$.
This is for instance the case of the natural order over real numbers.

In \mquery, we require $\le$ to be a total order (i.e.~a total, reflexive, transitive and antisymmetric binary relation) over literals.
However, it is common practice in databases or programming languages to treat some values with different types (e.g. some integers and floats) as equivalent for sorting and/or comparison purposes.
For instance, the three  \mongodb expressions
\verb|$eq:[1,1.0]|, 
\verb|$lte:[1.0,1]| and 
\verb|$gte:[1.0,1]| 
all evaluate to \texttt{true}.
Therefore the so-called “comparison order” described in \mongodb's documentation is technically not an order, but a preorder.

As a consequence, the literals of \mquery cannot literally coincide with the atomic BSON values of \mongodb.
Instead, they must be viewed as a slight abstraction, namely the equivalence classes induced by \mongodb's “comparison order”.
For instance, \mquery does not distinguish $1$ and $1.0$: for all intents and purposes, they are the “same” \dvalue.
We made this simplification because it makes our definitions more streamlined, 
and does not affect the (higher-level) properties that our study focuses on.

\subsubsection{Arrays and Objects}
\mquery lifts the order $\le$ over literals to arrays and objects, as specified in Appendix~\ref{sec:appendix_order}.
However, we emphasize that the choice of a specific natural ordering of \dvalues is irrelevant for our purpose,
as long as this is a total order. 
We only defined one for the sake of completeness,
as close as possible to what is described 
in the \mongodb's documentation.

First, the order $\le$ is extended to compare an array with an object or literal,
and an object with a literal.
By convention, literals precede objects, which in turn precede arrays.
\mongodb's documentation specifies instead a more complex precedence, which depends on concrete datatypes. 
Among others, numbers precede objects, which precede arrays, which precede booleans, which precede dates.
Because \mquery abstracts from concrete datatypes, we could not enforce such a fine-grained ordering,
and opted instead for a best possible approximation (objects still precede arrays).

Next, the order $\le$ is lifted inductively to complex \dvalues, in order to compare two arrays and two objects.
Two arrays are compared by means of a standard lexicographic product, analogously to strings:
compare the first element of each array w.r.t. $\le$,
then the second element in case of equality, etc.
The product of two total orders (resp.~preorders) being a total order (resp.~preorder),
this guarantees that \mquery's (resp. \mongodb's) arrays are totally ordered (resp. preordered).

As for objects, they are also compared lexicographically,
one key-value pair after the other (under the ordered semantics), in accordance with \mongodb's documentation.
One difference here is the way two key-value pairs are compared with each other:
\mongodb first compares the types of the values, then (if identical) the keys (as strings, lexicographically), and then only (if identical) the values.
Because \mquery abstracts from concrete datatypes, we used once again the best possible approximation:
first comparing the \dvalue types of values (literal, array or object), then comparing keys, and then values.
In addition, we specify a way to compare two objects under the unordered semantics (i.e. when the order of keys in an object is semantically irrelevant), which is not supported by \mongodb.

\subsubsection{Comparison Operators}
The behavior of the operators \texttt{\$lte}, \texttt{\$lt} \texttt{\$lte} and \texttt{\$lte} used
in \mongodb query predicates (see Section~\ref{sec:diff_queryPredicates})
seems based on an order (or more precisely preorder, as discussed above) over BSON values that is not total,
meaning that two values may not be comparable.
For instance, let us assume that the collection \texttt{bands} contains the following document:
\begin{lstlisting}
 { "_id": 1, "name": "Gorillaz", "formation": "January 1998" }
\end{lstlisting}
At first sight, the following query should retrieves all bands formed either before or strictly after 2000:
\begin{lstlisting}
db.bands.aggregate([{
    $match: { $or: [ { "formation": { $lte: 2000 } }, { "formation" : { $gt: 2000 } } ] },
}]);
\end{lstlisting}
However, this query does not retrieve the document above, which shows that in this context, the string “\textit{January 1998}” is not comparable with the integer 2000.
This contradicts \mongodb's documentation, where numbers are said to precede strings.
Besides, this latter interpretation apparently dictates how comparison operators  behave in \mongodb “expressions”.
For instance, let us use a \mongodb “expression” in place of a query predicate in the query above (thanks to the operator \texttt{\$expr},
as we also did in Section~\ref{sec:diff_expr}):
\begin{lstlisting}
db.bands.aggregate([{
    $match: { $expr: { $or: [ { $lte: ["$formation", 2000] }, { $gt: ["$formation", 2000] } ] } }
}]);
\end{lstlisting}
This query does retrieve the document above, which seems in line with \mongodb's documentation.

In \mquery, we opted for the latter behavior.
The reason is that the semantics of \mongodb (and \mquery) Boolean expressions is 2-valued (as opposed for instance to SQL),
so that the operator \texttt{\$lte} must evaluate to \texttt{true} or \texttt{false.}
With a 2-valued semantics, a total order (or total preorder) guarantees that elementary Boolean properties expected from a “less than or equal” operator are enforced.
For instance,
in \mquery, the expression $\neg (\texttt{formation}  > 2000)$ can be simplified as $\texttt{formation} \le 2000$,
as in most query and programming languages (and as one would intuitively expect).
However, 
over the document above,
the query predicate 
\verb|$not: {"formation": {$gt: 2000}}| 
evaluates to \texttt{true}
(since its negation \verb|"formation": {$gt: 2000}| evaluates to false),
whereas the seemingly equivalent query predicate 
\verb|"formation": {$lte: 2000}| 
evaluates to \texttt{false}.

There is another reason why the interpretation of comparison operators may differ in \mongodb query predicates and \mquery expressions,
namely the way query predicates are evaluated against arrays.
We already explained and illustrated this in Section~\ref{sec:diff_queryPredicates}.

\subsubsection{Sorting}
The \mongodb $\texttt{\$sort}$ operator is very similar to the sort operator of \mquery.
However, their behaviors differ when comparing objects based on arrays present in each object.
For instance, assume that the collection \texttt{bands} consists of the three following documents:
\begin{lstlisting}
{ "_id": 1, "name": "Patti Smith group", "tours": [1985, 1976] },
{ "_id": 2, "name": "Queen", "tours": [1979, 1983] },
{ "_id": 3, "name": "ABBA", "tours": [1982, 1978] },
\end{lstlisting}
The following aggregate query sorts bands according to the value of the attribute \texttt{tours}:
\begin{lstlisting}
db.bands.aggregate([
    { $sort : { "tours": 1 } }
])
\end{lstlisting}
The analogous \mquery is

\begin{minipage}[t]{0.5\textwidth}
\mquerybox{
$\texttt{bands} \pipeline \sort{\texttt{tours}^+}$
}
\end{minipage}

The semantics of \mquery dictates that the arrays in each document are compared lexicographically:
compare the first entries, then in case of a tie the second entries, etc.
So in this example, the \mquery returns the objects for Queen, ABBA and the Patti Smith group, in this order, because $1979 < 1982 < 1985$.

The behavior of \mongodb is different.
From our (empirical) observations, 
it seems like \mongodb in such a case compares the arrays \ei by least value.
For instance, in this example the aggregate query returns the objects for the Patti Smith group, ABBA and Queen, in this order,
because the least date for the Patti Smith group (1976) is inferior to the least date for ABBA (1978), which is inferior to the least date for Queen (1979).

Besides being harder to understand, the strategy applied by \mongodb violates elementary properties expected from a 
sorting procedure.
One of them is that a collection sorted in ascending order should be symmetric to the same collection sorted in descending order.
For instance, the aggregate query
\begin{lstlisting}
db.bands.aggregate([
    { $sort { "tours": -1 } }
]);
\end{lstlisting}
is identical to the previous one, but sorts in descending order.
It returns the objects for the Patti Smith group, Queen and ABBA, in this order.

So if $o_i$ is the object with $\id\ i$ and if $\preceq$ is the preorder induced by comparing the values of the attribute \texttt{tours},
then according to the first aggregate query:
\[
o_1 \preceq o_3 \preceq o_2
\]
Whereas according to the second aggregate query:
\[
o_1 \succeq o_2 \succeq o_3
\]
So in this example, $\preceq$ is not the inverse of $\succeq$.
% (if it were, then 
%
% By transitivity (e.g.~$o_1 \preceq o_2 \preceq o_3 \preceq o_1$),
% our precedence relation collapses, meaning that all three bands are equivalent w.r.t.~this ordering.

\subsection{Data-independent Semantics}
\label{sec:data-independent}
The semantic differences discussed above did not address another, 
more fundamental question about query evaluation:
\emph{can we always reason about query behavior without examining the actual data in collections?}

This question is crucial for query optimization and formal analysis.
Classical database theory strongly favors the idea that query semantics should be
determined solely by the query's logical structure, and not adapt dynamically based on the data
encountered during execution. The principle of \emph{data independence} allows optimizers
to apply algebraic transformations based purely on query structure,
without considering the content of a specific database instance.

\mquery adopts the data-independent approach,
whereas \mongodb's implementation exhibits some data-dependent behaviors,
where query semantics can change based on the actual values encountered during execution.
Specifically,
when evaluating a (syntactically valid) aggregate query,
\mongodb may throw a runtime exception if the database instance does not comply with assumptions made in this query.
This means that the query may or may not be semantically valid, depending on the current state of the database.

We provide a simple illustration where this fundamental difference may affect query reasoning and optimization.
% (which can improve usability at the cost of formal predictability).
%
The follow aggregate query retrieves bands for which the attribute \texttt{origin} is an array, and contains the value “Japan”.
\begin{lstlisting}
db.bands.aggregate([
    { $match: { $expr: { $in: ["Japan", "$origin"] } } }
])
\end{lstlisting}
As expected, it retrieves the following document:
\begin{lstlisting}
 { "_id": 1, "name": "Gorillaz", "formation": 1998, "origin": ["UK", "Japan", "US"] }
\end{lstlisting}
Now let us add to our collection another document, for the band Queen:
\begin{lstlisting}
 { "_id": 2, "name": "Queen", "formation": 1970, "origin" : "UK" } 
\end{lstlisting}
After adding this second document to the database, the query above does not retrieve the first one (for Gorillaz) anymore.
Instead, it throws an exception, because the attribute \valuefont{origin} in the second document does not evaluate to an array.

Note that in this example, the query is valid, and so is the data.
However, their combination is not.
This makes any attempt to formalize the semantics of aggregate queries significantly more complex: a query may or may not have a meaning,
depending on the context in which it is used.

We view such exceptions as potential obstacles to a good understanding of the formal properties of this language, notably how it compares (e.g. in terms of expressivity) to well-studied formalisms like (nested) relational algebra.
As a simple illustration, in most query languages (as well as data processing pipelines),
it is generally expected that two selections (also called filters) can permute or be combined.
For instance, consider the following query, which selects bands formed after 1990, and Japanese ones among these.

\begin{lstlisting}
db.bands.aggregate([
    { $match: { 
        $expr: { $ge: [ "$formation", 1990 ] }
    } },
    { $match: { 
        $expr: { $in: [ "Japan", "$origin" ] ] }
    } }
])
\end{lstlisting}
One may reasonably expect the order of these two filters to be irrelevant.
In other words, this query seems equivalent to the following one:
\begin{lstlisting}
db.bands.aggregate([
    { $match: { 
        $expr: { $in: [ "Japan", "$origin" ] ] }
    } },
    { $match: { 
        $expr: { $ge: [ "$formation", 1990 ] }
    } }
])
\end{lstlisting}
Besides, a more idiomatic syntax for these two filters would be a Boolean conjunction:
\begin{lstlisting}
db.bands.aggregate([
    { $match: { 
        $expr: { $and: [
            $ge: [ "$formation", 1990 ], 
            $in: [ "Japan", "$origin" ]
        ] }
    } }
])
\end{lstlisting}
However, these three queries behave differently over our collection.
The first one retrieves the document for Gorillaz, whereas the two latter throw an exception.
In contrast, the following three \mqueries are equivalent:

\begin{minipage}[t]{0.5\textwidth}
\mquerybox{
\begin{align*}
  \texttt{bands}\ \pipeline&\ \match{\ \textit{"Japan"}\, \in\, \texttt{origin}}\ \pipeline\ \match{\ \texttt{formation}\, \ge\, 1990}\\
  \texttt{bands}\ \pipeline&\ \match{\ \texttt{formation}\, \ge\, 1990}\ \pipeline\ \match{\ \textit{"Japan"}\, \in\, \texttt{origin}}\\
  \texttt{bands}\ \pipeline&\ \match{\ \texttt{formation}\, \ge\, 1990\ \land\ \textit{"Japan"}\, \in\, \texttt{origin} }
\end{align*}
}
\end{minipage}

\section{Algebraic Optimisation}
\label{sec:optimisation}
% This section focuses on optimizing the execution of \mqueries.
% In Section~\ref{sec:algebraic_opt},
% we list algebraic transformations that preserve the semantics of an \mquery, and discuss their benefits.
% Section~\ref{sec:joins} focuses instead on the relation between \mquery's lookup operator and more conventional join operations (which are the traditional focus of query optimisation in relational databases).

% as opposed to more costly (typically untractable, or even undecidable) procedures needed to analyze a non-normalized query.

% \subsection{Algebraic transformations}
% \label{sec:algebraic_opt}
We introduce in this section a variety of algebraic transformations that can be applied to an \mquery, while preserving its semantics.
Our purpose is to lay foundations for algebraic query optimisation.
For instance,
when evaluating a query,
one may want to filter documents as early as possible,
in order to reduce the size of the input to an expensive downstream operation.
% The transformed query may also take advantage of indexes declared over the source collections for operations like filtering or joins.
For the same reason, it is usually preferable to unnest arrays as late as possible.

Algebraic optimisation may not be needed for a simple query, especially if manually crafted by an experienced user:
one may expect the structure of such a query to be already close to optimal.
However, 
as the complexity of queries increases,
automated procedures to simplify or just reorganize them may become a necessity.
In particular, 
\mongodb being based on JSON, it is not unrealistic to foresee an increasing 
number of applications (let alone agentic workflows) rely on automatically generating and combining \mongodb aggregate queries.
And as is often the case with automatically generated "code", such queries may not be in a form suitable for efficient execution.

Besides, algebraic manipulations open the possibility to put queries into some \emph{normal form},
which can have multiple benefits for an evaluation engine.
For instance, this may ease the search for efficient query execution plans.
A normal form may also allow implementing efficient (albeit incomplete) strategies to identify redundancies or inconsistencies in a query (e.g.~unsatisfiable subqueries).
\mongodb \mversion already applies such transformations, some of which are (informally) documented.\footnote{
  \href{https://www.mongodb.com/docs/manual/core/aggregation-pipeline-optimization/}{https://www.mongodb.com/docs/manual/core/aggregation-pipeline-optimization/}
}

In this section, we provide a precise description of a set of transformations that are possible in \mquery,
focusing on the core fragment of Grammar~\ref{eq:mquery-grammar} (match, project, unwind, union, lookup and graphLookup).
Before we provide these transformations, 
Section~\ref{sec:transform_notation} introduces additional notation needed to formulate them.
The transformations themselves are described in the following sections,
each of which focuses on a type of \mquery stages (match , unwind, etc.).

\subsection{Notation}
\label{sec:transform_notation}
\paragraph{Equivalence}
Our algebraic manipulations transform a (sub)query into an equivalent one.
We will use the symbol $\equiv$ to denote query equivalence.
For instance, we write 
\[s_1 \pipeline s_2 \equiv s_2 \pipeline s_1\]
to state that the stage $s_1$ can freely be executed before or after the stage $s_2$.
Formally, this means that for any collection $O$ and DB instance $I$,
\[
  \evalcdb{s_1 \pipeline s_2} = \evalcdb{s_2 \pipeline s_1 }
\]
Because the semantics of \mqueries is compositional, this implies that these two stages can also commute within a more complex query.
% In other words, we could have written here
% \[
%   \evalcdb{q_1 \pipeline s_1 \pipeline s_2 \pipeline q_2}
% =
% \evalcdb{q_1 \pipeline s_2 \pipeline s_1 \pipeline q_2}
% \]
% with $q_1, q_2$ arbitrary pipelines.
More generally,
\[
  q \equiv q'\ \te{implies}\ 
  q_1 \pipeline q \pipeline q_2
  \equiv
  q_1 \pipeline q' \pipeline q_2
\]
for any \mqueries $q, q', q_1$ and $q_2$.

\paragraph{Paths and prefixes}
Most algebraic equivalences listed below are conditioned by properties of the paths used in the initial query (as well as prefixes of these paths).
For instance, as we will see, an unwind stage $\unwind{p}$ may be postponed after a match stage $\match{\varphi}$ if none of the paths used in $\varphi$ is a prefix of $p$ or extends $p$.
In order to express this, we introduce a dedicated notation to refer to paths and prefixes.

As a reminder, we say that two paths are compatible if none of them is a prefix of the other.
We will also say below that a path $p$ is \emph{compatible with a set} $P$ of paths if $p$ is compatible with every $p'$ is $P$.

Sometimes we will only need to refer to the prefixes of a path $p$, or only to its extensions.
We use $\prefixes{p}$ for the set of prefixes of $p$ (including the empty path $\varepsilon$ and $p$ itself).
For instance, $\prefixes{\mathtt{a.b.2.c}} = \{\varepsilon, \mathtt{a}, \mathtt{a.b} , \mathtt{a.b.2}, \mathtt{a.b.2.c} \}$.
We will also use $\extensions{p}$ for the (infinite) set of all paths that extend $p$.
For instance, $\extensions{\mathtt{a.b}}$ contains (among others) the paths $\mathtt{a.b.c}$, $\mathtt{a.b.c.2}$ and $\mathtt{a.b.1.d}$.
We technically do not need both notations, since $\extensions{p_1} = \{p_2 \mid p_1 \in \prefixes{p_2}\}$.
But one may be more convenient than the other in certain contexts.
We also lift these notations to a set $P$ of paths, i.e.
\begin{align*}
  \prefixes{P} & = \bigcup_{p \in P} \prefixes{p}\\
  \extensions{P} & = \bigcup_{p \in P} \extensions{p}
\end{align*}
Note that “$p$ is compatible with $P$” is a shortcut for $P \cap (\prefixes{p} \cup \extensions{p}) = \emptyset$.
% We will make repeated use of conditions of the form 
% "none of the paths in $P$ is a prefix of $p$ or has $p$ as prefix",
% where $p$ is a path and $P$ a set of paths.
% So in order to keep our notation concise, we introduce a dedicated notation $\nopref{p}{P}$ to indicate that this condition holds, i.e.:
% \[
%   \nopref{p}{P}
%   \quad 
%   \te{iff} 
%   \quad
%    P \cap (\prefixes{p} \cup \extensions{p}) = \emptyset
% \]

Next, if $\tau$ is a term (which, as a reminder, may be a Boolean expression),
we use $\paths{\tau}$ for the set of paths used in $\tau$.
For instance,
$\paths{\mathtt{a.b} = 5 \land \mathtt{c} \neq \mathtt{a.c} } = \{\mathtt{a.b}, \mathtt{c}, \mathtt{a.c} \}$.
However, we need to clarify what $\paths{\tau}$ means when $\tau$ contains the $\mapz$ or the $\filterz$ operator,
because this operator can use "relative" paths.
Intuitively, $\paths{\tau}$ is the set of absolute paths used in $\tau$, possibly reconstructed from relative ones.
The full inductive definition of $\paths{\tau}$ is mechanical, and provided in Appendix~\ref{sec:appendix_paths}.
We only reproduce here the case of the $\mapz$ operator (the case of $\filterz$ is analogous):
\[\paths{\map{p}{x}{\tau}} = \{p.p' \mid p' \in \paths{\tau} \}\]

% We also lift our notation for prefixes to a term $\tau$:
% \begin{align*}
%   \prefixes{\tau} & = \bigcup_{p\, \in\, \paths{\tau}} \prefixes{p}\\
% \end{align*}

Finally,
we use $\tau[p_1/p_2]$ for the term identical to $\tau$,
but where each occurrence of the path $p_1$ is replaced with the path $p_2$.

\paragraph{Paths definitions}
As a reminder, a path definition is a pair $p/\tau$, where $p$ is an index-free path and $\tau$ a term.
A special case is the one where $\tau = p$.
If $D$ is a sequence of path definitions, we use $\idle{D}$ for the set of such definitions in $D$,
each represented as a single path, i.e:
\[\idle{D} = \{p \mid p/p \in D\}\]
We will also use $\rightsplit{D}$ for the set of paths definitions in $D$,
% and $\rightsplit{D}$ for their definitions,
% We will also use $\leftsplit{D}$ for the set of paths defined in $D$,
% and $\rightsplit{D}$ for their definitions,
i.e.
\begin{align*}
  % \leftsplit{D} & = \{p \mid p/\tau \in D\}\\
  \rightsplit{D} & = \{\tau \mid p/\tau \in D\}
\end{align*}
And similarly for a set $X$ of variable definitions.

% (assuming that $p_1$ does not appear in $p_2$).

\subsection{Match}
\label{sec:transform_match}
Applying filters as early as possible is often desirable, for at least two reasons:
\begin{enumerate}
  \item when executed as the first operation of a pipeline, a filter many take advantage of indexes over the source collection,
  \item applying filters early can reduce the size of the collections input to other, potentially costly operations (such as grouping).
\end{enumerate}

\begin{figure}[htbp]
{
\renewcommand{\arraystretch}{1.4}
\[
\begin{array}{lll}
\toprule
\textbf{Pre-condition} & \textbf{Initial query} & \textbf{Transformed query} \\
\midrule
                       & \match{\varphi} \pipeline \match{\psi} & \match{\varphi \land \psi}\\

% [0.5pt/1pt]   % dotted line across columns 1‑3
\begin{aligned}
  % \nopref{p}{\paths{\varphi}}
  p \te{is compatible with} \paths{\varphi}
\end{aligned}
                       & \unwind{p} \pipeline \match{\varphi \land \psi} 
                       & \match{\varphi} \pipeline \unwind{p} \pipeline \match{\psi} \\
\begin{aligned}
  % \nopref{p}{\paths{\varphi}}
  p \te{is compatible with} \paths{\varphi}
\end{aligned}

                       & \lookup{X,t}{p} \pipeline \match{\varphi \land \psi} 
                       & \match{\varphi} \pipeline \lookup{X,t}{p} \pipeline \match{\psi} \\
\begin{aligned}
  \paths{\varphi} \subseteq  \extensions{\idle{D}}
\end{aligned}
                       & \project{D} \pipeline \match{\varphi \land \psi} 
                       & \match{\varphi} \pipeline \project{D} \pipeline \match{\psi} \\
\begin{aligned}
  \paths{\varphi} \subseteq  \extensions{\id}
\end{aligned}
                       & \group{g}{A} \pipeline \match{\varphi \land \psi} 
                       & \match{\varphi[\id/g]} \pipeline \group{g}{A} \pipeline \match{\psi} \\
\bottomrule
\end{array}
\]
\caption{Algebraic transformations enabling anticipated filtering.}
\label{tab:transform_match}
}
\end{figure}

Accordingly,
Figure~\ref{tab:transform_match} provides a series of simple algebraic transformations that allow anticipating the execution of filters.
To explain how to read this table, we focus on the second row.
% of the unwind operator (second line in Table~\ref{tab:transform_match}).
This row states that for any path $p$ and Boolean expressions $\varphi$ and $\psi$, 
if 
% $\nopref{p}{\paths{\varphi}}$ holds, 
$p$ is compatible with $\paths{\varphi}$,
then:
\[
  \unwind{p} \pipeline \match{\varphi \land \psi} 
  \equiv 
  \match{\varphi} \pipeline \unwind{p} \pipeline \match{\psi} 
\]

In practice, this means that if this pre-condition on paths and prefixes is satisfied, 
then the (sub)query
$\unwind{p} \pipeline \match{\varphi \land \psi}$
can be safely transformed into 
$\match{\varphi} \pipeline \unwind{p} \pipeline \match{\psi}$.
If $\unwind{p} \pipeline \match{\varphi \land \psi}$ was the beginning of the initial pipeline,
then the filtering condition $\varphi$ may now take advantage of indexes declared over the source collection.
If not, then this filter may be propagated further upstream, by applying the operations of this table inductively.
A specific case of this transformation is when the whole Boolean expression of the match stage satisfies the pre-condition
(or equivalently when $\psi = \mathtt{true}$).
In this case, the match and unwind stages fully commute i.e.~$\unwind{p} \pipeline \match{\varphi}$ can be rewritten $\match{\varphi} \pipeline \unwind{p}$.

We emphasize that the preconditions listed in this table (as well as in the following sections) are sufficient for the equivalence to hold, but not necessary.
For instance, in this example,
there may be $q, p, \varphi$ and $\psi$ such that 
\[q \pipeline  \unwind{p} \pipeline \match{\varphi \land \psi} \equiv q \pipeline \match{\varphi} \pipeline \unwind{p} \pipeline \match{\psi}\]
even though the pre-condition
% \[  \nopref{p}{\paths{\varphi}} \]
($p \te{is compatible with} \paths{\varphi}$)
is not met.
Deciding whether two consecutive stages can permute within a query is in general complex (the problem is intractable for most types of stages),
and may require analysing the whole query.
Instead, our objective here is to provide simple pre-conditions that can be verified locally (focusing on two stages only) and efficiently:
all these conditions can be verified in $O(n \log n)$, and this also holds in the following sections.

\subsection{Unwind}
It is usually preferable to unnest arrays as late as possible during the execution of a query,
so as to limit the size of intermediate collections.
Accordingly, Figure~\ref{tab:transform_unwind} provides transformations that postpone an unwind stage.
% This table is meant be read like the previous one, as explained in Section~\ref{sec:transform_match}.

% The last equivalence in this table is marked with a \dag\ symbol.
% This indicates (here and in further sections) that the transformation may alter the order of key-value pairs in each output object.
% Or in other words, this equivalence only holds under the unordered semantics.

\begin{figure}[htbp]
{
\renewcommand{\arraystretch}{1.4}
\[
\begin{array}{lll}
\toprule
\textbf{Pre-condition} & \textbf{Initial query} & \textbf{Transformed query} \\
\begin{aligned}
  p_1 \not\in \prefixes{p_2}
\end{aligned}
                       & \unwind{p_1} \pipeline \unwind{p_2}
                       & \unwind{p_2} \pipeline \unwind{p_1}\\
\begin{aligned}
  % \nopref{p}{\paths{\varphi}}
  p \te{is compatible with} \paths{\varphi}
\end{aligned}
                       & \unwind{p} \pipeline \match{\varphi \land \psi} 
                       & \match{\varphi} \pipeline \unwind{p} \pipeline \match{\psi} \\
\begin{aligned}
  % \nopref{p_1}{\paths{\rightsplit{X}} \cup \{p_2\}}
  p_1 \te{is compatible with} \paths{\rightsplit{X}} \cup \{p_2\} 
\end{aligned}
                       & \unwind{p_1} \pipeline \lookup{X,t}{p_2}
                       &  \lookup{X,t}{p_2} \pipeline \unwind{p_1}\\
\begin{aligned}
  % \nopref{p}{\paths{\rightsplit{D}}}
  p \te{is compatible with} \paths{\rightsplit{D}}
  \te{and} p \in \extensions{\idle{D}}
\end{aligned}
                       & \unwind{p} \pipeline \project{D}
                       & \project{D} \pipeline \unwind{p} \\
\bottomrule
\end{array}
\]
\caption{Algebraic transformations enabling postponed unnesting (the second row is reproduced from Figure~\ref{tab:transform_match}).
}
\label{tab:transform_unwind}
}
\end{figure}

\subsection{Project}
% Projections are typically postponed when possible,
% which may first come as a surprise.
% The reason is that the efficiency gain of an early projection is usually marginal (if any), 
% whereas a query with a unique (e.g. final) projection can be easier to read or analyze.
% Besides, for our specific purpose, pushing project stages down the pipeline has another benefit:
Figure~\ref{tab:transform_project} lists transformations that postpone (i.e.~push “down” the pipeline) a project stage.
This may enable pushing down other stages (e.g.~an unwind stage) that would be "blocked" by the projection otherwise.

\begin{figure}[htbp]
{
\renewcommand{\arraystretch}{1.4}
\[
\begin{array}{lll}
\toprule
\textbf{Pre-condition} & \textbf{Initial query} & \textbf{Transformed query} \\
\begin{aligned}
  \paths{\rightsplit{D_2}} \subseteq \extensions{\idle{D_1}}
\end{aligned}
                       & \project{D_1} \pipeline \project{D_2}
                       & \project{D_2} \\
\begin{aligned}
  \paths{\varphi} \subseteq  \extensions{\idle{D}}
\end{aligned}
                       & \project{D} \pipeline \match{\varphi \land \psi} 
                       & \match{\varphi} \pipeline \project{D} \pipeline \match{\psi} \\
{\setlength{\jot}{0pt}
\begin{aligned}
  % \nopref{p}{\paths{\rightsplit{D}}}
  & p \te{is compatible with} \paths{\rightsplit{D}} \te{and}\\
  &\qquad p \in \extensions{\idle{D}}
\end{aligned}
}
                       & \project{D} \pipeline \unwind{p}
                       & \unwind{p} \pipeline \project{D}\\
{\setlength{\jot}{0pt}
\begin{aligned}
% \nopref{p}{\paths{\rightsplit{D}}}
& p \te{is compatible with} \paths{\rightsplit{D}} \te{and}\\
& \qquad \paths{\rightsplit{X}} \cup \{p\} \subseteq  \extensions{\idle{D}} 
\end{aligned}
}
                       & \project{D} \pipeline \lookup{X,t}{p}
                       & \lookup{X,t}{p} \pipeline \project{D} \\
\begin{aligned}
\{A \cup \{g\}\} \subseteq \idle{D}
\end{aligned}
                       & \project{D} \pipeline \group{g}{A}
                       & \group{g}{A} \pipeline \project{(\id) \cdot A} \\
\bottomrule
\end{array}
\]
\caption{Algebraic transformations enabling postponed or simplified projections (the second row is reproduced from Figure~\ref{tab:transform_match}).
}
\label{tab:transform_project}
}
\end{figure}

Note that at first sight,
the third row (swapping project with unwind) 
seems to cancel out a transformation listed in the previous section (swapping unwind with project).
But this is not a contradiction: within a more general optimisation procedure, both transformations may be useful.
For instance, one may optimize a query by first pushing all project stages downstream (iteratively, until none can be),
then all unwind stages downstream, and then all match stages upstream.

\subsection{Lookup}

The case of the lookup stage is more nuanced.
Two possibly conflicting optimisation strategies may apply:
\begin{enumerate}
  \item Postponing the execution of a lookup stage is likely to reduce the size of intermediate query results.
  \item However, a lookup executed as the very first operation of a pipeline can benefit from indexes declared over a source collection.
\end{enumerate}

So if a query contains a lookup stage, one may first attempt to move it upstream, to see if it can be executed as the very first operation.
And if not, then try to postpone it instead.
For this reason, Figure~\ref{tab:transform_lookup} provides transformations that allow stages to permute with lookup in both directions.

\begin{figure}[htbp]
{
\renewcommand{\arraystretch}{1.4}
\[
\begin{array}{lll}
\toprule
\textbf{Pre-condition} & \textbf{Initial query} & \textbf{Transformed query} \\
{\setlength{\jot}{0pt}
\begin{aligned}
% \nopref{p_1}{\paths{\rightsplit{X_2}} \cup \{p_2\}}
&p_1 \te{is compatible with} \paths{\rightsplit{X_2}} \cup \{p_2\} \te{and}\\
% \nopref{p_2}{\paths{\rightsplit{X_1}}}
&p_2 \te{is compatible with} \paths{\rightsplit{X_1}}
\end{aligned}
}
                       & \lookup{X_1,t_1}{p_1} \pipeline \lookup{X_2,t_2}{p_2}
                       & \lookup{X_2,t_2}{p_2} \pipeline \lookup{X_1,t_1}{p_1} \\
\begin{aligned}
  % \nopref{p}{\paths{\varphi}}
p \te{is compatible with} \paths{\varphi}
\end{aligned}

                       & \lookup{X,t}{p} \pipeline \match{\varphi \land \psi} 
                       & \match{\varphi} \pipeline \lookup{X,t}{p} \pipeline \match{\psi} \\
\begin{aligned}
  % \nopref{p}{\paths{\varphi}}
p \te{is compatible with} \paths{\varphi}
\end{aligned}
                       & \match{\varphi} \pipeline \lookup{X,t}{p}
                       & \lookup{X,t}{p}  \pipeline \match{\varphi} \\
\begin{aligned}
  % \nopref{p_1}{\paths{\rightsplit{X}} \cup \{p_2\}}\\
p_1 \te{is compatible with} \paths{\rightsplit{X}} \cup \{p_2\}
\end{aligned}
                       & \unwind{p_1} \pipeline \lookup{X,t}{p_2}
                       &  \lookup{X,t}{p_2} \pipeline \unwind{p_1}\\
\begin{aligned}
  % \nopref{p_2}{\paths{\rightsplit{X}} \cup \{p_1\}}\\
p_2 \te{is compatible with} \paths{\rightsplit{X}} \cup \{p_1\}
\end{aligned}
                       &   \lookup{X,t}{p_1}  \pipeline \unwind{p_2}
                       &  \unwind{p_2}   \pipeline \lookup{X,t}{p_1} \\
{\setlength{\jot}{0pt}
\begin{aligned}
% \nopref{p}{\paths{\rightsplit{D}}}
&p \te{is compatible with} \paths{\rightsplit{D}} \te{and}\\
&\paths{\rightsplit{X}} \cup \{p\} \subseteq  \extensions{\idle{D}} 
\end{aligned}
}
                       & \project{D} \pipeline \lookup{X,t}{p}
                       & \lookup{X,t}{p} \pipeline \project{D} \\

{\setlength{\jot}{0pt}
\begin{aligned}
% \nopref{p}{\paths{\rightsplit{D}}}
&p \te{is compatible with} \paths{\rightsplit{D}} \te{and}\\
&\paths{\rightsplit{X}} \cup \{p\} \subseteq  \extensions{\idle{D}} 
\end{aligned}
}
                       &   \lookup{X,t}{p} \pipeline \project{D}
                       &  \project{D}  \pipeline  \lookup{X,t}{p}\\
\bottomrule
\end{array}
\]
\caption{Algebraic transformations enabling anticipated or postponed lookups (Rows 2, 4 and 6 are reproduced from previous figures).
}
\label{tab:transform_lookup}
}
\end{figure}

% \susubsection{GraphLookup}
%
% \subsubsection{Union}
% We have not identified simple conditions that allow a union stage to permute with another stage in a useful way.
%
% \subsubsection{Group}
% We only identified two useful transformations with simple and local pre-conditions that involve a group stage.
% They have already been provided in Figures~\ref{tab:transform_match} and~\ref{tab:transform_project}.

% Finally, observe that this table does not contain an entry for all sequences of to stages that involve a match stage.
% This is by design: missing combioations can have two reasons:
%
% \begin{itemize}
%   \item we did not see a practical benefit in certain stage permutations.
%     For instance, in this table, we did not consider  postponing a match stage.
%   \item no (non-trivial) pre-condition for permutatiuon exists that can be ckecked locally (i.e. involves only two stages).
%   For instance, the 
% \end{itemize}

% \subsection{Joins}
% \label{sec:joins}

%

%%% Local Variables:
%%% mode: latex
%%% TeX-master: "main-tr"
%%% fill-column: 79
%%% End:

\newpage
\bibliography{bibl}

\newpage
\appendix
\section{Appendix}
\label{sec:appendix}

% \subsection{Evaluation of a path, \nullvalue and \missingvalue}
% \label{app:path-null-missing}

\subsection{Summary of notation}
\nb{JC: consider presenting this table in landscape mode.}
\begin{figure}[htbp]
  \begin{center}
\begin{tabular}{cr|cr}
\toprule
\textbf{Symbol} & \textbf{Meaning} & \textbf{Symbol} & \textbf{Meaning} \\
\midrule
\hline
% \textbf{Item} & \textbf{Quantity} & \textbf{Price (\$)} \\ \hline
$\ell$        &  a literal  &
$\L$     &  universe of literals \\
$v$        &  a \dvalue &
$\V$        &  universe of \dvalues\\
$k$        &  a key  &
$\K$        &  universe of keys\\
$x$        &  a variable  &
$\X$        &  universe of variables\\
$a = \arraybr{v_1, \ldots, v_n} $      &  an array  &
$\A$        &  universe of all arrays\\
$o = \objectbr{k_1 \mapsto v_1, \ldots k_n \mapsto v_n}$        &  an object &
$\O$        &  universe of all objects\\
$O = \bag{o_1, \ldots, o_n}$        &  a collection (bag of objects) &
$i,j$ &  indices\\
$p$        &  a path &   $o(p)$    &
value of path $p$ in object $o$\\
$p_1 . p_2$ & path concatenation  & 
$\varepsilon$  & the empty path \\
$\tau$ & a term & $\eval[v]{\tau}$    &
value of term $\tau$ in \dvalue $v$ \\
$\varphi$ & a Boolean expression &
$v \models \varphi$    & condition $\varphi$ holds in \dvalue $v$\\
% $f$  & a function name & $f^*$ & a function\\
$C$    & a collection name&
$I$ & a database instance\\
$s$ & a stage &
$q = C \pipeline s_1 \pipeline \ldots \pipeline s_n $  & an \mquery/pipeline \\
$O' = \evalcdb{q}$    & evaluation of $q$ over $O$ &
$a' = \evalcdbl{q}$    & evaluation of $q$ over $a$\\
$p /\tau$ & a path definition & $D$ & a sequence of path definitions\\
$x /\tau$ & a variable definition & $X$ & a sequence of variable definitions\\
$g$ & a grouping key & $A$ & a sequence of keys\\
$t$ & a query template & $o[p/v]$ & overriding path $p$ with value $v$ in $o$\\
$p^+,p^-$ & object comparators & $S$ & a sequence of object comparators\\
$\project{D}$ & a project stage & $\unwind{p}$ & an unwind stage\\
$\match{\varphi}$ & a match stage & $\group{g}{A}$ & a group stage\\
$\lookup{X,t}{p}$ & a lookup stage & $\glookup{p,C,p,p}{p}$ & a graphLookup stage\\
$\unionwith{q}$ & a union stage & $\sort{S}$ & a sort stage\\
\bottomrule
\end{tabular}
  \end{center}
  % \caption{}\label{fig:}
\end{figure}

\subsection{Example Collections}

\begin{figure}[h]
  \caption{The $\valuefont{bands}^I$ collection.}
  \label{fig:bands-collection}
  \centering
\begin{lstlisting}
{   "_id": 2,
    "name": "Queen",
    "formation": 1970,
    "albums": [
        { "title": "Queen", "release": 1973 },
        { "title": "A Night at the Opera", "release": 1975, "length": "43:08" },
        { "title": "News of the World", "release": 1977, "labels": ["EMI", "Elektra"] }
    ],
    "members": [
        { "name": "Freddie Mercury", "role": ["lead vocals", "piano"] },
        { "name": "Brian May", "role": ["guitar", "vocals"] },
        { "name": "Roger Taylor", "role": ["drums", "vocals"] },
        { "name": "John Deacon", "role": "bass" }
    ]
},
{   "_id": 3,
    "name": "ABBA",
    "formation": 1972,
    "albums": [
        { "title": "Waterloo", "release": 1974, "length": "38:09" },
        { "title": "ABBA", "release": 1975, "labels": ["Polar", "Epic", "Atlantic"] }
    ],
    "members": [
        { "name": "Agnetta Faltskog", "role": "lead vocals" },
        { "name": "Björn Ulvaeus", "role": ["guitar", "vocals"] },
        { "name": "Benny Andersson", "role": ["keyboard", "vocals"] },
        { "name": "Anni-Frid Lyngstad", "role": "vocals" }
    ]
}  
\end{lstlisting}
\end{figure}

\begin{figure}
\centering
\begin{minipage}{0.58\linewidth}
  \begin{lstlisting}
{   "_id": 1,
    "title": "One night in Bangkok",
    "composers": ["Björn Ulvaeus"],
    "interprets": ["Murray Head"] 
},
{   "_id": 2,
    "title": "SOS",
    "composers": ["Björn Ulvaeus", "Benny Andersson"],
    "interprets": ["ABBA", "Portishead", "Cher"] 
},
{   "_id": 3,
    "title": "Gloria",
    "composers": ["Van Morisson"],
    "interprets": ["Them", "Patti Smith Group"] 
},
{   "_id": 4,
    "title": "Under pressure",
    "composers": ["David Bowie", "Freddy Mercury"],
    "interprets": ["Queen", "David Bowie"]
},
  \end{lstlisting}
\end{minipage}
\begin{minipage}{0.4\linewidth}
  \begin{lstlisting}[framexbottommargin=7mm]

{   "_id": 5,
    "title": "Ice Ice Baby",
    "interprets": ["Vanilla Ice"],
    "samples": 4
},

{   "_id": 6,
    "title": "Bambi",
    "interprets": ["BB TRickz"],
    "samples": 5
},

{   "_id": 7,
    "title": "7th Street",
    "interprets": ["Chinese Man"],
    "samples": 4
}
  \end{lstlisting}
\end{minipage}
\label{fig:coll-songs-appendix}
\caption{The $\valuefont{songs}^I$ collection.}
\end{figure}

\subsection{Natural Order}
\label{sec:appendix_order}
In this section, we instantiate the natural order $\le$ over the universe $\V$ of \dvalues,
which was left underspecified in the main body of this report.
This order comes in two flavours, one for the ordered semantics, and one for the unordered semantics.
The former is meant to comply with the order defined in the \mongodb documentation (unless stated explicitly).
And the latter is an adaptation of the former, with minimal modifications to accommodate unordered objects.
%
% Then we clarify how this order can be interpreted under the ordered and ordered semantics.

\subsubsection{Ordering Literals}
Our set $\L$ of literals is an abstraction.
In practice, a literal has a datatype (e.g. int, long, string or Boolean).
MongoDB's datatypes are called \emph{BSON types},
and the documentation of BSON types specifies how these literals are naturally ordered.
For instance, strings are ordered lexicographically (as expected), and numbers are ordered in the natural way.
In addition, the BSON specification indicates how these orders extend to literals of seemingly incompatible types.
For instance, a number always precedes a string.
% and $\nullvalue$ precedes all literals.
%

When it comes to \mqueries, we only assume some (underspecified) natural order $\le$ over $\L$,
the only requirement being that $\le$ is indeed a total order (i.e.~a total, reflexive, transitive and antisymmetric binary relation over $\L$).

\subsubsection{Ordering Arrays and Objects}
\label{sec:appendix_order}
The natural order $\le$ over the set $\L$ of literals can be lifted to the universe $\V$ of all \dvalue{s}.
To do so, we assume that the universe $\K$ of keys is also equipped with a total order $\preceq_\K$.
% with least element $\id$.
%
We extend $\le$ to a total order over $\V$ inductively,
as follows:
\begin{description}
  \item[Across types.] If $l \in \L$, $o \in \O$ and $a \in \A$,
    then $l < o < a$.
% \deltabox{
% In MongoDB, $[] < \nullvalue$
% }
  \item[Arrays.] $\A$ is ordered lexicographically: if $a, a' \in \A$, then $a < a'$ iff either
    \begin{itemize}
      \item $a[k] < a'[k]$, where $k$ is the least index such that $a[k] \neq a'[k]$, or
      \item $a$ is a strict prefix of $a'$, i.e. if $a$ has length $n$, then $a'$ has length at least $n+1$ and $a[i] = a'[i]$ for $i \in \{0, \ldots, n-1\}$.
    \end{itemize}

  \item[Objects.]
We first lift the order $\le$ to the set $\K \times \V$ of key/\dvalue pairs,
according to the behavior of MongoDB.
Precisely,
\begin{align*}
  (k_1, v_1) < (k_2, v_2) \te{iff either}\\
 & v_1 \te{and} v_2 \te{have different types} \te{(e.g.} v_1 \in \L \te{and} v_2 \in \O) \te{and} v_1 < v_2, \te{or}\\
 & v_1 \te{and} v_2 \te{have the same type and either}\\
 & \qquad k_1 \prec_K k_2, \te{or} \\
 & \qquad  k_1 =_K k_2 \te{and} v_1 < v_2.
\end{align*}
\medskip
Then we lift this order over $\K \times \V$ to an order over $\O$, for each of the two semantics:
\begin{itemize}
  \item 
    Under the ordered semantics (where a object is an \emph{sequence} of key/\dvalue pairs),
$\O$ is ordered lexicographically, analogously to $\A$.
\item 
  Under the ordered semantics (where a object is a \emph{set} of key/\dvalue pairs),
when comparing two objects,
we simply compare their sorted version (under the ordered semantics).
Precisely, if $o$ is an object under the unordered semantics,
let $\sorted{o}$ denote the sorted version of $o$,
i.e. the sequence that consists of the same key/\dvalue pairs as $o$,
sorted according to $\le$.
Then for $o_1$ and $o_2 \in \O$,
\begin{align*}
  o_1 &\le o_2 \te{under the unordered semantics iff}\\
  \sorted{o_1} &\le \sorted{o_2} \te{under the unordered semantics.}
\end{align*}

\end{itemize}

\end{description}

\subsection{Set of Paths in a Term}
\label{sec:appendix_paths}
We lift the notation $p_1 . p_2$ (for paths) to a set of path on the RHS, i.e.:
\[p.P = \{p.p' \mid p' \in P\}\]
for a path $p$ and set $P$ of paths.

We can now define the set $\paths{\tau}$ of paths in a term $\tau$, by induction on the structure of $\tau$:
\begin{itemize}
\item
$\paths{\arraybr{\tau_1,\dots,\tau_m}}= \bigcup_{i = 1}^m \paths{\tau_i}$
\item
 $\paths{\objectbr{k_1\mapsto\tau_1,\dots,k_m\mapsto\tau_m}} = \bigcup_{i = 1}^m k_i.\paths{\tau_i}$
\item 
$\paths{f(\tau)}= \paths{\tau}$
\item 
$\paths{\cond{\varphi}{\tau_1}{\tau_2}} = \paths{\varphi} \cup \paths{\tau_1} \cup \paths{\tau_2}$
\item
$\paths{\map{p}{x}{\tau}} = \{p.p' \mid p' \in \paths{\tau} \}$
\item
$\paths{\filter{p}{x}{\varphi}} = \{p.p' \mid p' \in \paths{\varphi} \}$
\end{itemize}

And similarly, we define $\paths{\varphi}$ for a Boolean expression $\varphi$:
\begin{itemize}
\item 
$\paths{p} = \{p\}$
\item 
$\paths{\exists p} = \{p\}$
\item
$\paths{\tau_1 \op \tau_2} = \paths{\tau_1} \cup \paths{\tau_2} \te{for}\op\in\{=,\leq, \in\}$
\item
$\paths{\neg \varphi} = \paths{\varphi}$
\item
$\paths{\varphi_1 \land \varphi_2} = \paths{\varphi_1} \cup  \paths{\varphi_2}$
\end{itemize}

% \subsection{Evaluation of a path inside an array}
% \label{app:path-inside-array}
%
%   TODO (Julien): explain how this differs from MongoDB's interpretation of paths/fields
%
%   Seems that a lot of weird behaviour of MongoDB wrt the comparisons stems from this flexible interpretations of fields.

%%% Local Variables:
%%% mode: latex
%%% TeX-master: "main-tr"
%%% fill-column: 79
%%% End:

\end{document}